\documentclass[preprint,showpacs,preprintnumbers,amsmath,amssymb,nofootinbib]{revtex4}

\usepackage{amsthm}

{Corollary}[section]

\usepackage{etex}
\usepackage{amscd,amsbsy,array}
\usepackage{bm}
\usepackage{amsmath,dsfont}
\usepackage{soul} 
\usepackage{graphics,graphicx,xcolor}
\usepackage[colorlinks=true, pdfstartview=FitV, linkcolor=blue, citecolor=blue, urlcolor=blue]{hyperref} %

\newcommand{\red}{\textcolor{red}}
\newcommand{\blue}{\textcolor{blue}}

\newcommand{\gb}{\colorbox{green}}

\newcommand{\dgreen}{\textcolor[rgb]{0,0.35,0}}
\newcommand{\cyan}{\textcolor{cyan}}

\newenvironment{redtext}{\color{red}}{\ignorespacesafterend}
\newenvironment{bluetext}{\color{blue}}{\ignorespacesafterend}

\newenvironment{magentatext}{\color{magenta}}{\ignorespacesafterend}
\newenvironment{orangetext}{\color{orange}}{\ignorespacesafterend}
\newenvironment{cyantext}{\color{cyan}}{\ignorespacesafterend}
\newcommand{\bblue}{\begin{bluetext}}
\newcommand{\eblue}{\end{bluetext}}
\newcommand{\bred}{\begin{redtext}}
\newcommand{\ered}{\end{redtext}}
\newcommand{\bmagenta}{\begin{magentatext}}
\newcommand{\emagenta}{\end{magentatext}}
\newcommand{\borange}{\begin{orangetext}}
\newcommand{\eorange}{\end{orangetext}}
\newcommand{\bcyan}{\begin{cyantext}}
\newcommand{\ecyan}{\end{cyantext}}

\numberwithin{equation}{section}
\renewcommand{\theequation}{\thesection.\arabic{equation}}
\let\ssection=\section
\renewcommand{\section}{\setcounter{equation}{0}\ssection}

\newcommand{\bA}{{\bf A}}
\newcommand{\cA}{{\mathcal{A}}}

\newcommand{\bb}{{\mathbf{b}}}
\newcommand{\bB}{{\mathbf{B}}}

\newcommand{\bbeta}{\boldsymbol{\beta}}

\newcommand{\fb}{\mathfrak{b}}
\newcommand{\fa}{\mathfrak{a}}

\newcommand{\fc}{\mathfrak{c}}
\newcommand{\fr}{\mathfrak{r}}

\newcommand{\bc}{{\mathbf{c}}}

\newcommand{\GW}{{gravitational wave\;}}
\newcommand{\GWs}{{gravitational waves\,}}
\newcommand{\SL}{{Sturm-Liouville\,}}

\newcommand{\diag}{\mathrm{diag}}

\newcommand{\cF}{{\mathcal{F}}}

\newcommand{\rg}{\mathrm{g}}

\newcommand{\bgamma}{\boldsymbol{\gamma}}

\newcommand{\bk}{\mathbf{k}}

\newcommand{\bomega}{{\boldsymbol{\omega}}}

\newcommand{\bchi}{{\boldsymbol{\chi}}}

\newcommand{\bp}{{\bf p}}

\newcommand{\bx}{{\bm{x}}}

\newcommand{\rot}{\vnabla\times}

\newcommand{\SO}{\mathrm{SO}}

\newcommand{\II}{\mathds{I}}

\newcommand{\Tr}{\mathrm{Tr}}

\newcommand{\hbx}{{\hat{\mathbf{x}}}}
\newcommand{\hu}{{\hat{u}}}

\newcommand{\hv}{\hat{v}}

\newcommand{\bX}{{\bf X}}

\newcommand{\dgamma}{\dot{\gamma}} 
\newcommand{\dchi}{\dot{\chi}}

\newcommand{\dx}{\dot{x}}
\newcommand{\da}{\dot{\fa}}
\newcommand{\db}{\dot{\fb}}
\newcommand{\ddx}{\ddot{x}}
\newcommand{\ddchi}{\ddot{\chi}}
\def\vnabla{{\bm{\nabla}}}
\def\bnabla{{\bm{\nabla}}}

\newcommand{\bY}{{\bf Y}}

\def\smallover#1/#2{\hbox{$\textstyle\frac{#1}{#2}$}} %

\def\Ort{{\rm O}}

\def\bp{{\bm{p}}}

\def\parag{\hfil\break} 
\def\kikezd{\parag\underbar}

\def\besub{\begin{subequations}}
\def\esub{\end{subequations}}
\def\benu{\begin{enumerate}}
\def\eenu{\end{enumerate}}
\def\beq{\begin{equation}}
\def\eeq{\end{equation}}
\def\beqa{\begin{eqnarray}}
\def\eeqa{\end{eqnarray}}
\def\nn{\nonumber}
\def\barray{\left(\begin{array}}
\def\earray{\end{array}\right)}
\def\barraynb{\begin{array}}
\def\earraynb{\end{array}}

\def\Ort{{\rm O}}

\def\IC{\mathfrak{C}}
\def\ID{\mathfrak{D}}
\def\IK{{\mathds{K}}}
\def\IP{{\mathfrak{P}}} 
\def\IQ{{\widetilde{\mathfrak{P}}}} 
\def\IS{{\mathfrak{S}}} 
\def\IIQ{{\widetilde{\mathfrak{Q}}}} 

\def\?{\quad{\gb{\fbox{\texttt{?}}\;}}\quad}
\def\p{{\partial}}

\def\v0{\mathbf{0}}

\def\Rarrow{{\quad\Rightarrow\quad}}

\def\beq{\begin{equation}}
\def\eeq{\end{equation}}
\def\bea{\begin{eqnarray}}
\def\eea{\end{eqnarray}}

\def\p{\partial}

\def \p{{\partial}}

\def\6{\partial}
\def\7{\tilde}
\def\8{\widehat}
 \def\bx{{\bf x}}

\def\G11{\Gamma_{11} }

\newcommand{\const}{\mathop{\rm const.}\nolimits}
\newcommand{\half }{\frac{1}{2}}

\def\smallover#1/#2{\hbox{$\textstyle\frac{#1}{#2}$}} %
\def\smallcirc{{\raise 0.5pt \hbox{$\scriptstyle\circ$}}}
\def\2{{\smallover1/2}}
\def\aand{{\quad\text{\small and}\quad}}
\def\where{{\quad\text{\small where}\quad}}
\def\for{{\;\;\text{\small for}\;\;}}
\def\with{{\quad\text{\small with}\;\;\;}}

\def\ie{{\;\text{\small i.e.}\;}}
\def\ie,{{\;\text{\small i.e.,}\;}}

\newcommand{\medbox}[1]{\fbox{%
\rule[-10pt]{0pt}{25pt}$\;\;\displaystyle{#1}\;\;$}%
}
\renewcommand{\theequation}{\thesection.\arabic{equation}}
\let\ssection=\section
\renewcommand{\section}{\setcounter{equation}{0}\ssection}

\begin{document}


\title{
Memory Effect \& Carroll Symmetry, \\
50 years later}

\author{
M. Elbistan$^{1}$\footnote{mailto: elbistan@itu.edu.tr},
P.-M. Zhang$^{2}$\footnote{mailto:zhangpm5@mail.sysu.edu.cn}
P. A. Horvathy$^{3}$\footnote{mailto:horvathy@univ-tours.fr},
}

\affiliation{
${}^1$ Physics Department,
Bo\u{g}azi{c}i University,
34342 \ Bebek / Istanbul, (Turkey)
\\
$^2$ School of Physics and Astronomy, Sun Yat-sen University, Zhuhai, China
\\
${}^3$ Institut Denis-Poisson CNRS/UMR 7013 - Universit\'e de Tours - Universit\'e d'Orl\'eans Parc de Grammont, 37200, Tours, (France)\\
\\}
\date{\today}


\begin{abstract}
Particles at rest before the arrival of a burst of gravitational wave move, after the wave has passed, with constant velocity along diverging geodesics. As recognized by Souriau 50 years ago and then forgotten,
 their motion is particularly simple in Baldwin-Jeffery-Rosen (BJR) coordinates (which are however defined  only in coordinate patches):
 they are determined when the  first integrals associated with the 5-parameter isometry group (recently identified as L\'evy-Leblond's
``Carroll'' group with broken rotations) are used. A global  description can be given instead in terms of Brinkmann coordinates,   however it requires to solve a Sturm-Liouville equation, whereas the relation between BJR and Brinkmann requires to solve yet another Sturm-Liouville equation.
The theory is illustrated by geodesic motion in a linearly polarized (approximate) ``sandwich'' wave proposed by Gibbons and Hawking for gravitational collapse, and by circularly polarized approximate sandwich waves with Gaussian envelope. 
\\
\vskip3mm
Annals Phys. \textbf{459} (2023), 169535
doi:10.1016/j.aop.2023.169535
\end{abstract}

\maketitle

\tableofcontents

\newpage

\section{Introduction}\label{Intro}

Gravitational waves were predicted by Einstein  in 1916 \cite{Einstein}, however doubt on their actual existence was cast  
soon after by Einstein and Rosen themselves \cite{EinsteinRosen}, 
 leading to a long-lasting controversy.
A breakthrough seemed to come at around 1970, when Joseph Weber  claimed to have \emph{observed} gravitational radiation using his resonant ``Weber bar''. His report \cite{Weber69} marked profoundly a whole generation of researchers of gravitational physics, despite remaining unconfirmed, and then falling progressively into embarrassed oblivion \cite{noise}. 

A different way to detect \GWs called now the \emph{Memory Effect} was proposed by Zel'dovich and Polnarev and their collaborators \cite{ZelPol,BraTho,BraGri}~:

\begin{quote}\textit{\narrower  \dots another,
 \underline{nonresonance}, type of detector,
 consisting of two noninteracting bodies (such as satellites). [\;\dots\;]
   {Although} [the] distance between free bodies will change, 
  their \underline{relative velocity will become vanishingly small} as the flyby event concludes.
 }
\end{quote}

A somewhat different proposal had been made earlier by  Ehlers and Kundt \cite{Ehlers} and (independently) by Souriau \cite{Sou73}~:
\begin{quote}\textit{\narrower  ``After a pulse has swept over the particles they have \underline{constant  velocities} ''}
\end{quote}

Although the verification has been out of experimental reach by the time when these proposals were made, they led ultimately to the LIGO/VIRGO gigaprojects. With the words of Rainer Weiss, one of the recipients of the Nobel prize awarded precisely for the detection of \GWs,

\begin{quote}\textit{\narrower  ``For as long as 40 years, people have been thinking about this, trying to make a detection, sometimes failing in the early days, and then slowly but surely getting the technology together to be able to do it.{}''}
\end{quote}

That of Zel'dovich et al \cite{ZelPol} (considered now as the ``standard'' one) will be called  \emph{displacement memory} effect (DM), and the one proposed by Ehlers-Kundt \cite{Ehlers}, and by  Souriau  \cite{Sou73}, will be referred to as  \emph{velocity memory} (VM) effect. 
Our paper mostly concerns VM in a plane gravitational wave \cite{Sou73,Grishchuk,GriPol,PolPer}. 
The intricate relation of the two types of memories raised by one of our referees will be discussed in a separate paper \cite{DMvsVM}.

\goodbreak

\begin{figure}[h]
\includegraphics[scale=.54]{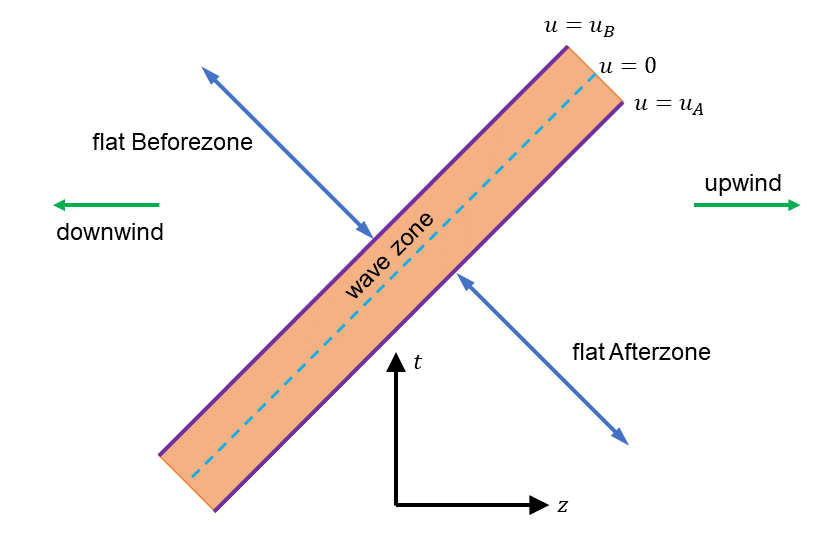}\vskip-3mm\caption{\textit{\small Schema of a sandwich wave (adapted from ref.\cite{BoPi89}) propagating downwind. $u=(z-t)/\sqrt{2}$ is a null coordinate. The space-time is flat in the yet undisturbed Beforezone $u < u_B$
and also in the Afterzone $u_A < u$ where the wave has already passed.
}}
\label{figinBoPi}
\end{figure}

In the early years which followed Weber's  sensational announcements, Gibbons \& Hawking \cite{GibbHaw71} considered, for example, a burst of a ``sandwich wave", which is non-flat only in a short interval $u_B \leq u \leq u_A$ of retarded time [called  Wavezone or Insidezone], and is flat both in the Beforezone $u < u_B$ that the wave has not  yet reached, and  in the Afterzone, $u_A < u$, where the wave has already passed, as illustrated in FIG.\ref{figinBoPi} adapted from ref.~\cite{BoPi89} \footnote{The null coordinate $u$ we use here has opposite sign with respect of that of Bondi and Pirani \cite{BoPi89}: it flows from the left to the right, whereas the wave advances from the right to the left.}.
\goodbreak
Our paper  studies  the geodesic motion of spinless test particles 
in a sandwich wave \cite{GibbHaw71} in a frame where 
those particles are at rest  before the wave reaches them,
\beq
\dot{\bx}(u) = 0\,
\for
u\leq u_B\, .
\label{Beforezone}
\eeq
We shall call them \emph{special geodesics}. (Bondi and Pirani \cite{BoPi89} call them ``basic geodesics''.)

50 years ago Jean-Marie Souriau has also studied particle motion in a \GW. He presented his pioneering results  {a year before} Zel'dovich et al. at a CNRS meeting held in Paris \cite{Sou73} in a paper   characteristic for its author:  he wrote it in French, used his personal notations and  ignored mainstream work.
His paper is hidden in an obscure conference proceedings and has never been published in any Journal. It would probably have got totally forgotten had Christian Duval, Souriau's former student, not remembered and resurrected it many years later \cite{Duval17,Carroll4GW}.
 The Souriau-Duval approach is summarized in sec.\ref{PlaneGW}.

Plane  \GWs have long been known to admit a generically 5-parameter isometry group \cite{Bondi57,BoPiRo,exactsol}. In the coordinate system proposed by Baldwin, Jeffery, and by Rosen (BJR) \cite{BaJeRo}, three of them are mere translations. The remaining two rather mysterious ones have only been known implicitly as solutions of a Sturm-Liouville (SL) equation \cite{Torre}. The multiple r\^oles played by Sturm-Liouville is emphasised, e.g., in \cite{SLC,ZCEH,zzh}.

Those two ``missing'' isometries had been identified in  \cite{Sou73} using a remarkable tool we refer to as the \emph{Souriau matrix}, $\IS$ in \eqref{Smatrix}. His results have long remained confidential and ignored, though.

Another seminal result concerns the \emph{group structure} related to another recently resurrected topic that young Jean-Marc L\'evy-Leblond had called, with tongue in cheek, the ``Carroll group'' \cite{Leblond,Alice}. See \cite{Carrollvs,DH-NC,Bekaert,Morand} for further developments.

\smallskip
\noindent\textbf {Theorem}~:
\textit{\small The isometries of a plane gravitational wave span the Carroll group without transverse rotations, implemented through the Souriau matrix ${\IS}$, see eqns. \eqref{Smatrix}-\eqref{genCarract}. }
\smallskip
\goodbreak

The  geodesics are determined by the conserved quantities associated with the isometries (see sec. \ref{MemoGeoSec})  \footnote{
L\'evy-Leblond's ``Carroll'' action appears {explicitly} in Souriau's talk \cite{Sou73}, however the relation of the work of the two colleagues has been recognized many years later only  \cite{Duval17,Carroll4GW}.}, using the Souriau matrix ${\IS}$, which is a key ingredient  \cite{Sou73,Carroll4GW}.

In sec.\ref{Illust} the general theory is illustrated first
by a simple Gaussian wave, and then by linearly \cite{GibbHaw71}, and finally circularly  polarized \cite{exactsol,SLC} sandwich waves.
For simplicity, we restrict our attention at Beforezones which are flat Minkowski space.

A concept which plays a fundamental role in our investigations is that of a \emph{Bargmann manifold} %
 \footnote{Souriau, and Duval et al \cite{DBKP},  argued that the 1-parameter central extension of the Galilei group studied before by V. Bargmann \cite{Barg54} is the best studied in an extended Kaluza-Klein-type framework. Such a framework was proposed by Eisenhart \cite{Eisenhart}, then forgotten and then reproposed independently in refs. \cite{Eisenhart,DBKP,DGH91}.
}, which is a manifold  with a metric with Lorentz  signature,  endowed also with a covariantly constant null vector \cite{Eisenhart,DBKP,DGH91,CDGH}, see the Appendix \ref{Appendix} for a summary. 

\section{Plane gravitational waves}\label{PlaneGW}

\subsection{Baldwin-Jeffery-Rosen (BJR) coordinates}\label{BJRSec}

Much of this section is taken from refs. \cite{Duval17,LongMemory}.
Following Souriau \cite{Sou73}, we write the metric in Baldwin-Jeffery-Rosen (BJR)   coordinates $(\bx,u,v),\, \bx=(x^1,x^2)$ \cite{BaJeRo},
\begin{equation}
\rg=a_{ij}(u)dx^i dx^j+2 du\,dv\,,
\label{BJRmetric}
\end{equation}
where the symmetric  positive $2\times2$ matrix
 $\fa=\big(a_{ij}\big)$ depends only on (retarded) time,
$u=(z-t)/\sqrt{2}$. Both $u$ and $v=(z+t)/\sqrt{2}$ are null coordinates. The vector $\xi=\partial/\partial v$ is covariantly constant. Thus \eqref{BJRmetric} is the metric of a \emph{Bargmann manifold} \cite{Eisenhart,DBKP,DGH91,ZHAGK,CDGH}. 
The only nonzero components of its Riemann tensor are 
\begin{equation}
R_{uiuj}= -\half r_{ij}
\where
r_{ij} = \Big(\ddot{\fa}-\half\dot{\fa}{\fa}^{-1}\dot{\fa}\Big)_{ij}\,
\label{RiemBJR}
\end{equation}
($\da=d{\fa}/du$).
The metric is thus determined by the symmetric $2\times2$ matrix
$\fr=\big(r_{ij}\big)$.
The only nonzero component of the Ricci tensor is thus,
\begin{equation}
R_{uu}= 
-\half\Tr\left(\dot{\fb}+\half{}{\fb}^2\right)
\where
\fb={\fa}^{-1}\da\,.
\label{Ricci}
\end{equation}


In the ``Before'' and ``Afterzones'', $u\leq u_B$ and $u_A\leq u$, respectively, the metric is flat.
The most general  {flat} metrics of the form (\ref{BJRmetric}) is found by setting $\fr=0$ in  (\ref{RiemBJR})
which implies,
\begin{equation}
\dddot{\fa}=0\,,
\label{r=0}
\end{equation}
hence $\fa(u)$ is at most quadratic in $u$ and is determined by the  initial conditions at some   $u_0$,
$ 
\fa_0=\fa(u_0)$ and $\da(u_0)=\da_0\,,
$ 
\beq
\fa(u)=
\fa_0+\da_0(u-u_0)+\frac{1}{4}\da_0\fa_0^{-1}\da_0(u-u_0)^2\,.
\label{Ricciflata}
\eeq
Denoting a square-root of the matrix $\fa_0$ by~${\IP}_{0}$, 
${\IP}_{0}^{\dagger}{\IP}_{0} =\fa_0$ \footnote{Various choices of the square root matrices of $\fa$ will be studied in sec.\ref{gaugesec}.}, we can write  \cite{Sou73},
\begin{equation}
\medbox{
\fa(u)={\IP}_{0}\Big(\II+(u-u_0)\fc_0\Big)^2{\IP}_{0}
\qquad
\text{where}
\qquad
\fc_0=\half{}{\IP}_{0}^{-1}
\da_0\,{\IP}_{0}^{-1}
\,.
}
\label{BJRa}
\end{equation}

In the Beforezone where $\fa_0=\II=\diag(1,1)$ is the $2\times2$ unit matrix and $\da_0=0$, we duly get $\fa(u)=\II$, but in the Afterzone the with non-trivial initial velocity $\fc_0\neq0$  our flat metric \eqref{BJRa} is clearly \emph{more general} than mere Minkowski space.

Flat BJR coordinates $(\hbx,\hu,\hv)$ can be introduced outside the Wavezone i.e., for $u\leq u_B$ and for $u_A \leq  u$ \cite{Duval17}. In both regions,
\begin{equation}
\begin{array}{c}
\bx
\\
{u}\\
{v}
\end{array}
\quad\longmapsto\quad
\begin{array}{l}
\hbx=\Big(\II+(u-u_0)\fc_0\Big){\IP}_{0}
\bx
\\
\hu=u
\\
\hv=v-\half\bx^\dagger{}{\IP}_{0}
\fc_0
\Big(\II+(u-u_0)\fc_0\Big){\IP}_{0}
\bx
\end{array}\,
\label{NewCoordinates}
\end{equation}
brings the metric (\ref{BJRmetric}) to the flat form,
\beq
\rg=d\bx^\dagger{}\fa(u)d\bx+2du\,dv
=d\hbx^\dagger{}d\hbx+2d\hu\,d\hv\,.
\label{Minkoform}
\eeq

In the Beforezone where $\fa_0=\II$ and $\fc_0=0$,
\eqref{NewCoordinates} is the identity transformation.
However in the Afterzone $u\geq u_A$ the initial velocity does \emph{not} vanish in general, $\fc_0\neq0$. Therefore the transition between the flat Before and Afterzones may be non-trivial, -- and this is precisely the crux of the story for scattering by a \GW.


Let us now turn to the metric inside the  Wavezone $[u_B,u_A]$. By (\ref{Ricci}), Ricci flatness is equivalent to
\begin{equation}
\Tr\left(\db+\half{}\fb^2\right)=0\,.
\label{Ricci=0}
\end{equation}
Setting \cite{Sou73}
\begin{equation}
\chi=\big(\det{\fa}\big)^{\frac{1}{4}}>0\,
\aand
\gamma=\chi^{-2}\fa
\label{chigamma}
\eeq
yields
$\fb=\gamma^{-1}\dgamma+2\chi^{-1}\dchi\,\II$\,.
Then $\det(\gamma)=1$ implies that $\Tr(\gamma^{-1}\dgamma)=0$;  thus we end up with a \emph{Sturm-Liouville equation} \cite{Sou73,Carroll4GW}
\begin{equation}
\medbox{
\ddchi+\omega^2(u)\chi=0,
\qquad
\omega^2 =\frac{1}{8}\Tr\left((\gamma^{-\half}\dgamma\gamma^{-\half})^2\right)\,,
}
\label{chiSL}
\end{equation}
which guarantees that the  vacuum Einstein equations are satisfied for an otherwise arbitrary choice of the unimodular symmetric $2\times2$ matrix $\gamma(u)$.

Outside the Wavezone, the frequency $\omega$ vanishes (as it will be illustrated  in FIG.\ref{omegafig}), and the Sturm-Liouville equation \eqref{chiSL} reduces to the \emph{free} equation,
\beq
\ddot{\chi}\approx 0 \Rightarrow \chi = c u + d\,,
\label{flatchieq}
\eeq
which vanishes at most once unless it is identically zero, as it happens in the Before, but \emph{not} in the Afterzone.
Moreover, as argued by Souriau \cite{Sou73}, eqn. \eqref{chiSL} implies that $\chi$  is \textit{concave}, $ \ddchi<0$, and therefore \emph{the determinant of the metric necessarily vanishes} at some $u_1>u_B$ \footnote{In the linearly polarized case this  was noticed, later, also by Bondi and Pirani \cite{BoPi89} by proving that at least one of the components changes sign.}
\beq
\chi(u_1)=0 \Rarrow \det{\fa}(u_1)=0\,.
\label{chi0}
\eeq

At $u_1$ the BJR expression \eqref{BJRmetric}  becomes singular, allowing us to conclude  that \emph{the BJR coordinates are necessarily  \emph{local} i.e. defined  only in coordinate patches, defined by the subsequent zeros of $\fa$} ---
 whereas the waves themselves are perfectly well-defined and regular (as it will be seen below using Brinkmann coordinates). The apparent singularity comes from the choice of the coordinates, not from the wave itself. See also \cite{Wang:2018iig} for a recent discussion.
 
Had Souriau's insight been known and appreciated, it could  have shortened the controversy \footnote{The BJR-type coordinate singularity is  analogous to the ``Dirac string'' of monopoles \cite{Dirac31,WuYang75}. Souriau has also included an analogous discussion into the  ``Prequantization'' chapter \cite{SouPrequant} of his the planned but never completed and let alone published revision of his book \cite{SSD}. He may have been stimulated by a lecture given by C. N. Yang at the CPT in Marseille roughly at around that time.}. 

\subsection{Plane \GWs in Brinkmann coordinates}\label{BrinkSec}

An exact plane gravitational wave can also be presented in Brinkmann coordinates \cite{Brink} $(\bX,U,V)$,
\besub
\begin{align}
ds^2=&\;\delta_{ij} dX^i dX^j + 2dUdV + K_{ij}(U) X^i X^j dU^2 \,,
\label{Bmetric}
\\[4pt]
K_{ij}(U){X^i}{X^j}=&\;
\half{{\mathcal{A}_{+}}}(U)\big((X^1)^2-(X^2)^2\big)+{\mathcal{A}_{\times}}(U)\,X^1X^2\,,
\label{Bprofile}
\end{align}
\label{genBrink}
\end{subequations}
where ${\mathcal{A}_{+}}$ and ${\mathcal{A}_{\times}}$ are the $+$ and $\times$ polarization-state amplitudes. The symmetric and traceless matrix $\IK=(K_{ij})$ is the Brinkmann profile matrix.
The  only  non-vanishing components of the Riemann tensor are, up to symmetry,
\beq
R_{iUjU} (U)  = -  K_{ij}(U) 
\label{BRtensor}
\eeq
(cf. \eqref{RiemBJR}). unlike their BJR counterparts, Brinkmann  coordinates are global \cite{Bondi57,BoPiRo}.

 The ``vertical'' vector $\p_V$ is covariantly constant, --- thus we have
a Bargmann manifold \cite{Eisenhart,DBKP,DGH91}, with the quadratic coefficient of the $dU^2$ term in \eqref{Bprofile} interpreted, in the Bargmann framework, as an effective oscillator potential, see the Appendix \ref{Appendix}.

The relation of the two coordinate systems is established in three steps \cite{Gibb75}.
\begin{enumerate}
\item
Starting with the Brinkmann profile, we
first  solve (generally only numerically) the Sturm-Liouville equation with a supplementary condition,
\beq\medbox{
\ddot{{\IP}}= {\IK}{\IP},
\qquad
{\IP}^\dagger\dot{{\IP}}=\dot{{\IP}^\dagger}{\IP}\,}
\label{SL+cond}
\eeq
for some  $2\times2$ matrix
$
{\IP}={\tiny \barray{ll}
P_{11} & P_{12}
\\
P_{21} & P_{22}
\earray}
$;
here $\dot{\{\,\}}= d/dU$.

\item
Then the symmetric $2\times2$ matrix
\beq
\fa = {\small \barray{rl}a_{11} &a_{12}
\\
a_{21} & a_{22}\earray}
= {\IP}^\dagger{}{\IP}\,
={\small \barray{ccc}P_{11}^2+P_{21}^2 &&P_{11}P_{12}+P_{21}P_{22}
\\[2pt]
P_{11}P_{12}+P_{21}P_{22} &&P_{22}^2+P_{12}^2
\earray}
\label{afromP}
\eeq
is the BJR profile.

\item
Brinkmann and BJR coordinates, $(\bX,U,V)$ and $(\bx,u,v)$, respectively, are related according to,
\beq
{\bX} ={\IP}(u)\bx,
\qquad
U=u,
\qquad
V=v-\frac{1}{4}\bx\cdot\da(u)\bx\,.
\label{BJRBrinkP}
\eeq

Conversely, the Brinkmann profile ${\IK}=(K_{ij})$ can be recovered  from the BJR profile $\fa=(a_{ij})$ by,
\beq
{\IK}=\half{}{\IP}\left(\dot{\fb}+\half{\fb}^2\right){\IP}^{-1}\,,
\qquad \fb=\fa^{-1}\da\,,
\label{KfromP}
\eeq
 cf. \eqref{Ricci}.
\end{enumerate}

Now $\det (\fa) = 
 \det ({\IP}^{\dagger}) \det({\IP}) = (\det {\IP})^2$ implies that
  BJR coordinates are (unlike their Brinkmann counterparts) defined only on coordinate patches between adjacent zeros
 $u_{k}$ and $u_{k+1}$ of the determinant of  ${\IP}$ (or of $\fa$), as said earlier. 

Saying that the Beforezone is undisturbed amounts to saying that the BJR profile in \eqref{BJRmetric} is $\fa(u)= \II$\,
for $u < u_B$, or symbolically,
\beq
\fa(-\infty)=\II\,.
\label{aboundary}
\eeq
%
\subsection{An internal structure~?}\label{gaugesec}

Our framework has a curious extra freedom: the 
``square root matrix'' ${\IP}$ can be changed by an internal rotation which may depend on $U$ but not on $\bx$.  Duval \cite{Duval17} has observed that   if ${\IP}$ is a
square root, $\fa={\IP}^{\dagger}{\IP}$, and $R\in\Ort(2)$
is an arbitrary orthogonal matrix, $R^{\dagger}R = RR^{\dagger} = \II$, then
$(R{\IP})^{\dagger} R{\IP}
 ={\IP}^{\dagger}{\IP} = \fa$ and thus
\beq
 \widetilde{\IP} = R\,{\IP}
 \label{RPgauge}
\eeq
is another square root, $\widetilde{\IP}^{\dagger}\widetilde{\IP}=\fa$. 
Eqns \eqref{afromP} and \eqref{BJRBrinkP} imply that the matrix $R$ can be at most a function of $U =u$, $R = R(U)$.  

What is the effect of a rotation  by  $R \in \Ort(2)$~?
Firstly, 
 our gauge freedom \eqref{RPgauge} can be use to ``dress'' ${\IP}$ in the Beforezone~: a suitable $R$ brings ${\IP}(-\infty)$ to the form,
\beq
 {\IP}(u)=\diag(1,1)
\for u\leq u_B\,,
\quad\text{\small symbolically }\quad
{\IP}(-\infty) = \diag (1,1)\,,
 \label{specPinBz}
\eeq
\goodbreak

Secondly, the redefinition \eqref{RPgauge} affects the BJR $\leftrightarrow$ B correspondence, which is in fact multivalued : a  BJR metric can be mapped into many different Brinkmann metrics by changing $R$.

\goodbreak

$\bullet$ When $R$ is a constant matrix then 
 \eqref{BJRBrinkP} becomes,
\begin{equation}
X^i = (R^{\dagger} \widetilde{\IP})^i_{\,j} x^j, \quad   U = u,  \quad V = v- \frac{1}{4} x^i x^j \dot{a}_{ij}\,.
\label{tildeXBJR-B}
\end{equation}
Note that the last term here does not change because 
$\widetilde{\IP}$ is a square root of $\fa$.

The SL problem \eqref{SL+cond} becomes in turn,
\beq
\ddot{\widetilde{\IP}}_{ij}= 
\big(R{\IK}R^{\dagger}\big)_{ik}\widetilde{\IP}^k_{\,j}\,,
\qquad
(\widetilde{\IP})^\dagger\dot{\widetilde{\IP}}
=
(\dot{\widetilde{\IP}})^\dagger\,\widetilde{\IP}\,.
\label{tildeSL+cond}
\eeq
Then, in terms of the new (transverse) coordinate 
\beq
\widetilde{X}^i = R^i{}_j X^j\,,
\label{tildeX}
\eeq 
 the  Brinkmann metric \eqref{Bmetric} is written as,
\begin{equation}
\label{RcBmet}
ds^2 = \delta_{ij}d\widetilde{X}^i d\widetilde{X}^j + 2 dU dV + \widetilde{K}_{ij}\widetilde{X}^i \widetilde{X}^j dU^2\,,
\qquad
\widetilde{K}_{ij} 
 = \big(R K R^\dagger\big)_{ij}\,.
\end{equation}

$\bullet$ When the $\Ort(2)$ matrix $R$ is $U\equiv u$-dependent, $\dot{R} \neq 0$, the situation gets more subtle. 
The B $\Leftrightarrow$ BJR coordinate transformation generates a gauge potential, (see Appendix \ref{Appendix}).  

We illustrate this point on a \emph{circularly polarized periodic} (CPP) \cite{PolPer} gravitational wave, whose Sturm-Liouville  equation (\ref{SL+cond}) was solved in \cite{Zhang:2018msv, Elbistan:2022umq}. 
The Brinkmann metric (\ref{genBrink}) is \eqref{Bmetric} with profile,
\beq
 K_{ij}(U) X^i X^j =\frac{A_0}{2} \Big(\cos(\omega U)  \big( (X^1)^2 - (X^2)^2\big)+2\sin(\omega U) X^1 X^2\Big)\,,
 \label{CPPprof}
\eeq  
where $A_0$ is the amplitude and $\omega$ is the frequency of the wave. 
Then, by following a suggestion of Kosinski \cite{Kosinski} and further developed in refs.\cite{PolPer,ZHAGK,AndrPrenc}, we switch to rotating coordinates $\widetilde{\bX}$ \eqref{tildeX} but now with a $U$-dependent transverse rotation  i.e., $\widetilde{X}^i = R^i{}_j (U) X^j$ , 
\begin{equation}
\label{introrot}
\begin{pmatrix}
 \tilde{X}^1 \\ \tilde{X}^2
\end{pmatrix}
= \begin{pmatrix}
\cos({\half}\omega U)  &  \sin({\half}\omega) 
\\[4pt]
-\sin({\half}\omega U)  & \cos({\half}\omega U)\,
\end{pmatrix}
\begin{pmatrix}
 X^1 \\ X^2
\end{pmatrix}\,,
\end{equation}
where $U$ and $V$ are kept fixed. The $U$-dependent rotation (\ref{introrot}) belongs to $\SO(2)$ which does not change neither the associated BJR metric $\fa$ and consequently nor the geodesic motion.

When $R=R(U)$, then
$ 
\label{introrotg}
d\tilde{X}^i = R^i{}_j dX^j + \dot{R}^i{}_j X^j dU
$ 
and the metric picks up additional terms. 
For our CPP wave we find, 
\besub
\begin{align}
\label{rotCPP}
ds^2 = &\;\delta_{ij} d\widetilde{X}^i d\widetilde{X}^j + 2 dUdV + \omega \epsilon_{ij} \widetilde{X}^i d\widetilde{X}^j dU + \Big(\Omega_+^2 (\widetilde{X}^1)^2 + \Omega_-^2 (\widetilde{X}^2)^2  \Big) dU^2,
\\[6pt]
\Omega_\pm^2 = &\;\frac{\omega^2}{4} \pm \frac{A_0}{2}\,.
\label{shiftfreq}
\end{align}
\label{tildemetricbis}
\esub
From the Bargmann point of view \cite{Eisenhart,DBKP,DGH91,ZHAGK}, 
(see the Appendix \ref{Appendix} for a summary) this metric describes an anisotropic oscillator with shifted but $U$-independent frequencies \footnote{This explains also the extra ``screw'' symmetry of a CPP \GW \cite{exactsol,Carroll4GW,ZHAGK,PolPer}.}, augmented by
 a new, linear-in-$\widetilde{\bX}$ gauge potential term,
\beq
2\bA{\cdot}d\widetilde{\bX} =
\omega \epsilon_{ij}\widetilde{X}^i \; d\widetilde{X}^j dU\,,
\label{gaugepot}
\eeq
[where $\epsilon_{ij}$ is the antisymmetric Levi-Civita symbol],
whose curvature  is 
\beq
\cF = d\bA = \omega\,d\widetilde{X}^1\wedge d\widetilde{X}^2\,.
\label{rotpot}
\eeq   
Thus for $\omega\neq0$ $\bA$ is \emph{not a pure gauge} and \emph{does} therefore have an effect on the geodesic dynamics~: --- it is indeed the source of the Coriolis force
\cite{ZHAGK}. The potential can be either  attractive or repulsive, depending on the relative strength of the two terms in \eqref{shiftfreq}. The middle, Coriolis term is always attractive, though, and if it is strong enough, it can bound all motions, as in FIGs.\ref{A2omaga0omega1} and \ref{A1omaga2omega4}.  

\begin{figure}[h]
\includegraphics[scale=.285]{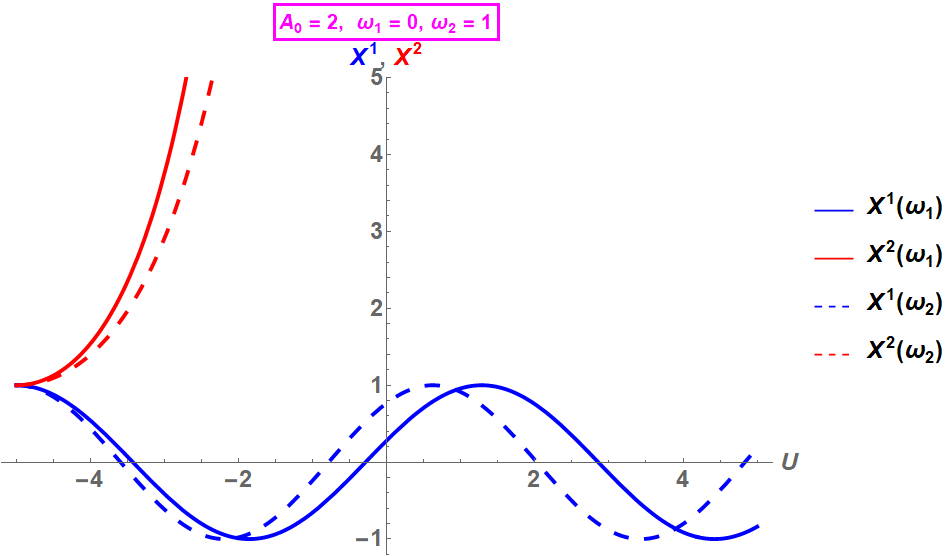}\\
\caption{\textit{\small Trajectories for 
 $A_0=2$, and for either $\omega_1=0$ or for $\omega_2=1$\,. 
 When $\bomega_1=0$, one of the oscillators in the last term is attractive and the is other repulsive; the tractories (shown in full lines) are, accordingly, \blue{\bf bounded} or \red{\bf unbounded}.
The trajectory (shown in dotted line) is different is however qualitatively similar  for 
$\bomega_2=1$,%
}}
\label{A2omaga0omega1}
\end{figure}
\begin{figure}[h]
\hskip-2mm\includegraphics[scale=.26]{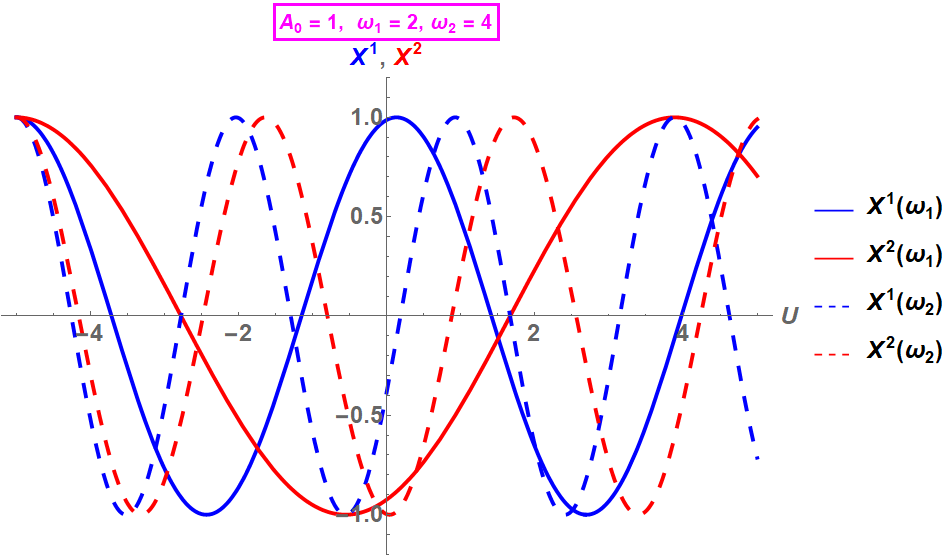}
\,
\includegraphics[scale=.22]{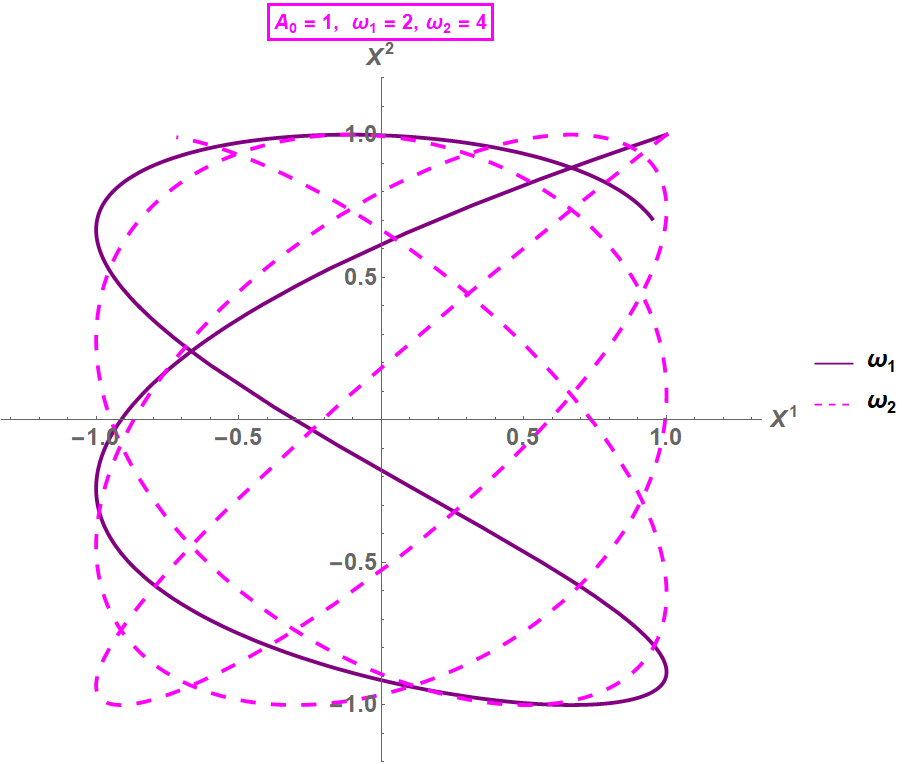}
\caption{\textit{\small 
Trajectories for $A_0=1,\,\omega_1=2,\,\omega_2=4$\,.
For (relatively) large $\omega$ and small $A_0$, both frequencies $\Omega_{\pm}> 0$ but are dominated by the strong Coriolis force:
 motions are bounded.
}}
\label{A1omaga2omega4}
\end{figure}
 The new term \eqref{gaugepot} changes also the Sturm-Liouville equation \eqref{SL+cond} which swaps
the Brinkmann and BJR metrics,
\begin{equation}
\label{SLpu}
\ddot{\widetilde{\IP}} =  \Omega\,\widetilde{\IP}+
\omega \epsilon\, \dot{\widetilde{\IP}}\,, 
\qquad 
\Big(\widetilde{\IP}^\dagger \dot{\widetilde{\IP}} - \dot{\widetilde{\IP}}^\dagger \widetilde{\IP} \Big) = 
\omega \widetilde{\IP}^\dagger \epsilon \widetilde{\IP}\,, 
\end{equation}
where $\epsilon$ is the anti-symmetric Levi-Civita matrix and 
\begin{equation}
\Omega = 
\begin{pmatrix}
\Omega_+^2 & 0 \\
0 & \Omega_-^2
\end{pmatrix}\,
\end{equation}
with $\Omega_\pm$ given in \eqref{shiftfreq}.
\goodbreak

 \kikezd{Example: Lukash  metric}:  
At this point, it is instructive to recall what happens for the Lukash  metric  \cite{Lukash,LukashI}. Switching provisionally to the notations of ref. \cite{LukashII}, for $U > 0$ the metric (to be compared with the CPP case \eqref{CPPprof}) is, 
\begin{eqnarray}
ds^2_{L} &= &d\bX^2 + 2 dU dV  \nn
\\[6pt]
& -&\left\{\frac{C}{U^{2}}\cos\Big(2\kappa\ln(U)\Big)
\,
\big[(X^1)^2-(X^{2}]^2\big)
\,-\, 
\frac{2C}{U^{2}}\sin\Big(2\kappa\ln(U)\Big)
 \,X^1X^2\right\}\,dU^2\,.\qquad\quad
\label{BLukash}
\end{eqnarray}
Then the coordinate transformation \cite{zzh}
\begin{equation}
U = e^{T}\,, 
\quad
\bm{X}=e^{T/2}\bm{\xi},
 \quad V \to V,
 \where \bm{\xi} = (\xi,\eta)
\label{lnchange}
\end{equation}
 yields
which is conformal, $ds_L^{2}= e^{\dagger}d\Sigma^2\,,
$ to  
\begin{eqnarray}
d\Sigma^2&=&d\xi^{2}+d\eta^{2}+2dTdV+\big({\xi} d\xi+{\eta} d\eta\big)dT + \nn
\\[6pt]
&&\left(\frac{1}{4}\left(\xi^{2}+\eta^{2}\right) -C\left(\xi^{2}-\eta ^{2}\right) \cos 2\kappa T+2C\xi \eta
\sin 2\kappa T\right) dT^{2}\,.
\label{Sigmametric}
\end{eqnarray}
(Remember that conformal metrics have identical null geodesics and are also simply related massive ones \cite{ZCEH,nonlocal}).
Our clue is that the $d\xi^idT$ term is a \emph{pure gauge}, 
\beq
\bA = A_i d\xi^i = \xi^i d\xi^i
=
\frac{1}{2} d\big(\xi^i \xi^i\big)
\Rarrow \cF = d\bA=0
\,.
\label{puregauge}
\eeq 
The associated magnetic field \emph{vanishes}, and therefore the vector potential $\cA$ has  no effect on the geodesic motion. It can in fact be gauged away by a redefinition of the coordinate $V$,
$ 
V \to v- \frac{1}{4} \xi^i \xi^i
$  
\cite{LukashII}.

\smallskip 
The difference between the two cases is also understood by observing that, unlike as for CPP, for Lukash the component oscillators in \eqref{shiftfreq} have \emph{no relative minus} sign and therefore can be gauged away by a redefinition of the $V$-coordinate.

We note for completeness that rotation with an $U$-dependent matrix $R(U)\in \Ort(2)$ would yield a modified Sturm-Liouville equation of rather complicated form, 
\begin{equation}
\label{SLpO}
\ddot{\widetilde{\IP}} = (R \IK R^\dagger) \widetilde{\IP}+ \dot{R} (\dot{R^\dagger} \widetilde{\IP} + 2 R^\dagger \dot{\widetilde{\IP}}),   
\qquad  
\widetilde{\IP}^\dagger \dot{\widetilde{\IP}} - \dot{\widetilde{\IP}}^\dagger \widetilde{\IP} = 2 \widetilde{\IP}^\dagger \dot{R} R^\dagger \widetilde{\IP}\,,  
\end{equation}
cf. (\ref{SLpu}). 
In conclusion, the $U$-dependent rotation \eqref{RPgauge} alters the Brinkmann metric (\ref{tildemetric}) and consequently the geodesic motion, while keeping the BJR metric  unchanged, $\fa = \widetilde{\IP}^\dagger \widetilde{\IP} = \IP^\dagger \IP$. In other words, the B $\leftrightarrow$ BJR correspondence is
many-to-one.

\section{Carroll symmetry of plane gravitational waves}\label{GWsymmetry}

It has long been known that the metric \eqref{BJRmetric} has a
(generically $5$-parameter) isometry group,
composed of manifest $\bx$ and $v$-translations, $\bx \to \bx+\bc$ and $v\to v+d$, respectively, augmented by a rather mysterious 2-parameter subgroup \cite{BoPiRo,exactsol}. Apparently ignoring the existing results, the latter has been identified  by Souriau in his ``well hidden'' conference paper \cite{Sou73} -- but the r\^ole and its relation to  Carroll symmetry \cite{Leblond} were
 established only recently \cite{Duval17,Carrollvs,Carroll4GW,ZHAGK}.
The isometry group \cite{BoPiRo} is generated by the Killing vectors, written, in BJR coordinates, 
\begin{equation}
Y^v = \frac{\partial}{\partial v}\,,  
\qquad 
Y^i_c = \delta^{ij}\frac{\partial}{\partial x^j}\,,   
\qquad 
Y^i_b = S^{ij}(u) \frac{\partial}{\partial x^j} - x^i \frac{\partial}{\partial v}\,. 
\label{infCarr} 
\end{equation}
Anticipating our results to come, the $Y^i_b,\, i=1,2$ will be referred to as \emph{Carroll boosts}. A crucially important role is played here by the symmetric $2\times 2$ \emph{Souriau matrix}
\beq
{\IS}(u)\equiv (S_{ij})=\displaystyle{\int^u_{u_0}}\!\!\fa^{-1}(t) dt\,,
\qquad 
u_k < u_0,\, u < u_{k+1}\,,
\label{Smatrix}
\eeq
which depends only on $\fa = {\IP}^{\dagger}{\IP}>0$. Therefore  we get,  for all choices of the orthogonal matrix $R$, the same  Souriau matrix:
the ``gauge freedom" $\IP \to R\,\IP$ in \eqref{RPgauge} disappears --- just as it happens when  passing from a gauge potential to the field strength tensor.

Each component  of the Souriau matrix satisfies the same  Sturm-Liouville equation   with the appropriate auxiliary condition  \eqref{SL+cond} as $\IP$ does \cite{SLC,ZCEH}.
Keeping this in mind, let us pull back the Carroll Killing vectors (\ref{infCarr}) to Brinkmann coordinates $X^\mu = \{X^i, U, V\}$ using  (\ref{BJRBrinkP}).
The vector fields (\ref{infCarr}) become, 
\begin{equation}
Y = Y^\nu (X(x)) \frac{\partial X^\mu}{\partial x^\nu}  \frac{\partial}{\partial X^\mu}\,.
\end{equation}

$\bullet$ For  $v$-translation (\ref{infCarr}) we get the covariantly constant null Killing vector written in Brinkmann coordinates,
\begin{equation}
Y^v = \frac{\partial}{\partial v} = \frac{\partial U}{\partial v} \frac{\partial}{\partial U} +  \frac{\partial X^i}{\partial v} \frac{\partial}{\partial X^i} +  \frac{\partial V}{\partial v} \frac{\partial}{\partial V}  =   \frac{\partial}{\partial V}\,. 
\end{equation}  

$\bullet$ $x$-translations  (\ref{infCarr}) become\,,
\begin{equation}
Y^i_c = \frac{\partial}{\partial x^i}
= \IP^{ji} \frac{\partial}{\partial X^j} - \frac{1}{2} \dot{\fa}_{ij}x^j  \frac{\partial}{\partial V}
=  \IP^{ji} \frac{\partial}{\partial X^j} - \dot{\IP}_j{}^i X^j \frac{\partial}{\partial V}\,,
\label{Ptrans}
\end{equation}  
where  the subsidiary condition  \eqref{SL+cond} was used. Thus the $\IP$-matrix (\ref{SL+cond}) generates Carroll translations in Brinkmann coordinates. 

$\bullet$  At last, Carroll boosts $Y^i_b$ (\ref{infCarr}) are expressed, in Brinkmann coordinates, as,
\begin{equation}
\label{PSboost}
Y^i_b = (\IP \IS)^{ji} \frac{\partial}{\partial X^j} - \dot{(\IP \IS)}_j{}^iX^j \frac{\partial}{\partial V}\,,
\end{equation}
where the matrix $\IP \IS$ that shares the same properties with the matrix $\IP$, generates Carroll boosts. 
In concusion, Torre's  expression \cite{Torre} is recovered.   
\eqref{Ptrans} and \eqref{PSboost} can be unified into a
a compact form,
\begin{equation}
Y^i = Q^{ji} \frac{\partial}{\partial X^j} - \dot{Q}_j{}^i X^j \frac{\partial}{\partial V} \,,
\label{Bboosttrans}
\end{equation}
where $Q_{ij} = P_{ij}$ for translations  and $Q_{ij} = ({\IP}{\IS})_{ij}$ for boosts. Conversely, the 
$\dot{Q}_j{}^i$ factor here gets absorbed under the inverse transformation $V \to v$, \eqref{infCarr}  \cite{LukashI}, as recalled above.

Eqn. (\ref{Bboosttrans}) is just (\ref{infCarr}) translated to  Brinkmann coordinates, however while (\ref{infCarr}) is simple and valid for any BJR profile, (\ref{Bboosttrans}) requires to solve the Sturm-Liouville equation for $\IQ=(Q_{ij})$. This underlines the advantage of using BJR coordinates to determine the symmetries of plane waves. 

Henceforce we restrict our investigations to an interval  $I_k=[u_k, u_{k+1}]$   chosen between two adjacent zeros of $\det(\fa)$ [equivalently, of $\det({\IP})$] and we pick a point $u_0\equiv u_0^{(k)} \in I_k$. The three translation are trivial, therefore we focus our attention at the 2-parameter subgroup generated by the C-boosts, $Y^i_b$ in \eqref{infCarr}. 
In BJR terms, the latter acts on space-time at fixed $u$, according to
\beq
\bx\to\bx+{\IS}(u)\bb\,,
\qquad
u\to u,
\qquad
v\to v-\bb\cdot\bx - \2\bb\cdot{}{\IS}(u)\bb\,.
\label{genCarract}
\eeq
In Minkowski space, for example, $\fa$ is the unit matrix $\II$. Now  $\dot{\IS}={\fa}^{-1} =\II$, hence 
\beq
 {\IS}(u)=(u-u_0)\II\,,
\label{MinkSmatrix}
\eeq
so that \eqref{genCarract} 
reduces to Galilei boosts in $2+1$ dimensions, lifted to Minkowski space written in light-cone coordinates, viewed as a particular Bargmann manifold with trivial profile $(K_{ij}) \equiv 0$ \cite{DBKP,DGH91}.

At the points $u_k$ where $\big(\det(\fa)\big)(u_k)=0$ the matrix $\fa$ is not invertible and ${\IS}$ is undefined.
Boosts in Brinkmann coordinates are obtained by using \eqref{BJRBrinkP}, which, when the profile $(K_{ij})$ is non-trivial, have a distorted implementation, as it will be illustrated in  FIG.\ref{Smatrixfig} in sect.\ref{Illust}. 
\besub
\begin{align}
&\bX \to  \bX+ {\IQ}\, \bb\,,
\label{BXsymimp}
\\
& V \to V
- \bX\cdot\dot{\IQ}\,\bb
-\frac{1}{2} \IQ\,\bb\cdot \dot{\IQ}\,\bb\,,
\label{BVsymimp}
\end{align}
\label{Bboost}
\esub\vspace{-5mm}
where
\beq
\IQ(u) = {\IP}(u){\IS}(u)\,.
\label{Qdef}
\eeq
The matrix ${\IQ}$  is $u$-dependent, however  boosts  leave each ``vertical'' slice $U=u=\const$ invariant, as depicted conceptually in FIG.\ref{CarrSlices}.
Examples will be presented in sec.\ref{Illust}.
\goodbreak
\begin{figure}[h]
\hskip-3mm
\includegraphics[scale=.36]{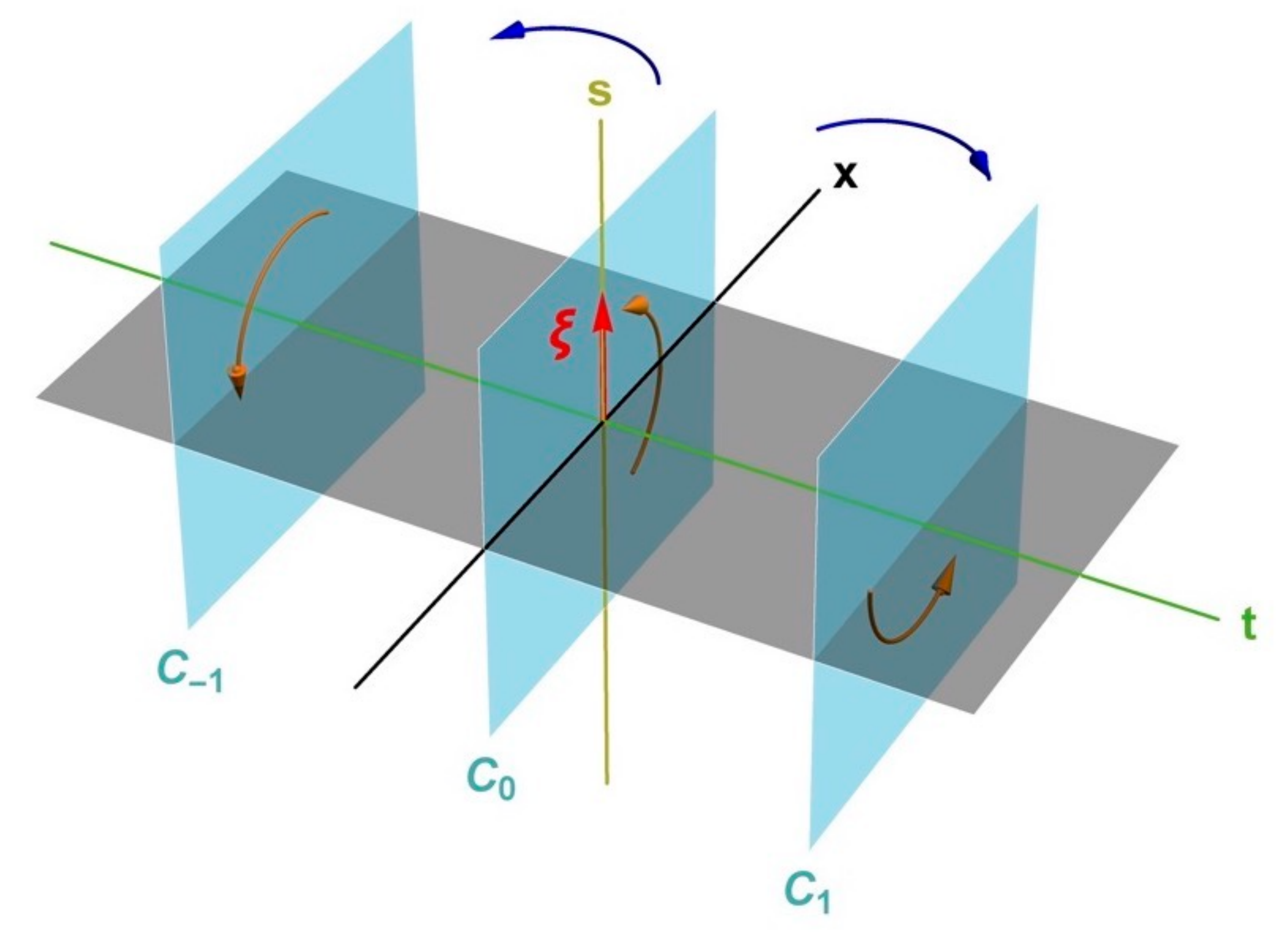}
\\
\vskip-6mm
\caption{\textit{\small  The distorted boosts \eqref{Bboost} leave the $u=\const$ slices invariant, upon which they
are implemented through the matrix ${\IQ}(u)={\IP}(u){\IS}(u)$\,.
}}
\label{CarrSlices}
\end{figure}
\goodbreak

The matrix ${\IQ}$ here satisfies the same  SL plus auxiliary equation \eqref{SL+cond} as ${\IP}$ does \cite{SLC},
\beq
\ddot{\IQ}={\IK}\,{\IQ}\,,
\qquad
{\IQ}^{\dagger}\dot{\IQ}-\dot{{\IQ}^{\dagger}}{\IQ}=0\,
\label{SLS}
\eeq
and then we could repeat our analysis for ${\IP}$.
It is however more convenient to combine our previous results obtained for ${\IP}$ and ${\IS}$.
In the Beforezone we have \eqref{MinkSmatrix} and the usual lifted Galilean action is recovered. 
The behavior in the Afterzone is more subtle, as it will be seen in sec.\ref{Illust}.
The boosted BJR profile  is related to the initial $\fa$ by a similarity transformation,
\begin{equation}
\fa = 
(a_{ij}) \to (\IP\IS)^{\dagger}_{\;ik} (\IP\IS)^k_{\;j} = 
(\IS^{\dagger} \fa \,\IS )_{ij}\,,
\label{SLSa}
\end{equation}
($\IS{\dagger} = \IS$).
\goodbreak

For the flat Minkowski metric, for example, the boosted BJR metric 
\eqref{Bboost} with \eqref{Qdef} reduces to the Souriau matrix, $\IS =u\II$, which plainly solves the free SL equation $\ddot{\IP}=0$. Accordingly in Brinkmann coordinates, 
$ 
\IQ = \IP\IS = u\II
$ 
yields the BJR metric 
\begin{equation}
g_{\mu\nu}dx^{\mu}dx^{\nu} = u^2 \delta_{ij}dx^i dx^j + 2 du dv\,,  
\label{flatQ}
\end{equation}
whose conform-flatness follows from \eqref{RiemBJR} and (\ref{Ricci}),  which imply that the Ricci tensor vanishes. 
This metric is degenerate at $u = 0$ ;  the Souriau matrix is $\sim u^{-1}$. Choosing $u_0 \neq 0$ would not change the description. 
 
\section{Memory Effect for Geodesics}\label{MemoGeoSec}

In globally defined transverse Brinkmann coordinates the geodesics are solutions of
\beq
\dfrac{d^2 \bX}{dU^2} - \half\barray{lr}
\cA_{+} &\cA_{\times}\\{\cA_{\times}} & -\cA_{+} \earray
\,\bX = 0\,,
\label{ABXeq}
\eeq 
while $V$ satisfies a  complicated equation, \# (II.3b) in \cite{PolPer}, we reproduce for completeness,
\beqa
\dfrac{d^2V}{dU^2}
&+ \dfrac{1}{4}\displaystyle{\frac{d\cA_+}{dU\,}}\big((X^1)^2 - (X^2)^2 \big)
+\cA_{+}\big(X^1\dfrac{dX^1}{dU\,}-X^2\dfrac{dX^2}{dU\,} \big)
\nn
\\[4pt]
&+ \dfrac{1}{2} \dfrac {d\cA_{\times}}{dU\,} X^1 X^2
+ \cA_{\times}\big(X^2\dfrac{dX^1}{dU\,} + X^1\dfrac{dX^2}{dU\,} \big) \, = \, 0\,\qquad
\label{ABVeq}
\eeqa
which is indeed a consequence of the transverse ones. 

Eqn \eqref{ABXeq} is in general a pair of coupled \SL equations that can be solved only numerically.
For a linearly polarized wave $\cA_{\times}=0$ and we denoting simply $\cA_+=\cA$, \eqref{ABXeq} reduces to
\emph{two uncoupled SL equations},
\beq
\dfrac {d^2\! \bX}{dU^2} - \half\diag(\cA,-\cA)\,\bX
= 0\,.
\label{Bgeoeq}
\eeq
In the Bargmann picture \cite{Eisenhart,DBKP,DGH91} (see Appendix \ref{Appendix}), the effective linear force  $\cA\bX$  is attractive or repulsive depending on its sign.

\kikezd{Velocity effect}: Returning to \eqref{ABXeq}, 
the total increment of the velocity along the transverse trajectory $\bX(U)=\big(X^1(U), X^2(U)\big)$ between the Before and Afterzones is given by  eqn. \# (3.4) of \cite{PolPer},
\beq
\Delta\dot{\bX}=
\int_{U_B}^{U_A}\!\! {\IK}(U)\,\bX(U) dU\,
=
\half\barray{l}
\displaystyle\int_{U_B}^{U_A}\!\!
{\small \big({\cA_+}(U)X^1(U)+\cA_{\times}(U)X^2(U)\big)\, dU}
\\[14pt]
\displaystyle\int_{U_B}^{U_A}\!\!
{\small\big(
{\cA_{\times}}(U)X^1(U)-{\cA_+}(U)X^2(U)\big)\, dU}
\earray\,
\label{genveljump}
\eeq
which, for a linearly polarized special geodesic
 \eqref{genveljump} reduces to
\beq
\Delta\dot{\bX} 
= {\small
\half\barray{r}
\quad\;\displaystyle\int_{U_B}^{U_A}\!\!
 {\cA}_+(U)P_{11}(U)X_0^1\, dU
\\[16pt]
-\!\displaystyle\int_{U_B}^{U_A}\!\!
{\cA}_+(U)P_{22}(U)X_0^2\, dU\,
\earray\,} 
= {\small
\half\barray{r}
\quad\;\displaystyle\int_{U_B}^{U_A}\!\!
 {\cA}_+(U)X^1(U)\, dU
\\[16pt]
-\!\displaystyle\int_{U_B}^{U_A}\!\!
{\cA}_+(U)X^2(U)\, dU\,
\earray\,}.
\label{linpolveljump}
\eeq

The variation of the relative (euclidean) distance
${\Delta}_X(\bX,\bY)= |\bX-\bY| =\sqrt{(X^1)^2+(X^2)^2}$
and of the relative velocity ${\Delta}_{\dot{X}}= |\dot\bX-\dot\bY|$, shown in FIG.\ref{Deltavel} \emph{confirm} the \emph{velocity effect}: a particle initially at rest is \emph{not} simply displaced but has a constant  nonvanishing residual relative velocity after the wave has passed.  This could in principle be observed through the Doppler effect \cite{BraGri} and is consistent with the Newtonian behavior in transverse space which follows from the Bargmann framework.
Illustrations, (confirmed analytically \cite{Chakraborty}), will be presented in secs.\ref{Illust}, \ref{LinkingSec} and \ref{ImpSec}.
\goodbreak


Further insight is gained by switching to BJR coordinates. First we note that the quantity called the Jacobi invariant,
\begin{eqnarray}
e=\half\rg_{\mu\nu}\,\dot{x}^\mu\dot{x}^\nu
\label{econst}
\end{eqnarray}
is conserved for geodesic motion; it
 is negative/zero/positive for a timelike/null/spacelike geodesic.
Here we deal with timelike geodesics and require therefore $e<0$.

The conserved quantities associated with the isometries of the metric are determined  by Noether's theorem. Choosing $u$ as parameter
the invariant $e$ in \eqref{econst} is
$ e=\half a_{ij}(u)\dot{x}^i\dot{x}^j-\dot{v}\,$
and the Noether quantities are, within the domain of the BJR coordinates \cite{BoPi89},
\besub
\begin{align}
\bp=&\;\fa \,\dot{\bx}\,\qquad &\text{momentum $\sim\bx$-translation}
\label{pmom}
\\[4pt]
m=&\;1\qquad &\text{mass $\sim v$-translation}
\label{mmass}
\\[4pt]
\bk=&\;\bx-{\IS}\bp\,\qquad &\text{boost momentum}
\label{Cboostmom}
\end{align}
\label{CarCons}
\esub
Conversely, the conserved quantities determine the geodesics,
\besub
\begin{align}
\bx(u)=& \quad {\IS}(u)\,\bp\; +\; \bk\,,
\label{xuS}
\\
v(u)=& \;-\half \bp\cdot {\IS}(u)\,\bp + e\,u+ v_0\,,
\label{vuS}
\end{align}
\label{CarGeo}
\esub
where $v_0$ is yet another integration constant.
Thus the only quantity to calculate is the Souriau matrix, ${\IS}(u)$ in \eqref{Smatrix}.
The formulae \eqref{CarGeo} make sense where  ${\IS}$ is well-defined i.e., between two adjacent zeros $u_k$ and $u_{k+1}$ of $\fa$. 
We note for later reference that the $v$-equation \eqref{vuS} is in fact determined by the null lift of 
\eqref{xuS} and by the Jacobi invariant $e$ in \eqref{econst}.
For massless geodesics $e =0$ \eqref{econst} the coordinate $v$ is indeed the classical action for the underlying NR system \cite{Carroll4GW}.

Special geodesics  are in rest in the Beforezone,
$
\bx(u)=\bx_0, u\leq u_B\,$.
Therefore  $\bp=0$  by \eqref{pmom} --- and momentum conservation then implies that
\beq
\bp=0 
\label{pvanish}
\eeq
for \emph{all} $u$ in the domain of definition of the coordinates. Thus
 by \eqref{CarGeo} we have,
\begin{equation}
\bx=\bx_0 =\bk\,
\qquad
\&
\qquad
v = e\,u +v_0\,,
\end{equation}
i.e., simple ``vertical'' motion \emph{with fixed transverse position} --- and that
\emph{for an arbitrary BJR profile $\fa$~!}
We stress 
that this extreme simplicity is due to our using BJR coordinates.
\goodbreak

Returning to Brinkmann coordinates, our special geodesics are instead \footnote{Remember that $u=U$ and $\bX_0=\bx_0$ by choice. Eqn. \eqref{XPX0} implies that $\bX(u)$ satisfies the same {SL} eqn \eqref{SL+cond} as ${\IP}$ does. In the flat Afterzone  $(K_{ij}) = 0$, and the motion becomes approximately free, as will be illustrated in sec.\ref{Illust}.},
\besub
\begin{align}
\bX(U)= &\;{\IP}(u)\bX_0\,,
\label{XPX0}
\\[4pt]
V(U)= &\;v - \half \bX_0 ({\IP}^\dagger\dot{{\IP}})\bX_0 + eU+ V_0\,,
\end{align}
\label{specBrinkGeo}
\esub
$V_0=\const$
Unlike their BJR ancestors, these trajectories can be very complicated due to the complicated forms of the ${\IP}$ and ${\IS}$ matrices, as will be seen in sec.\ref{Illust} (see FIG. \ref{BspecGeo3d} for example).

A particular set of special geodesics in Brinkmann coordinates is obtained by choosing the initial position $\bx^{(1)}_0={\tiny \barray{c}1\\0\earray}$ and $\bx^{(2)}_0={\tiny \barray{c}0\\1\earray}$ in the Beforezone, for which we get the \emph{columns of the ${\IP}$-matrix},
\beq
\bX^{(1)}(u) = {\IP}(u) {\tiny \barray{c}1\\0\earray}=
{\tiny \barray{c}P_{11}(u)\\[5pt] P_{21}(u)\earray}
\aand
\bX^{(2)}(u) = {\IP}(u){\tiny \barray{c}0\\1\earray}=
{\tiny  \barray{c}P_{12}(u)\\[5pt] P_{22}(u)\earray}\,.
\label{Pcolumns}
\eeq

Special geodesics are at rest in the Beforezone, ${\IP}=\II$, $\bX(U)=\bX_0\equiv\bx^0$ for $U\leq U_B$; Our plots in sec.\ref{Illust} then confirm that they will be focused, as argued by Bondi and Pirani \cite{BoPi89}.

Let $\bX_a(U)$ and $\bX_b(U)$ be indeed two special geodesics; we show that for appropriate initial conditions
 there exists an $U_1$ such that the two geodesics meet,
$
\bX_a(U_1)=\bX_b(U_1)
$
for some finite and universal $U_1> U_B$. Setting  $\bX_{a,b}(U)=\bX_{a}(U)-\bX_{b}(U)$ we have
$\bX_{a,b}(U) = {\IP}(u) \bx_{a,b}^0$, so we need
 $\bx^0=\bx_a^0-\bx_b^0 \neq 0$ in the kernel of ${\IP}(u_1)$,
\beq
{\IP}(u_1) \bx^0 = 0\,.
\label{Pu0}
\eeq
Then all such geodesics
meet at the same $U_1\equiv u_1$. This requires ${\IP}(u_1)$ to be a singular matrix with a nontrivial kernel, detected by the vanishing of its determinant,
\beq
\det\big({\IP}(u_1)\big) = 0\,.
\label{detP0}
\eeq
But $\det(\fa) = (\det {\IP})^2 = \chi^4$ and therefore
$\det {\IP} = 0$ iff $\chi=0$; then Souriau's proof shows that $\chi$ necessarily vanishes at some $u_1\geq u_B$.

More generally, consider, for any $\bx_0$
$
\bx_0^* = \bx_0 + \bx^{\perp},
$
where
$
{\IP}(u_1)\bx^{\perp}=0\,.
$
Then for the geodesic issued from
$\bx_0^*$ we have,
\beq
\bX(u_1)={\IP}(u_1)\bx_0^* = {\IP}(u_1)\bx_0\,,
\eeq
i.e., all  special geodesics whose initial positions  differ by a vector $\bx^{\perp}$ end at the same point $\bX(u_1) ={\IP}(u_1)\bx_0$ \cite{BoPi89}.

The ${\IP}$-matrix of a linearly polarized wave is, for example,  diagonal,
${\IP}=\diag(P_{11},P_{22})$ and the $X^1(U)$ (resp. $X^2(U)$) components meet at those critical points where  $P_{11}(u_1)=0$ (resp. $P_{22}(u_2)=0$),
 as it will be seen in FIG.\ref{collapsegeofig} in sec.\ref{Illust}.

In conclusion, we have two different ways to find the  geodesics: either by using global Brinkmann coordinates which yield regular but complicated trajectories found by solving (numerically) the \SL equations \eqref{Bgeoeq}.

 Alternatively, we can find first our simple geodesics in BJR with help of the Carroll symmetry \eqref{CarGeo}
and then carry them to Brinkmann by \eqref{BJRBrinkP} by using the ${\IP}$ matrix. However BJR coordinates are defined only on coordinate patches between adjacent zeros
 $u_{k}$ and $u_{k+1}$ of the determinant of ${\IP}$ (or of $\fa$). Our plots in sect.\ref{Illust} show, moreover, that those BJR solutions diverge at the contact points  $u_{k}$  
 and then we have to glue together the bits of trajectories obtained by pulling back to Brinkmann coordinates from the adjacent intervals $I_k=[u_k, u_{k+1}]$ --- yielding, remarkably, \emph{smooth curves}. 
FIG.\ref{fig3C} confirms that we end up with the same result.

\subsection{Hamiltonian and Lagrangian aspects}\label{LagHamSec}

We complete our investigations with an outline of the Lagrangian and Hamiltonian aspects.
The geodesic Lagrangian resp. Hamiltonian for spinless test particle in a GW background are given by, 
\begin{equation}
L_{BJR} = \frac{1}{2} a_{ij} \dot{x}^i \dot{x}^j + \dot{u} \dot{v}    \aand  H_{BJR} = \frac{1}{2} a^{ij} p_i p_j + p_u p_v\, ,
\end{equation}
where $p_i$ and $p_v$ are constants. The transverse geodesic motion can be integrated using the conserved  \eqref{xuS}, and we end up with,
\begin{equation}
x^i(u) = x^i_0 + S^{ij}(u) \frac{p_j}{p_v}\,,
\label{xtraj} 
\end{equation}
where $p_v$ is identified with the NR mass, $M$.
In conclusion, the  motion is  determined by the Souriau matrix. 

Now we show that $\IS$  plays a key role also for the underlying NR motion obtained via null projection. The constraint 
\begin{equation}
\label{nullcond}
    H_{BJR} \equiv 0\, ,  
\end{equation}
yields indeed the (potentially $u$-dependent) lower-dimensional NR Hamiltonian
\begin{equation}
    H_{BJR}^{NR} = -p_u = \frac{a^{ij}p_ip_j}{2 M}\, .
\end{equation}
The momentum $p_u$ is identified with the 
 Hamiltonian $H_{BJR}^{NR}$, and the coordinate $v$ is related to the NR action.  Written in terms of velocities, we see that 
\begin{equation}
H^{NR}_{BJR} = L^{NR}_{BJR} = \frac{1}{2} M a_{ij} (x^i)' (x^j)' 
\end{equation}
where the ``prime" denotes derivation w.r.t. NR time, $u$. The null condition \eqref{nullcond}
implies, 
\begin{equation}
v' = - \frac{1}{2} a_{ij} (x^i)' (x^j)' = - \frac{1}{M} L_{NR},
\end{equation}
whose integration yields, 
\begin{equation}
v (u) = - \frac{1}{M} \int^u\!\! L_{NR} dt = - \frac{1}{M} S^{NR}_{BJR} = -\frac{p_i p_j}{2 M^2}\int^u \!\!a^{ij} dt = -\frac{S^{ij}(u)p_i p_j}{2M^2}\, ,   
\label{vActS}
\end{equation}
where $p_i =M a_{ij}(x^j)' $. 
Thus  for when $e=0$ in \eqref{econst},  the Hamiltonian classical action and the Souriau's matrix ${\IS}$ are interrelated~:  $v$ is indeed (minus) the classical Hamiltonian action of the underlying NR system
(identified  with Hamilton's principal function)
  obtained by projecting a null geodesic.

The Souriau's matrix determines the classical NR action (identified  with Hamilton's principal function):
\begin{equation}
v(u) = - \frac{1}{M} S^{NR}_{BJR} = -\frac{S^{ij}(u)p_i p_j}{2 M^2}\,.
\end{equation}

The Hamiltonians in Brinkmann and BJR coordinates are related by a canonical transformations \cite{Elbistan:2023pqb}. The coordinate transformation 
$ 
V = v - \frac{1}{4} (a_{ij})' x^i x^j 
$ 
implies that 
\begin{equation}
L^{NR}_{B} = L^{NR}_{BJR} +  \frac{d F}{dU}\,,
\where
F = \frac{M}{4} (a_{ij})' x^i x^j \,.
\end{equation} 
Therefore the Hamilton principal function is, 
\begin{equation}
S_{B}^{NR}(U) = \int^U\!\! L_{B}^{NR} d\tilde{U} = \frac{S^{ij}(u) p_i p_j}{2M} + F(u) . 
\end{equation}
The Brinkmann geodesics are determined by using 
$ 
X^i = P^i_j x^j\, ,
$ 
\begin{equation}
X^i(U) = P^i_j \big(x^j_0 + S^{jk}(u) \frac{p_k}{M} \big)\,. 
\end{equation}
Both matrices  $\IP$ and $\IP\IS$  satisfy the same SL equation, 
\eqref{SLS}, therefore, $X^i(U)$ obeys the geodesic equation $(X^i)'' = K^{i}_{\;j} X^j$. 

\subsection{Multiple B versus  BJR correspondence}\label{MultiBBJR}

Let us now return to the coordinate transformations defined by the matrix ${\IP}$ \eqref{BJRBrinkP} which swaps the Brinkmann and BJR metrics. 
Starting with a given Brinkmann expression, we may consider another BJR metric using  instead of $\IP$, $\widetilde{\IP}=\IP\IS$ 
which however satisfies the same aumented SL system as ${\IP}$ does,
\begin{equation}
\ddot{\widetilde{\IP}} = \IK\,\widetilde{\IP},  
\qquad (\widetilde{\IP})^{\dagger}\dot{\widetilde{\IP}} = \dot{\widetilde{\IP}}^{\dagger} \widetilde{\IP}\,,    
\end{equation}
Accordingly, the BJR metric defined by,
\begin{equation}
\label{ctBBJRalt}
{X}^i = (\widetilde{\IP})^i_j(U) x^j,   
\quad  U = u, \quad 
V = v - \frac{1}{4} x^i \dot{\tilde{\fa}}_{ij} x^j 
\end{equation}
is connected to the old one by a similarity transformation,
\begin{equation}
d{\tilde{s}}^2 = \tilde{\fa}_{ij} dx^i dx^j + 2 du dv 
\with 
\tilde{\fa} = (\widetilde{\IP})^{\dagger} (\widetilde{\IP}) = \IS \,\fa\, \IS\,.
\label{tildemetric}
\end{equation}
 
Thus to a given Brinkmann metric we can associate two BJR metrics, defined by $\IP$ and  by $\widetilde{\IP}=\IP\IS$,
respectively. 
The new Souriau matrix, Lagrangian and geodesic equations,
\beq
\big(\widetilde{\IS}\big)^{ij} = \int \tilde{a}^{ij} d\tilde{u}\,,
\quad
L = \frac{1}{2} \tilde{a}_{ij} \dot{x}^i \dot{x}^j + \dot{u} \dot{v}\,, 
\quad
\ddot{x}^i = \tilde{a}^{ij} \dot{\tilde{a}}_{jk} \dot{x}^k
\label{S-L-eqmot}
\eeq
 change accordingly.
\goodbreak
 
Note that switching from $\IP \to\IP\IS$ is  \emph{different} from the (rigid) gauge transformation (\ref{RPgauge}) considered in Section \ref{gaugesec}, where a single BJR profile  corresponds to many Brinkmann metrics related by an $\Ort(2)$ rotation. Here we associate, conversely,   multiple BJR metrics to a given Brinkmann metric. 

Consistently with \eqref{infCarr}, the symmetries involve the associated Souriau matrix,
\begin{equation}
\label{Souriaub}
\widetilde{\IS}^{ij} = \int \!{\widetilde{\fa}}^{ij}(\tilde{u}) d\tilde{u}\,. 
\end{equation}
Infinitesimal boosts are implemented, in particular, by $Y^i_b = \widetilde{\IS}^{ij}\partial_j - x^i\partial_v$. Pulling it back to Brinkmann coordinates via (\ref{ctBBJRalt}) comes with a surprise: we get  \emph{translations} (\ref{Ptrans}) instead of the expected boosts. The pull back of \emph{translations} $\frac{\partial}{\partial x^i}$ is, conversely,  a boost (\ref{PSboost}).

This procedure can be repeated in ever increasing order. Combining  the  square-root $\IP \IS$ of $\tilde{\fa}$ (\ref{ctBBJRalt}) with the Souriau matrix $\widetilde{\IS}$ (\ref{Souriaub}), we set 
\beq
\IIQ = \IQ\widetilde{\IS} =\IP \IS\,\widetilde{\IS}\,.
\eeq 
The new matrix satisfies again the augmented SL equation \eqref{SL+cond}. Therefore the transformation
\begin{equation}
\label{ctBBJRalt2}
X^i = (\IIQ)^i_j(U) x^j,   \quad  U = u , \quad V = v - \frac{1}{4} x^i \dot{\fc}_{ij} x^j 
\end{equation}
yields yet a new BJR metric, 
\begin{equation}
\label{BJRc}
\tilde{g} = \fc_{ij} dx^i dx^j + 2 du dv,
\qquad
\fc = \IIQ^\dagger \IIQ = 
 \widetilde{\IS}\, \,\tilde{\fa} \,\widetilde{\IS} \,. 
\end{equation}

Then one can continue and introduce further new matrices 
\beq
\IP \to \IP\IS \to (\IP \IS) \widetilde{\IS}\, ...\, ,
\label{PSS}
\eeq
 with associated Souriau matrices $\IS \to \widetilde{\IS} \to ...$  
 All such matrices in \eqref{PSS} satisfy the augmented SL equation and thus define multiple BJR metrics.

This scheme reminds us Lipkin's zilches \cite{Lipkin}, obtained from optical helicity by replacing the fields with their  curls, see e.g. \cite{EHZduality,CCL}  and references in them.

\section{Illustrating examples}\label{Illust}

\subsection{Linearly polarized burst with Gaussian profile}\label{linpolGaussEx}

Our simplest illustration is a linearly polarized (approximate) sandwich wave, \GW $\cA_{\times}=0$, with \underline{Gaussian profile}
(In Bargmann terms \cite{DBKP,DGH91,ZHAGK}, an anisotropic oscillator with time-dependent frequency (Appendix \ref{Appendix}),
\beq
K_{ij}(U)X^iX^j=\frac{e^{-U^2}}{\sqrt{\pi}}\big((X^1)^2-(X^2)^2\big)\,,
\label{Gaussprof}
\eeq
shown in FIG.\ref{linGaussfig}.
\begin{figure}[h]
\includegraphics[scale=0.45]{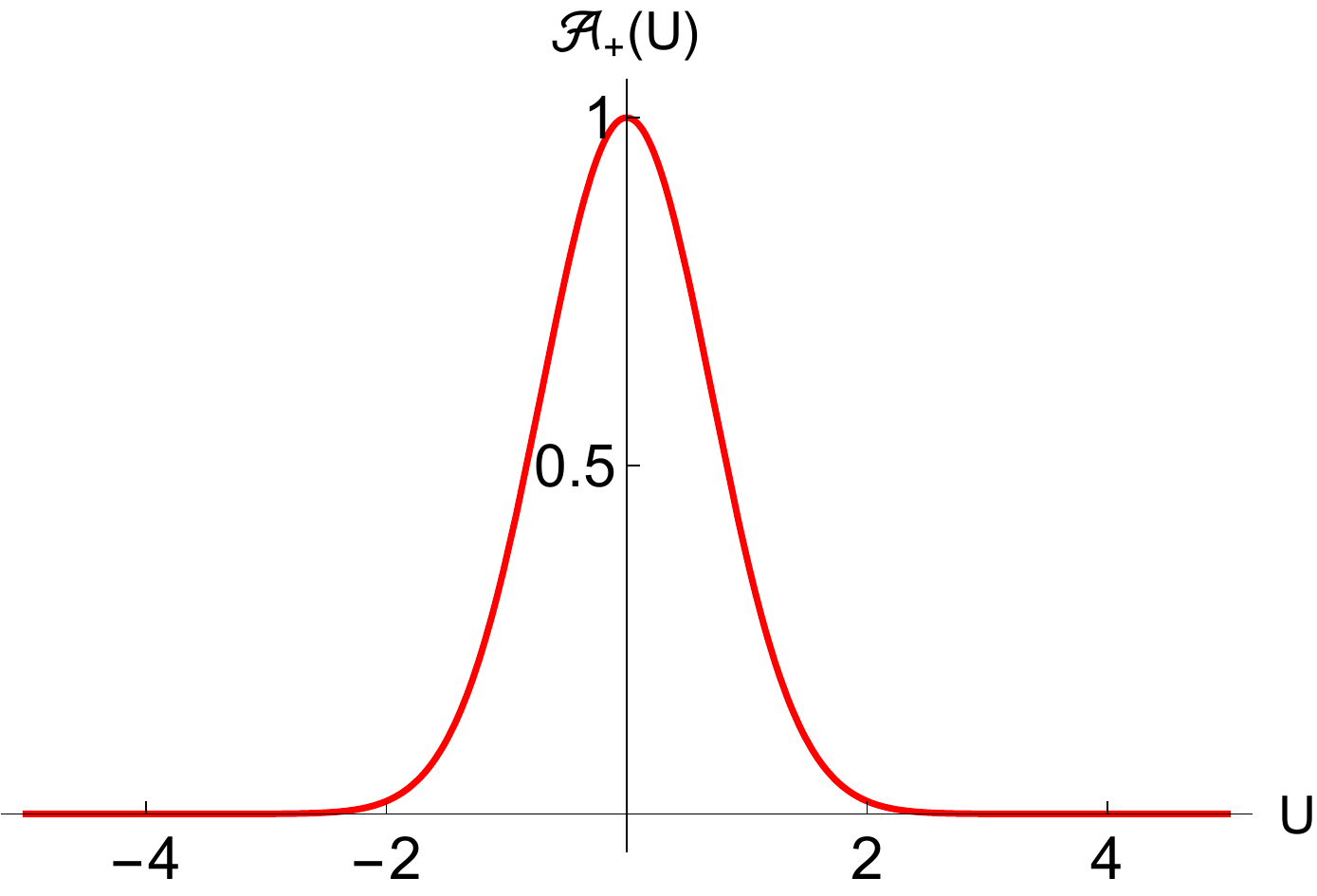}
\\
\caption{\textit{\small Linearly polarized \GW with \underline{Gaussian profile}, $\cA_+(u)=\exp(-u^2),\, \cA_{\times}=0$\,.
}}
\label{linGaussfig}
\end{figure}
The geodesics are solutions of eqn. \eqref{ABXeq}--\eqref{ABVeq}.
The $X^{1,2}$-components are decoupled from the $V$-component:
 the projection of the $4d$  worldline to the transverse $(X^1-X^2)$ plane is independent of  $V(U_0)$ and of $\dot{V}(U_0)$
 \footnote{For a massless particle, the $V$ component is a horizontal lift of the transverse motion $\bX(u)$, supplemented by a term which is linear in $u$ in the massive case \cite{ZCEH,nonlocal,Elbistan:2023pqb}.}.
 By assumption, the particle is at rest in Beforezone:
\beq
\bX(U)=\bX_0,\quad\dot{\bX}(U)=0\, \for
U \leq U_B\,.
\label{Grestcond}
\eeq
Numerical integration then yields FIG.\ref{linGaussgeo}.
\begin{figure}[h]
\hskip-2mm
\includegraphics[scale=.195]{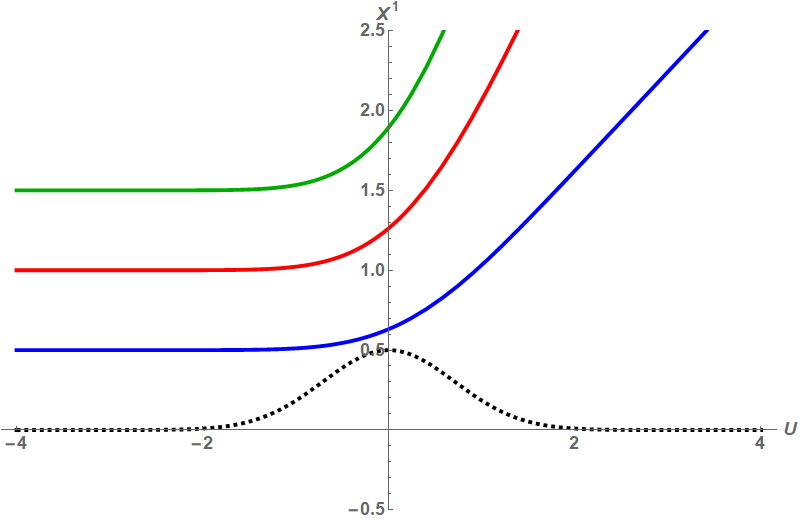}\!\!
\includegraphics[scale=.195]{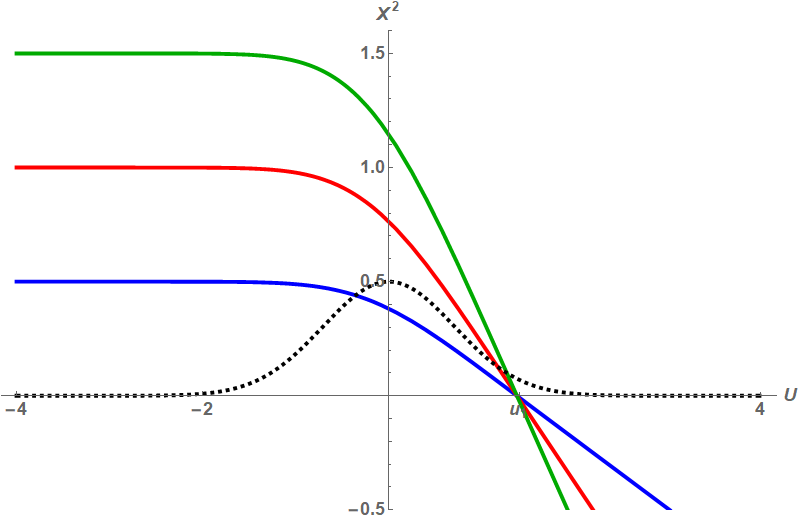}\!\!
\includegraphics[scale=.195]{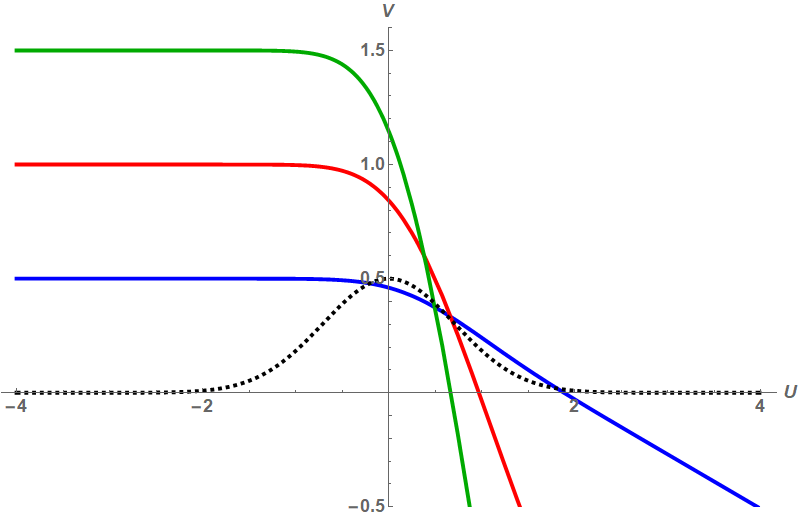}
\\{}\vskip-4mm
\caption{\textit{\small Geodesics in a Gaussian burst issued from  various (\blue{blue}/\red{red}/\dgreen{green})~ initial rest positions in the Beforezone. By \eqref{Bgeoeq} the effective force is repulsive in the $X^1$ coordinate and is attractive in that of $X^2$  where it  focuses, for all initial positions at $(0, X_0^2)$, at ${\mathbf u_1}= 1.37653$. In the ``far-After zone'' i.e., beyond ${\mathbf u_1}$, the trajectories follow diverging straight lines.
}
\label{linGaussgeo}
}
\end{figure}
The variation of the relative (euclidean) distance and  the relative velocity,
\beq
{\Delta}_X(\bX,\bY)= |\bX-\bY|\aand
{\Delta}_{\dot{X}}= |\dot\bX-\dot\bY|,
\label{relposvel}
\eeq
shown in FIG.\ref{Deltavel} \cite{LongMemory} are consistent with the  \emph{velocity effect} discussed in sec.\ref{MemoGeoSec}.
\begin{figure}[h]
\hskip-3mm
\includegraphics[scale=.19]{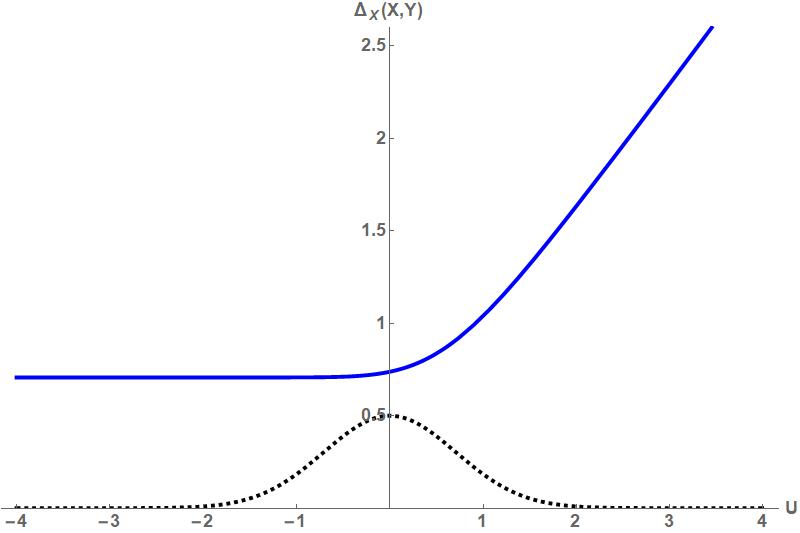}\;
\includegraphics[scale=.2]{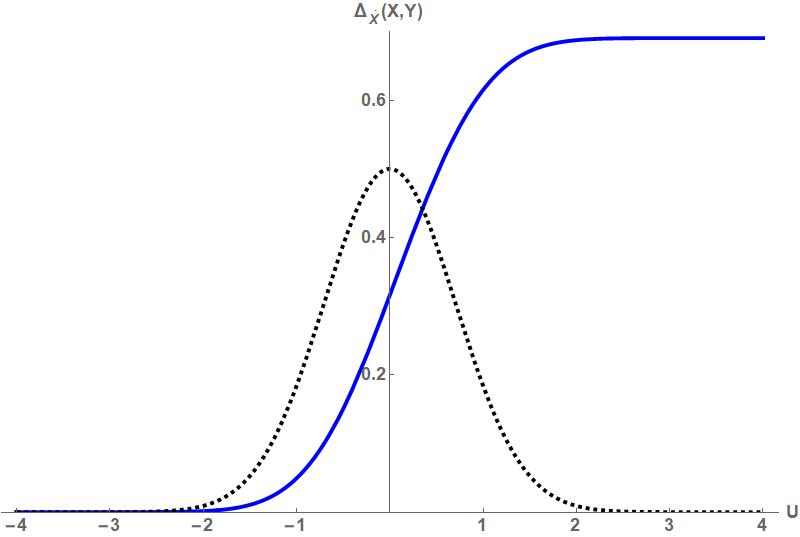}
\\
\vskip-3mm
\caption{\textit{\small The relative velocity of particles initially at rest tend to a residual nonvanishing constant value after the wave has passed.
}}
\label{Deltavel}
\end{figure}
\goodbreak

\subsection{Linearly polarized sandwich wave, modelling  flyby}\label{flybySec}

In \cite{GibbHaw71} Gibbons and Hawking proposed to describe \emph{flyby} by a Brinkmann profile  proportional to  \emph{first derivative} of a  Gaussian, 
\beq
\cA_{\times}=0 \aand
\cA(U) =
\frac{1}{2}\,\frac{d(e^{-U^2})}{dU}\,.
\label{flybyprof}
\eeq 
\begin{figure}[h]
\hskip-4mm
\includegraphics[scale=.19]{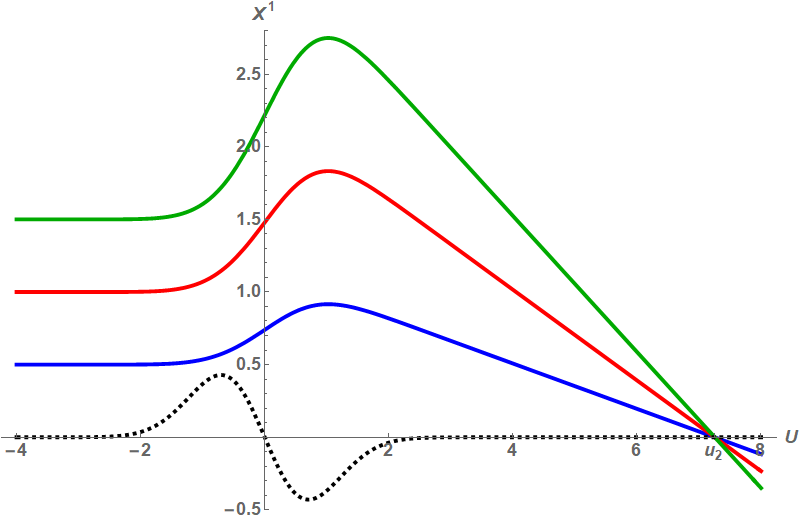}\!
\includegraphics[scale=.19]{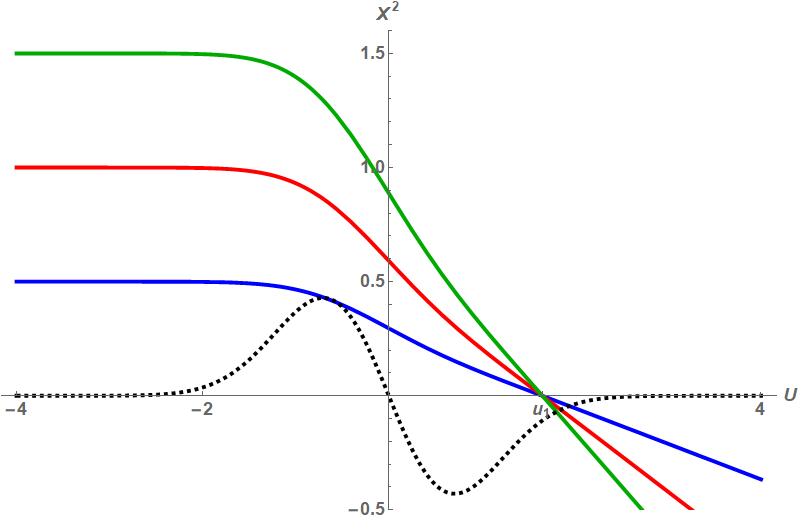}\!
\includegraphics[scale=.19]{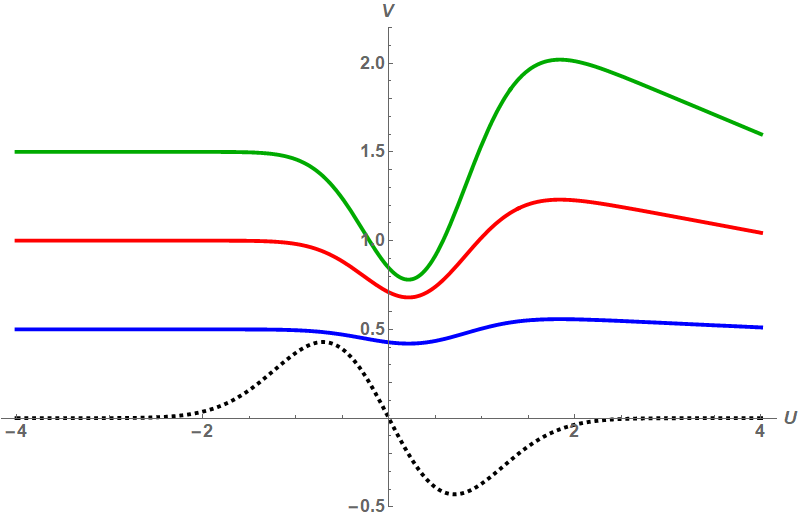}\vskip-2mm
\caption{\textit{\small Evolution of geodesics with various resting positions for the \underline{flyby}  profile \eqref{flybyprof} proposed in \cite{GibbHaw71} shown in the background. The singular points where all components are focused are ${\bf u_2}= 7.25437$ for $X^1$, and ${\bf u_1}= 1.65034$ for $X^2$.}
\label{flybygeo}
}
\end{figure}

Further insight can be gained from the Tissot diagram \cite{Tissot}:
The evolution of (sorts of) ``smoke rings'' indicate the distorsion  of a circle of particles which are at rest in the Beforezone due to the \GW.

\null\bigskip
\begin{figure}[ht]
\includegraphics[scale=.48]{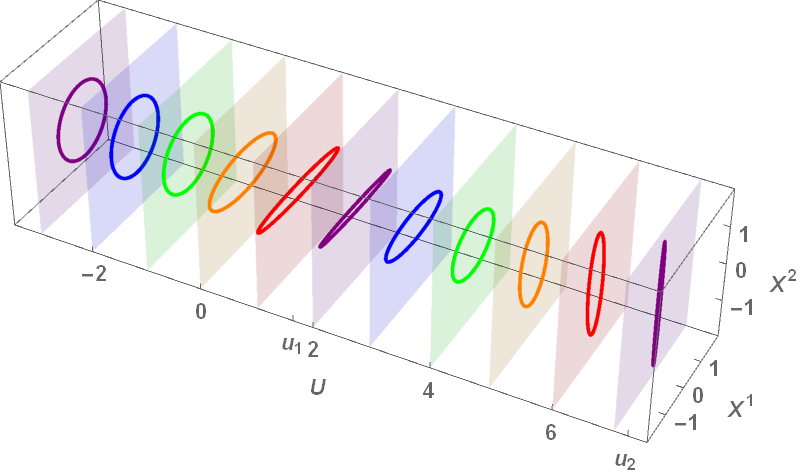}
\vskip-3mm
\caption{\small \textit{Tissot  diagram  for  the  \underline{flyby profile}  \eqref{flybyprof}, to be compared with fig.\ref{flybygeo}. The initial Tissot ring degenerates to a line segment at the values ${\bf u_2}$ and ${\bf u_1}$ where one or the other component vanishes.
}
\label{flybyTissot}
}
\end{figure}
\noindent
The numerical results \cite{LongMemory,Elbistan:2023pqb} are  reproduced in FIGs.\ref{flybygeo} and \ref{flybyTissot}.
\goodbreak

\subsection{The Braginsky -- Thorne system}\label{BragThornSystem} 
 
 The ``burst with memory" scenario proposed by Braginsky and Thorne  \cite{BraTho} to observe \GWs has a linearly polarized profile which involves  the \emph{second derivative} of a Gaussian, 
\beq
\cA_{+}(U)= 
 \frac{1}{2}\,\frac{d^2(e^{-U^2})}{dU^2}
\,.
\label{Kd2}
\eeq
The geodesics are depicted in FIG.\ref{Gauss2Geo}.  
\begin{figure}[h]
\hskip-3mm
\includegraphics[scale=.195]{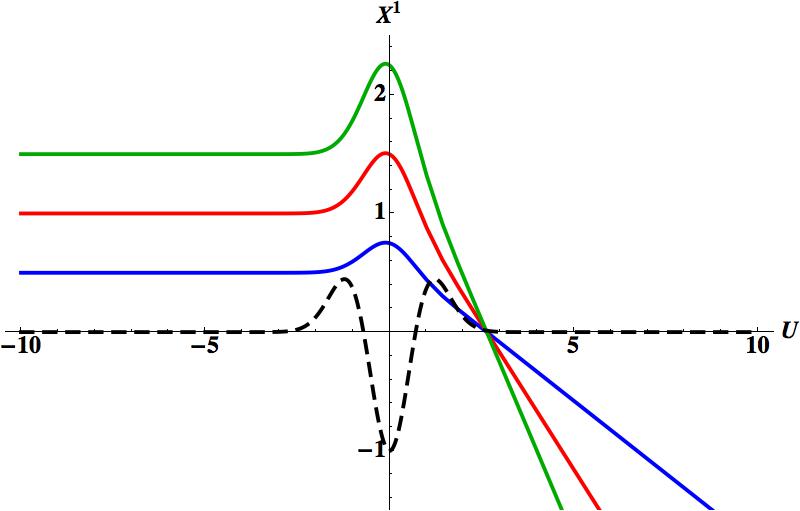}
\includegraphics[scale=.195]{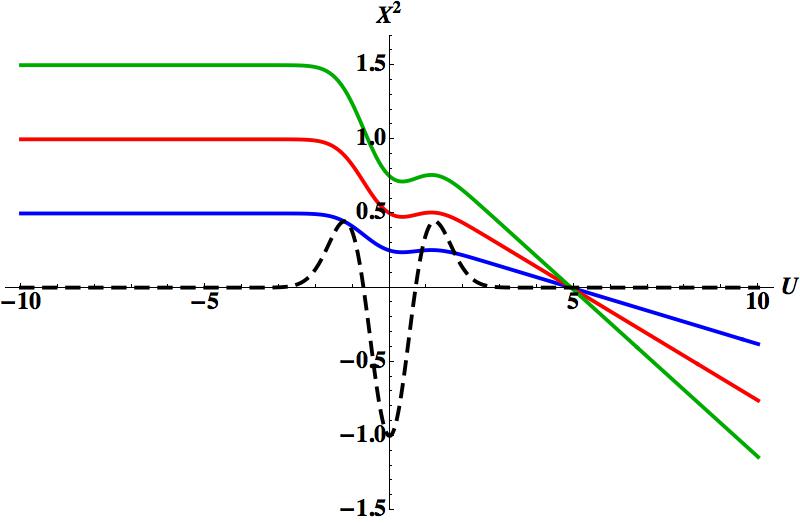}
\includegraphics[scale=.195]{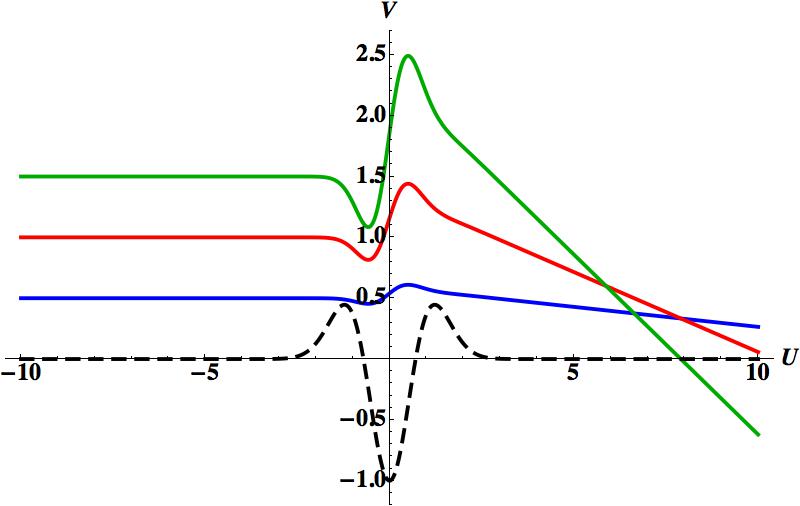}\\
\vskip-2mm
\caption{\textit{The \underline{burst with memory wave} proposed by Thorne and Braginsky \cite{BraTho} corresponds to the second derivative of a Gaussian, (\ref{Kd2}). All transverse components are focused, however at different points for the components.}}
\label{Gauss2Geo}
\end{figure}

\subsection{Linearly polarized sandwich wave, modelling  gravitational collapse}\label{CollapseSec}

Next we focus our attention at  the {linearly polarized} (approximate) sandwich wave, $\cA_{\times}=0$, with Brinkmann profile
\eqref{genBrink} proposed in \cite{GibbHaw71} to model \emph{gravitational collapse}, with profile
\beq
{\IK}(U)=\cA(U)\diag(1,-1)\,, \qquad
\cA(U)
 =\frac{1}{2}\,\frac{d^3(e^{-U^2})}{dU^3}\,.
\label{collapseprofile}
\eeq
The numerical solution of the Brinkmann geodesic eqn. (\ref{Bgeoeq}) for the collapse profile proposed in \cite{GibbHaw71} is shown in FIG.\ref{collapsegeofig}. 
\begin{figure}[h]
\hskip-3mm
\includegraphics[scale=.25]{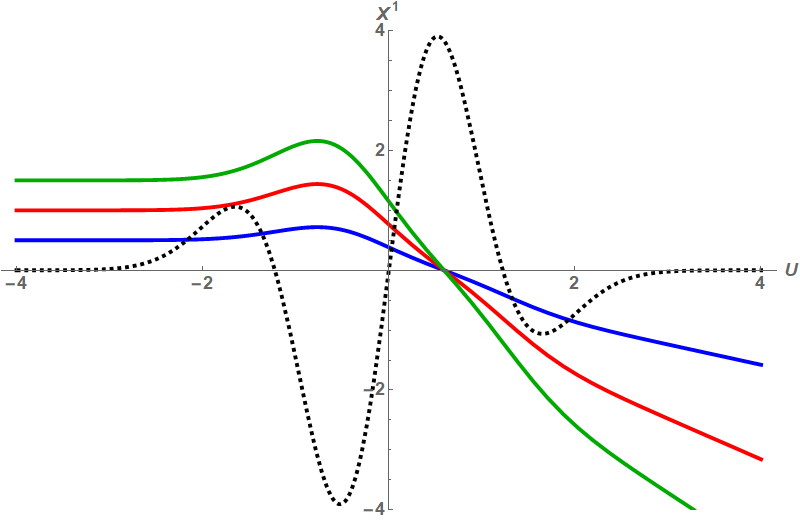}\quad
\includegraphics[scale=.25]{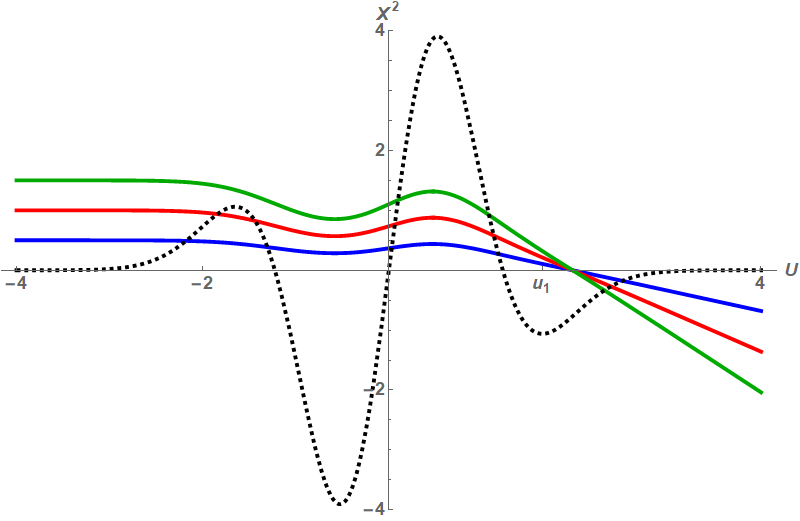}
\\
\vskip-3mm
\caption{\textit{\small The special (Brinkmann) geodesics at rest at various (\blue{\bf blue}-\red{\bf red}-\dgreen{\bf green}) positions in the Beforezone of the in the \underline{gravitational collapse} wave \cite{GibbHaw71},  (\ref{collapseprofile}) (whose profile is dotted in the background). They
 converge where the effective potential in \eqref{Bgeoeq} is attractive and diverge where it is repulsive. The components are focused for ${\bf u_1}=0.593342$ (resp. at ${\bf u_2} = 1.97472$)
 which separate  the regions where attraction turns into repulsion.
 In the flat Afterzone the motion is along straight lines with constant non-zero velocity.}}
\label{collapsegeofig}
\end{figure}
\begin{figure}[h]
\hskip-6mm
\includegraphics[scale=.3]{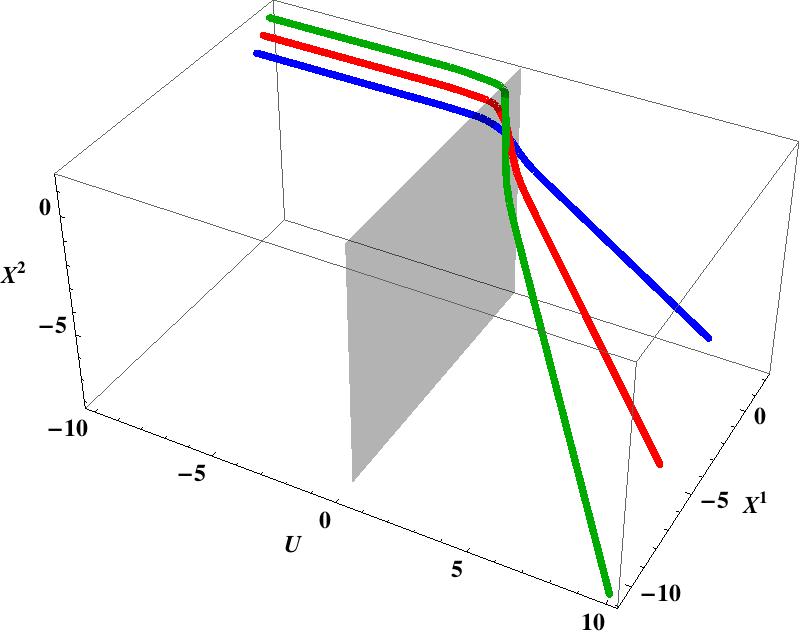}
\\
\vskip-3mm
\caption{\textit{\small The trajectories in the gravitational collapse metric \eqref{collapseprofile}, unfolded to $(2+1)$D,
diverge along straight lines in the Afterzone\,.
}
\label{3dtrajectory}
}
\end{figure}
Consistently with \eqref{specBrinkGeo}, all {special} geodesics  $\bX(u)$  whose initial position $\bX_0$ belongs to the kernel of ${\IP}(u)$ are focused at a caustic point \cite{BoPi89}. This happens where $\det {\IP}(u_i)=0,\, {i=1,2}$, or equivalently,
where $\chi(u_i)=0$.
 FIG.\ref{collapsegeofig}  shows the focusing  at $u_1$ (resp. at $u_2$) for the $X^1$ (resp. ($X^2$) component, cf. \cite{BoPi89}.
Figs. \ref{collapsegeofig} and \ref{3dtrajectory} hint also at that
  the motions become again free  in the flat Afterzone. From the Bargmann point of view \cite{Eisenhart,DBKP,DGH91,ZHAGK}: the 4d geodesics project to those of a free non-relativistic particle in $(2+1)$d which moves consistently with Newton's~1st~law.
\goodbreak

\medskip

\begin{figure}[h]
\includegraphics[scale=.57]{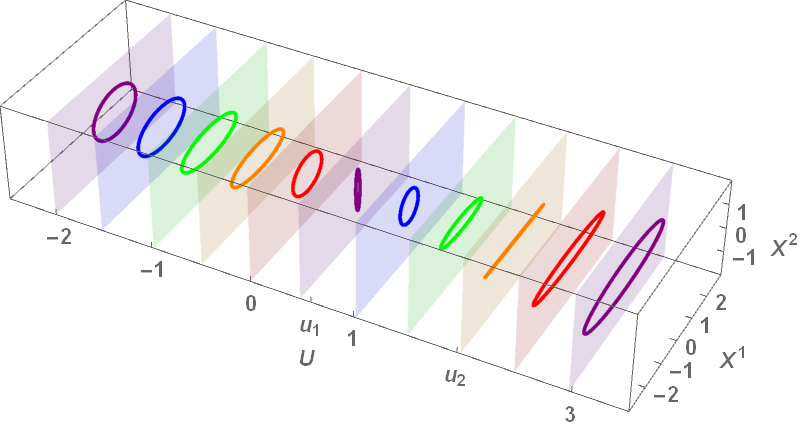}
\vskip-4mm
\caption{\textit{\small Tissot diagram for the
linearly  polarized sandwich wave \eqref{collapseprofile}.
At the focal values ${\bf u_1}=0.593342$ and  at ${\bf u_2} = 1.97472$, respectively, the initial ring degerates to a vertical resp. horizontal  segment, corresponding to the vanishing of one or the other column of the ${\IP}$-matrix.
}
\label{collapseTissot}
}
\end{figure}

\goodbreak
\kikezd{In Brinkmann coordinates}.
Outside the (approximate) Wavezone the Brinkmann profile vanishes, ${\IK} \approx 0$, and
the {SL} equation \eqref{SL+cond}  reduces, in both the Before and the After zones, to the free equation,
\beq
\ddot{\IP} \approx 0 \Rarrow
\IP(u) \approx \IC\,u + \ID\,
\where \IC=\diag(C_1,C_2)\,,\; \ID=\diag(D_1,D_2)
\label{PSL}
\eeq
\!with   constants $C_1, C_2$ and $D_1, D_2$.
Remember that, by \eqref{XPX0}, the columns of the ${\IP}$-matrix can also be viewed as particular geodesics, $\bX_1(u)= {\tiny \barray{c}P_{11}(u)\\ 0\earray}$  and $\bX_2(u)= {\tiny \barray{c} 0\\ P_{22}(u) \earray}$.
The  initial conditions are cf. \eqref{specPinBz},
\beq
\left\{\barraynb{llll}
 P_{11}(u_0^B)= 1,\quad P_{22}(u_0^B) &=& 1
\\[4pt]
\dot{P}_{11}(u_0^B)=\:0,\quad \dot{P}_{22}(u_0^B) &=& 0
\earraynb\right.
\;\for u_0^B < u_B\,,
\label{BzonePinit}
\eeq
in the Beforezone, implying that
$
C_1=C_2=0
$
there. In the Afterzone we have instead,
\beq
\left\{\barraynb{lll}
 P_{11}(u_0^A) = D_1\,,\quad P_{22}(u_0^A) &=& D_2
 \\[4pt]
\dot{P}_{11}(u_0^A)=C_1\,, \quad \dot{P}_{22}(u_0^A) &=&\;C_2
\earraynb\right.
\;\for u_0^A > u_A\,.
\label{AzonePinit}
\eeq
Thus
\besub
\begin{align}
&
{\IP}(u) = \diag(1,1)
&\for u < u_B\,,
\label{11BeforeP}
\\
&
{\IP}(u) = \diag(C_1u+D_1,C_2u+D_2)
&\for u > u_A\,.
\label{11AfterP}
\end{align}
\label{Outsidesol}
\esub
\goodbreak
\vskip-10mm
\kikezd{BJR}.
The BJR  matrix $\fa=(a_{ij})$ \eqref{BJRmetric} is obtained by following the recipe of sec.\ref{BrinkSec}. By \eqref{PSL},
\beq
\fa = {\IP}^2  =
\left\{\barraynb{cll}
\diag(1,1) &\quad \text{in Beforezone} &u< u_B
\\[6pt]
\diag\Big((C_1u+D_1)^2,(C_2u+D_2)^2\Big)
&\quad \text{in Afterzone} &u_A < u\quad
\earraynb\right.\;,
\label{alinpol}
\eeq
shown in FIG.\ref{linpol-a}.

By \eqref{CarGeo} the
trajectories in BJR coordinates are determined by the conserved quantities $\bp$ and $\bk$, and by the Souriau matrix ${\IS}$ in \eqref{Smatrix}.

The determinant \eqref{chigamma} satisfies the Sturm-Liouville equation \eqref{chiSL}, whose
behavior  depends on the frequency $\omega(u)$, depicted in FIG.\ref{omegafig}. The determinants $\det{(\IP)}$ and $\det({\fa})$ vanish when either $P_{11}$ or $P_{22}$ does, i.e., at ${\mathbf u_1}$ and ${\mathbf u_2}$, respectively.
%
\begin{figure}[h]
\includegraphics[scale=.34]{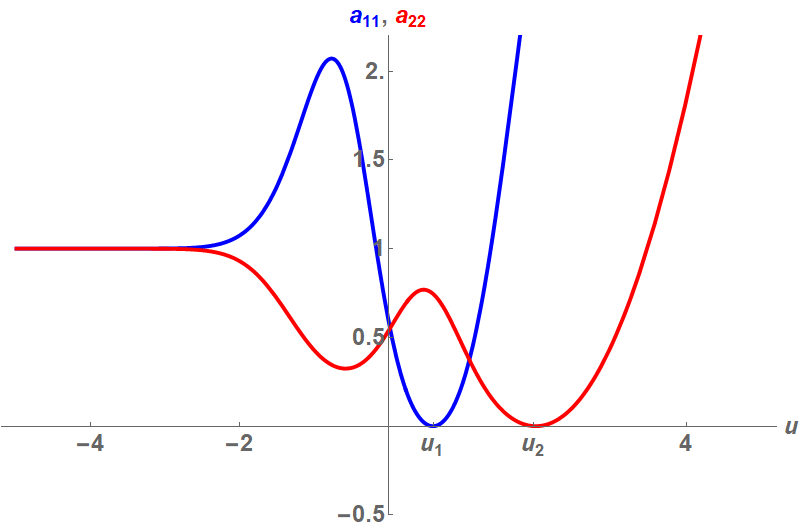}
\vskip-5mm
\caption{\textit{\small  The BJR profile $\fa=(a_{ij})$ of the
collapse wave, whose Brinkmann profile is \eqref{collapseprofile}.
\blue{$a_{11}$} = \red{$a_{22}$} $=1$ in the Beforezone.  The zeros  ${\mathbf u_i}, i=1,2$ are where one or the other component vanishes (and where the Tissot ring degenerates to a segment, cf. FIG.\ref{collapseTissot}).}
\label{linpol-a}
}
\end{figure}

\begin{figure}
\includegraphics[scale=.34]{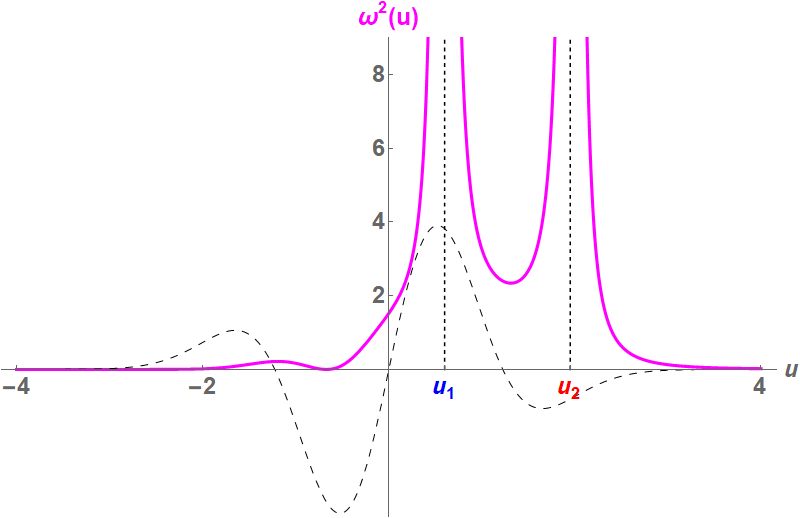}
\\
\vskip-3mm
\caption{\textit{\small In BJR, the frequency \blue{$\bomega$} in the Sturm-Liouville equation \eqref{chiSL} for the linearly polarized collapse wave (with Brinkmann profile \eqref{collapseprofile}  of
 \cite{GibbHaw71} shown in the background) diverges at the  points ${\bf u_i}, i=1,2$ where the determinant of the metric  vanishes, $\chi(u_i)$.
}}
\vskip-2mm
\label{omegafig}
\end{figure}
\goodbreak

From eqn \eqref{chiSL} we infer that to keep $\ddot{\chi}$ finite  the frequency $\omega$ must diverge to infinity at a zero of the determinant,  $\lim_{u\to u_i} \omega = \infty$ when $\chi(u_i)=0,\, i=1,2$.
Outside the Wavezone the frequency $\omega$ is approximately zero and the SL eqn \eqref{chiSL} reduces to  $\ddot{\chi}\approx 0$, implying that  $\chi$ is approximately linear,
\beq
 \chi\approx cu+d\,.
 \eeq
The condition
$c=0$ in the Beforezone changes to $c\neq0$ in the Afterzone, due precisely to the wave.
Therefore $\chi$ vanishes there at most once, consistently with  FIGs.\ref{omegafig} and~\ref{linpolchi}.

\begin{figure}
\includegraphics[scale=.27]{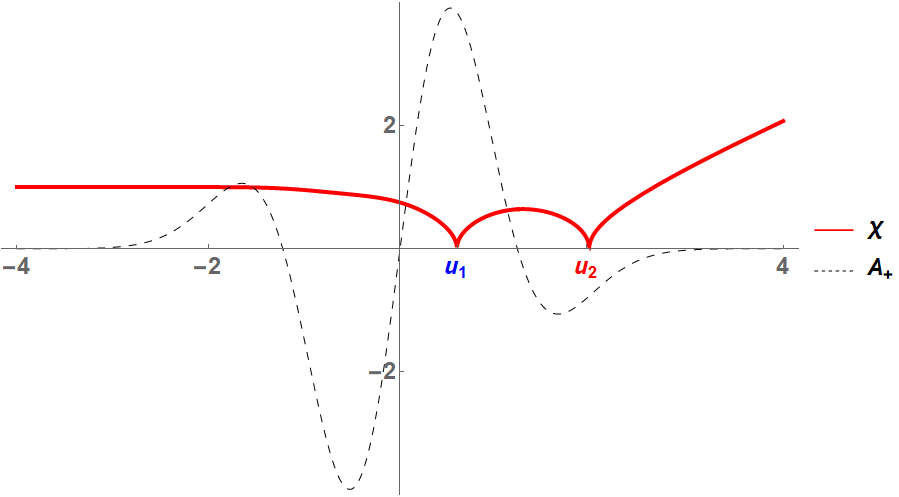} \\
\null\vskip-12mm
\caption{\textit{\small  \red{$\bchi=(\det(\fa))^{1/4}$} for the linearly polarized ``collapse'' wave \eqref{collapseprofile}.
The zeros of \red{$\bchi$} coincide with the points ${\bf u_i},\,i= 1,2$, where the Brinkmann trajectories are focused, and also where the frequency \blue{$\bomega$} diverges, cf. (FIG.\ref{omegafig}).
In the flat Outside regions \red{$\bchi$} is approximately linear.
}}
\label{linpolchi}
\end{figure}

\begin{figure}[h]
\includegraphics[scale=.42]{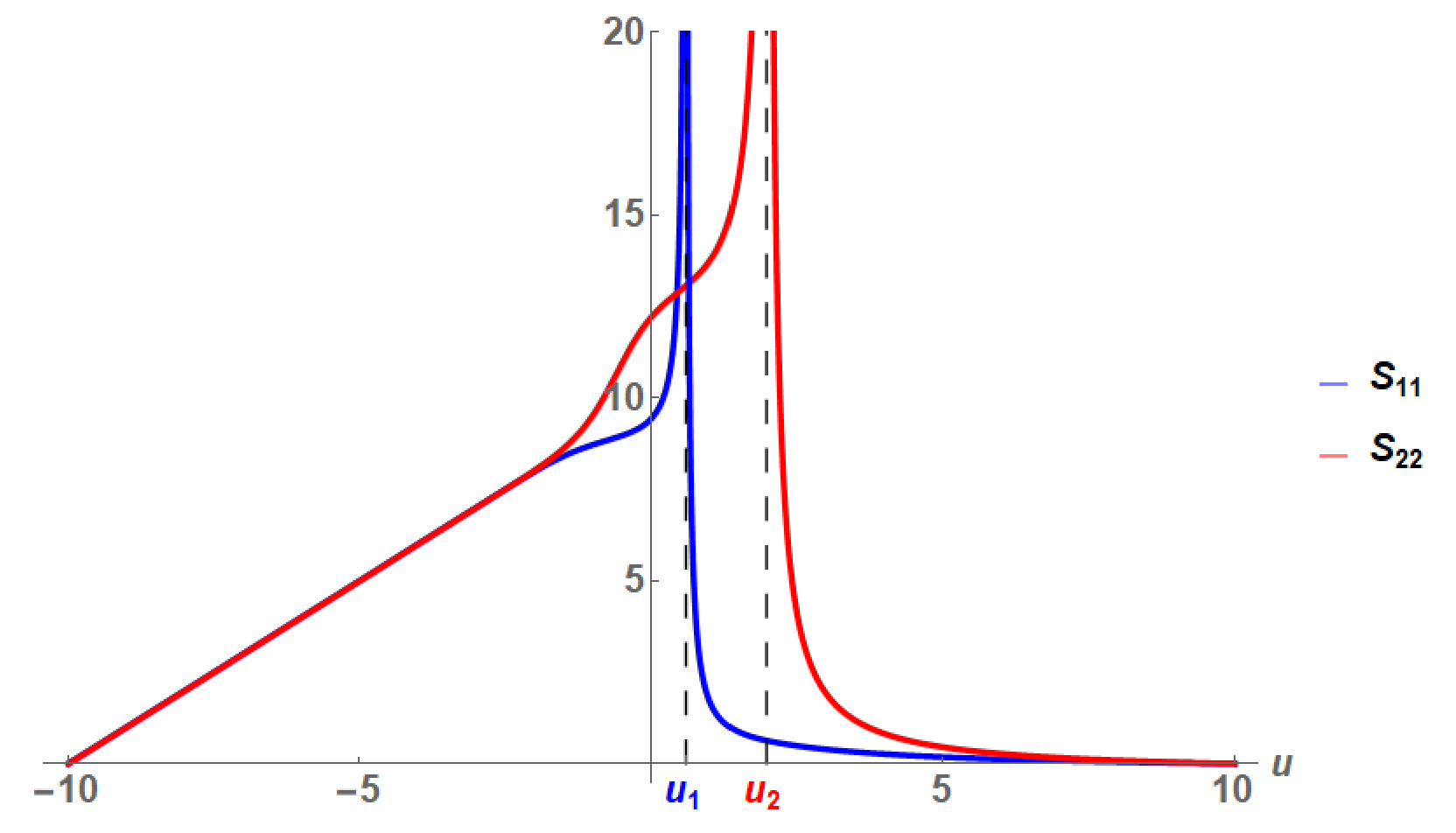}
\vskip-3mm
\caption{\textit{\small  The  Souriau matrix ${\IS}$  (\ref{Smatrix}) provides us, in BJR coordinates, both with the geodesics \eqref{CarGeo} and also with (distorted) boosts, implemented as in \eqref{genCarract}.
${\IS}(u)\approx u{\,}\II$ in the Beforezone, then diverges at \blue{${\mathbf u_1}$} and at \red{${\mathbf u_2}$} and  falls off rapidly in the Afterzone.
}}
\label{Smatrixfig}
\end{figure}
%
The behavior of the Souriau matrix we read off FIG.\ref{Smatrixfig} has remarkable consequences. In the Beforezone, it behaves as in the Minkowski/Galilei case, but in the Afterzone  ${\IS}\approx 0$,
and \eqref{CarGeo} reduces to
\beq
\bx \approx \bk = \const
\qquad
 v(u) \approx  e\,u+ v_0\,,
\label{afterboost}
\eeq
so that  the geodesic is ``vertical''.  The transverse  coordinates are fixed, confirming that in BJR coordinates \emph{``Carroll particles do not move"} [in transverse space].
\begin{figure}
\includegraphics[scale=.3]{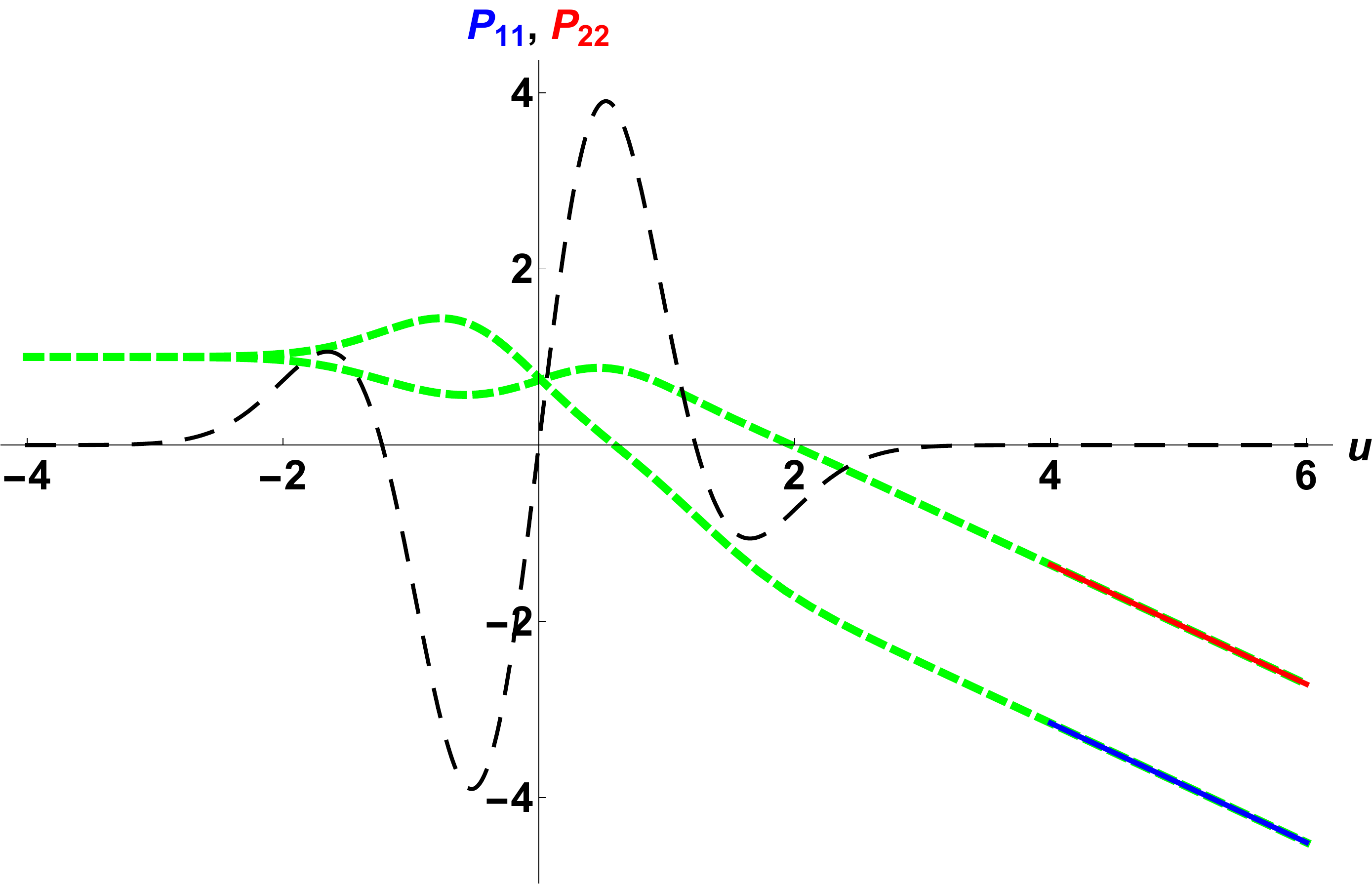}\\
\null\vskip-12mm
\caption{ \textit{\small In a linearly polarized ``gravitational collapse wave'' \eqref{collapseprofile}
the numerically found Brinkmann trajectories in the flat Afterzone are  straight lines, which overlap perfectly with those obtained in BJR coordinates and then carried to Brinkmann by \eqref{BJRBrinkP}.
\\
}}
\vskip-7mm
\label{fig3C}
\end{figure}
\smallskip

FIG.\ref{fig3C} 
 confirms that carrying the BJR solutions to Brinkmann by the map \eqref{BJRBrinkP} removes the divergences. The pieces fit together and we end up with the same regular curve obtained before numerically \cite{SLC,ZCEH,ShortMemory}.
 \goodbreak

\kikezd{Distorted Carroll symmetry}.
 According to \eqref{genCarract}, a boost with parameter $\bb$ acts, in BJR coordinates, as,
\beq
\bx \to \bx + {\IS}(u)\,\bb\,
\label{BJRboost}
\eeq
where ${\IS}(u)$ is the Souriau matrix \eqref{Smatrix}.
Expressed in Brinkmann coordinates, this  becomes \eqref{Bboost} \ie,
\beq
\bX \to \bX+ {\IQ}(u)\bb\,,
\where
{\IQ}(u)={\IP}(u){\IS}(u)\,.
\eeq

$\bullet$ For a special geodesic in particular, $\fa={\IP}^2=\II$ in the Beforezone and  the Souriau matrix  is simply
$
{\IS}(u)=(u-u_0)\II \for u, u_0 < u_B\,.
$
The usual free (Galilean) implementation is thus we recover 
\beq
{\IQ}= (u-u_0)\,\II\,
 \for u < u_B\,.
\label{QBefore}
\eeq

$\bullet$
In the Afterzone the Souriau matrix ${\IS}=\diag(S_{11},S_{22})$ can be calculated by using   $\fa$ in \eqref{alinpol},
\beq
S_{11} =\frac{1}{C_1(C_1u+D_1)} \aand S_{22} = \frac{1}{C_2(C_2u+D_2)}\,
\;\for u_A < u\,.
\label{Safter}
\eeq
Combining  with  \eqref{PSL}  we find that in the Afterzone
${\IQ}={\IP}{\IS}$ is an (approximately) constant diagonal matrix,
\beq
{\IQ} = {\IP}{\IS} \approx  -\,\diag(\frac{1}{C_1},\frac{1}{C_2})\,  \;\for u_A < u\,.
\label{Qafter}
\eeq
Thus, as confirmed numerically in FIG.\ref{PHfig},
 boosts act in the far-right Afterzone as (generically anisotropic) \emph{translations},
\beq
\bX \to \bX +
\,\big(\frac{b_1}{C_1},\frac{b_2}{C_2}\big) \,.
\label{Afterboostact}
\eeq

Outside the Wavezone our results are consistent with the numerical ones shown in  FIG.\ref{PHfig}.

\begin{figure}[h]
\includegraphics[scale=.48]{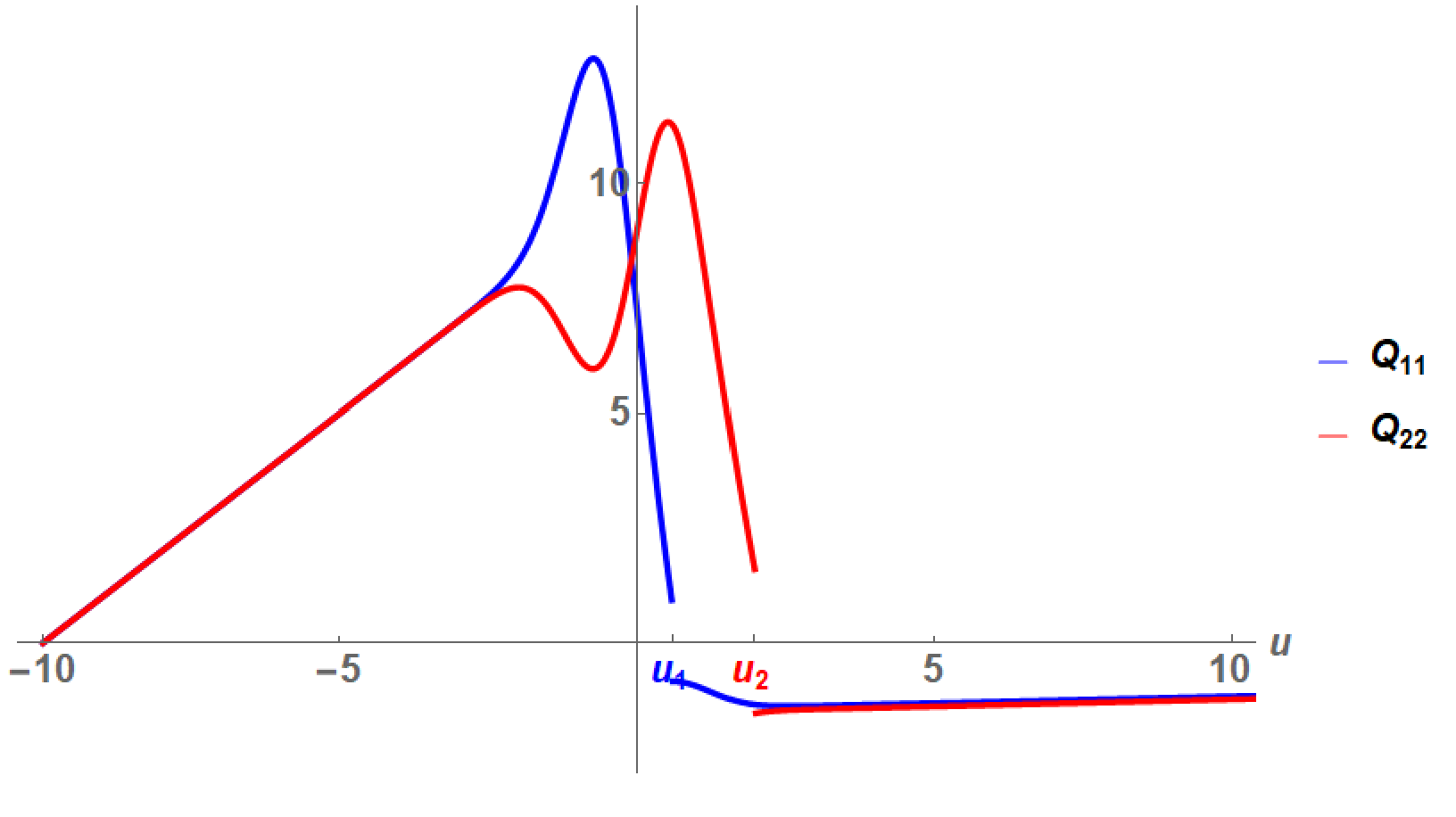}\\
\vskip-6mm
\caption{\textit{\small  The matrix ${\IQ}={\IP}{\IS}$ reduces to the usual Galilean expression in the Beforezone but is approximately  a \underline{constant} matrix, \eqref{Qafter}, in the Afterzone. At the singular points $u_i$  where  the ${\IP}$ and $\fa$
matrices vanish so that the Souriau matrix ${\IS}$ diverges we have a numerical uncertainty.
}}
\label{PHfig}
\end{figure}
Translations,
\beq
\bx \to \bx +\bc\,,
\eeq
which are manifest symmetries of the  BJR metric \eqref{BJRmetric}, behave in the opposite way.
In Brinkmann coordinates they are implemented as
\beq
\bX \to \bX +{\IP}(u)\,\bc\,.
\eeq
 Thus while they are usual translations in the Beforezone where ${\IP}(u)=\II$, in the Afterzone we find, by \eqref{11AfterP},
\beq
\bX \to \bX + \diag\Big(C_1u + D_1)c_1, (C_2u + D_2)c_2\Big)\,.
\eeq

\subsection{Circularly polarized sandwich waves}\label{CPSSec}

As yet another illustration, we study a \emph{circularly polarized (approximate) sandwich wave},
\eqref{genBrink}, with  \cite{PolPer,SLC},
\beq
{\mathcal{A}_{+}}(U) =
\frac{\lambda}{\sqrt{\pi}}\,e^{-\lambda^2U^2} \cos(\Omega U)\red{\,},
\qquad
{\mathcal{A}_{\times}} (U) =
\frac{\lambda}{\sqrt{\pi}}\,e^{-\lambda^2U^2}
\sin(\Omega U)\, .
\label{polgaussprof}
\eeq
depicted in FIG.\ref{CPPSProf}.
We shall choose $\lambda = 0.1$ and $\Omega=\sqrt{2}$. The behavior for various (small and large) values of $\lambda$ is studied in \cite{PolPer}.
\begin{figure}[h]
\includegraphics[scale=.3]{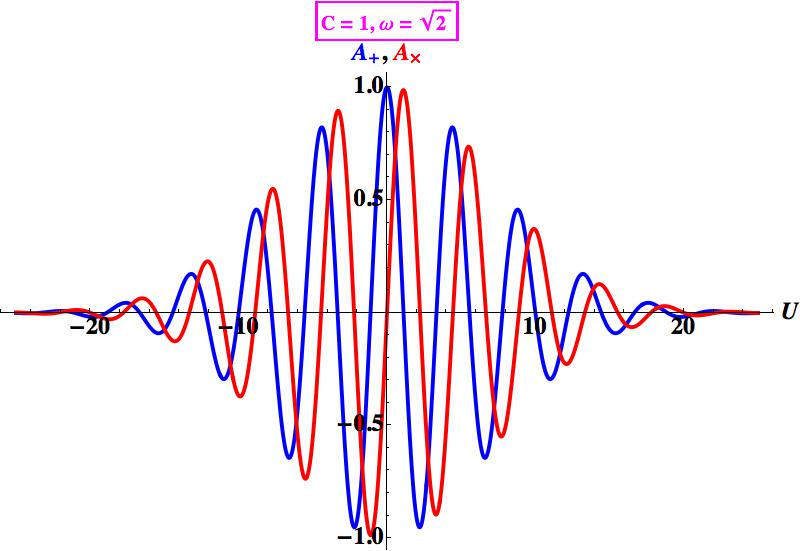}\\
\null\vskip-13mm
\caption{\textit{\small Circularly polarized (approximate) sandwich wave in Brinkmann coordinates with Gaussian envelope, (\ref{polgaussprof}).
The colors refer to the $\bf{\blue{\mathcal{A}_{+}}}$ and the \red{${\mathcal{A}_{\times}}$} polarisation components.
}}
\label{CPPSProf}
\end{figure}

The metric in Brinkmann coordinates
 is perfectly regular for all $U$.
The BJR coordinates  given by \eqref{BJRBrinkP} are instead defined only in \emph{coordinate patches between adjacent zeros} of the determinant of  ${\IP}$ (or of $\fa$), plotted in FIG.\ref{aPdet}.
\begin{figure}[h]
\includegraphics[scale=.36]{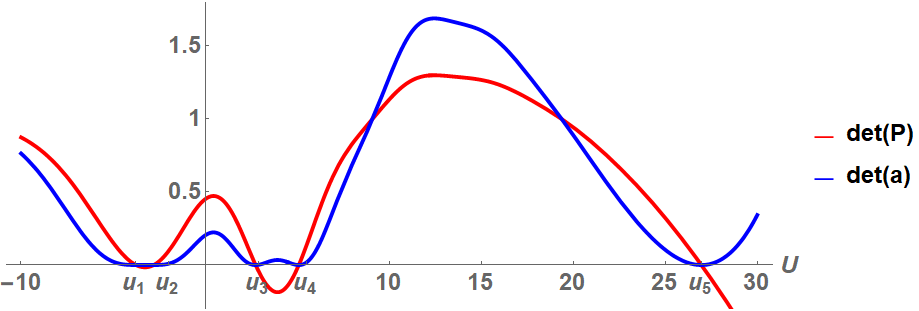}\quad
\\
\null\vskip-10mm
\caption{\textit{\small The BJR coordinates are well-defined only between between adjacent zeros $\mathbf{u_i}$ of the determinant of the BJR profile, $\blue{{\mathbf \chi^{4}}=\det(\fa)}>0$ or equivalently, the zeros of \red{${\mathbf \det({\IP})}$}~: $\mathbf{u_1} \approx -3.75,\, \mathbf{u_2} \approx-2.80,\, \mathbf{u_3} \approx 2.74,\, \mathbf{u_4} \approx 5.09, \,\mathbf{u_5} \approx 26.97$, where the BJR coordinates become singular.
 \blue{$\det(\fa) \approx 1$} in the Beforezone (far left), and  then steadily increasing  in the Afterzone (on the far right). The 5 zeros of $det({\IP})$ are  $U = -3.75, -2.80, 2.74, 5.09, 26.97$  therefore  6 patches are necessary to cover the entire $u$-axis.
 }}
\label{aPdet}
\end{figure}
 The entries of the matrix ${\IP}=(P_{ij})$, obtained by solving numerically the Sturm-Liouville equation \eqref{SL+cond} are shown in FIG.\ref{BspecP}.
\begin{figure}[h]
\medskip
\includegraphics[scale=.28]{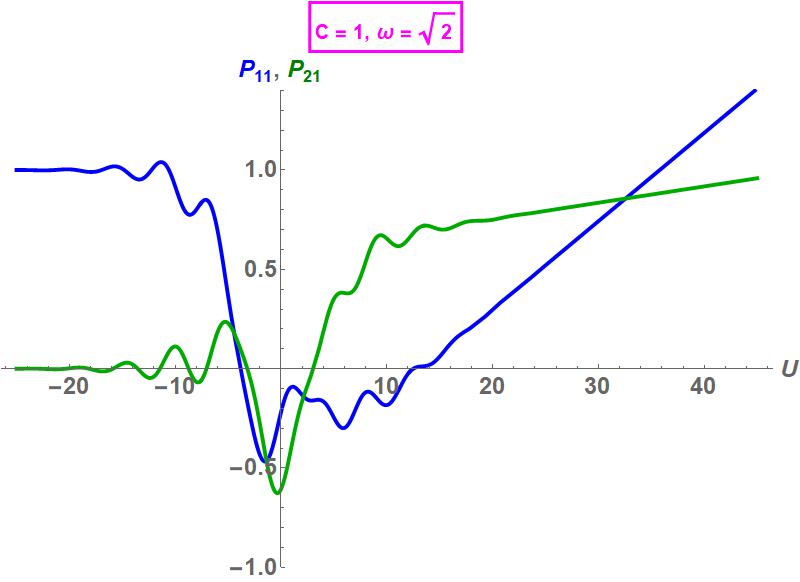}
\;
\includegraphics[scale=.28]{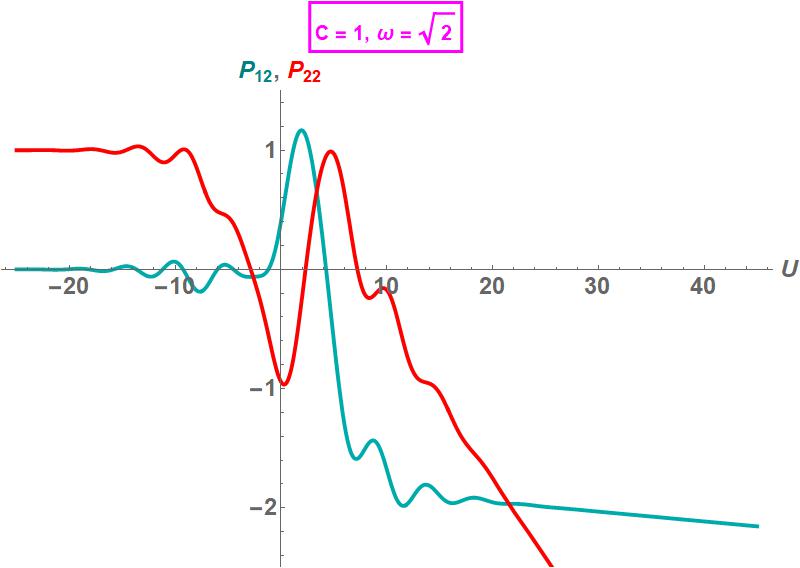}
\\\vskip-8mm
\null \hskip-23mm(a) \hskip 78mm (b)\\
\null\vskip-11mm
\caption{\textit{\small
The  ${\IP}$-matrix is obtained by solving (numerically) the Sturm-Liouville eqn. \eqref{SL+cond}. $\dgreen{\bf{P}_{21}}$, and $\cyan{\bf{P}_{21}}$ and $\red{\bf{P}_{22}}$ can also be viewed as components of special geodesics for particular choices of initial position.}
}
\label{BspecP}
\end{figure}
The BJR profile $\fa=\big(a_{ij}\big)$  \eqref{afromP} deduced from the ${\IP}$-matrix is  shown  FIG.\ref{BJRaprof}.
\begin{figure}[h]
\includegraphics[scale=.32]{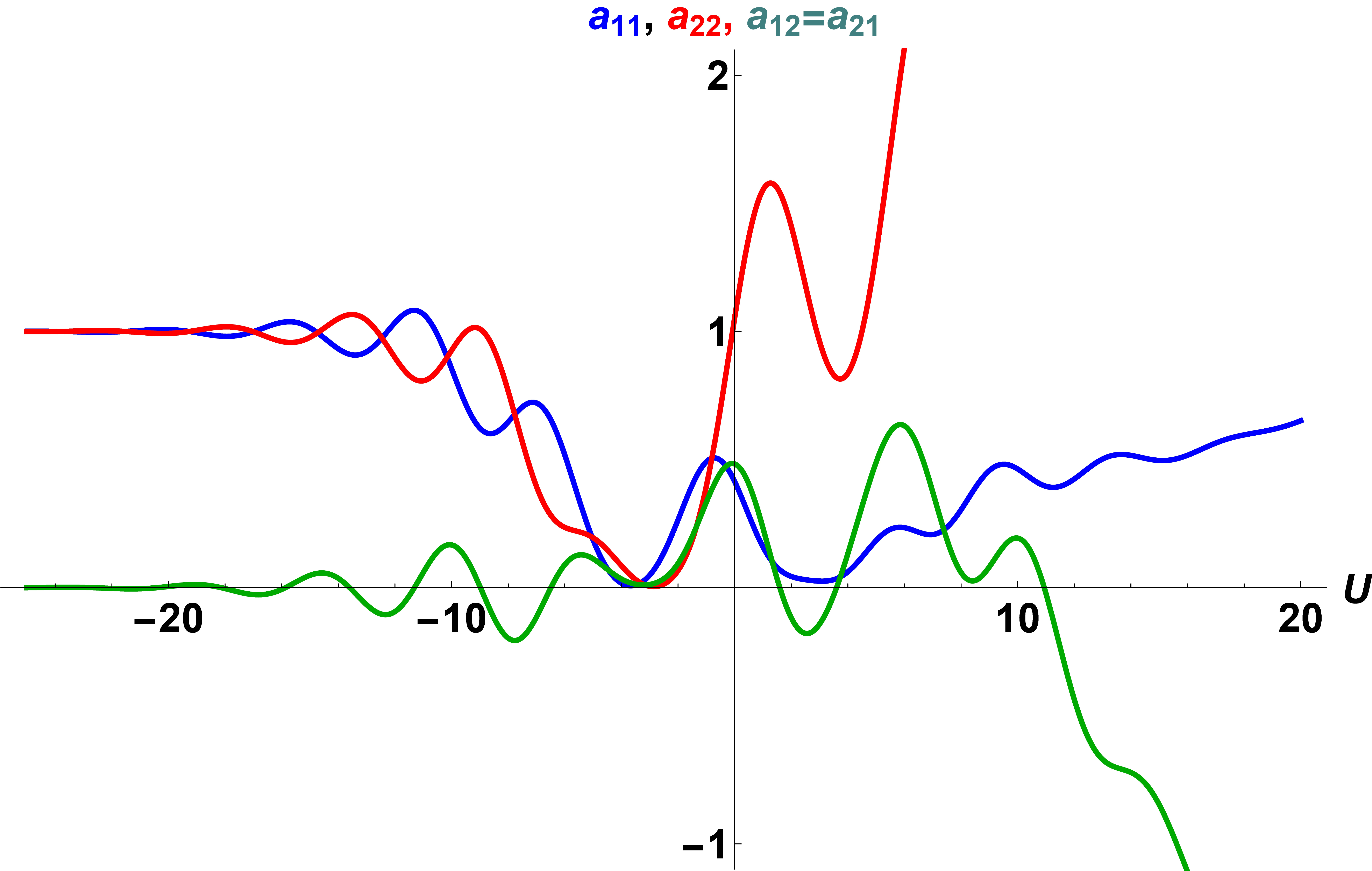}\vskip-4mm
\caption{\textit{\small The  BJR profile of the
  \GW \eqref{genBrink}-\eqref{polgaussprof} is given by a symmetric matrix
${\bf a}=\big(a_{ij}\big)={\IP}^{\dagger}{\IP}$.}}
\label{BJRaprof}
\end{figure}

\kikezd{Special geodesics}.
The two columns of the ${\IP}$-matrix unfolded to (2+1)d,
\beq
\hat{\bX}^{(1)}(U) =\left[\small {\barray{c}
P_{11}(U) \\ P_{21}(U)
\earray}, U\right]
\aand
\hat{\bX}^{(2)}(U) =\left[\small {\barray{c}
P_{21}(U) \\ P_{22}(U)
\earray}, U\right]\,,
\label{specX1X2}
\eeq
($U=u$) are special Brinkmann-form  geodesics at rest for $u_0<u_B$
at
\beq
\bX^{(1)}_{0}={\small {\barray{c}
\blue{\bf1} \\ \dgreen{\bf0} \earray}}
\aand
\bX^{(2)}_{0}={\small {\barray{c}
\cyan{\bf0} \\ \red{\bf1}\earray}}\,,
\label{X1X2init}
\eeq
respectively. FIGs.\ref{BspecP} - \ref{BspecGeo3d} show that the complicated motions in the Wavezone are straightened out in the Afterzone, consistently with the Bargmann interpretation. Tissot's indicatrix  is plotted in FIG.\ref{newTissot}.

\begin{figure}[h]
\includegraphics[scale=.33]{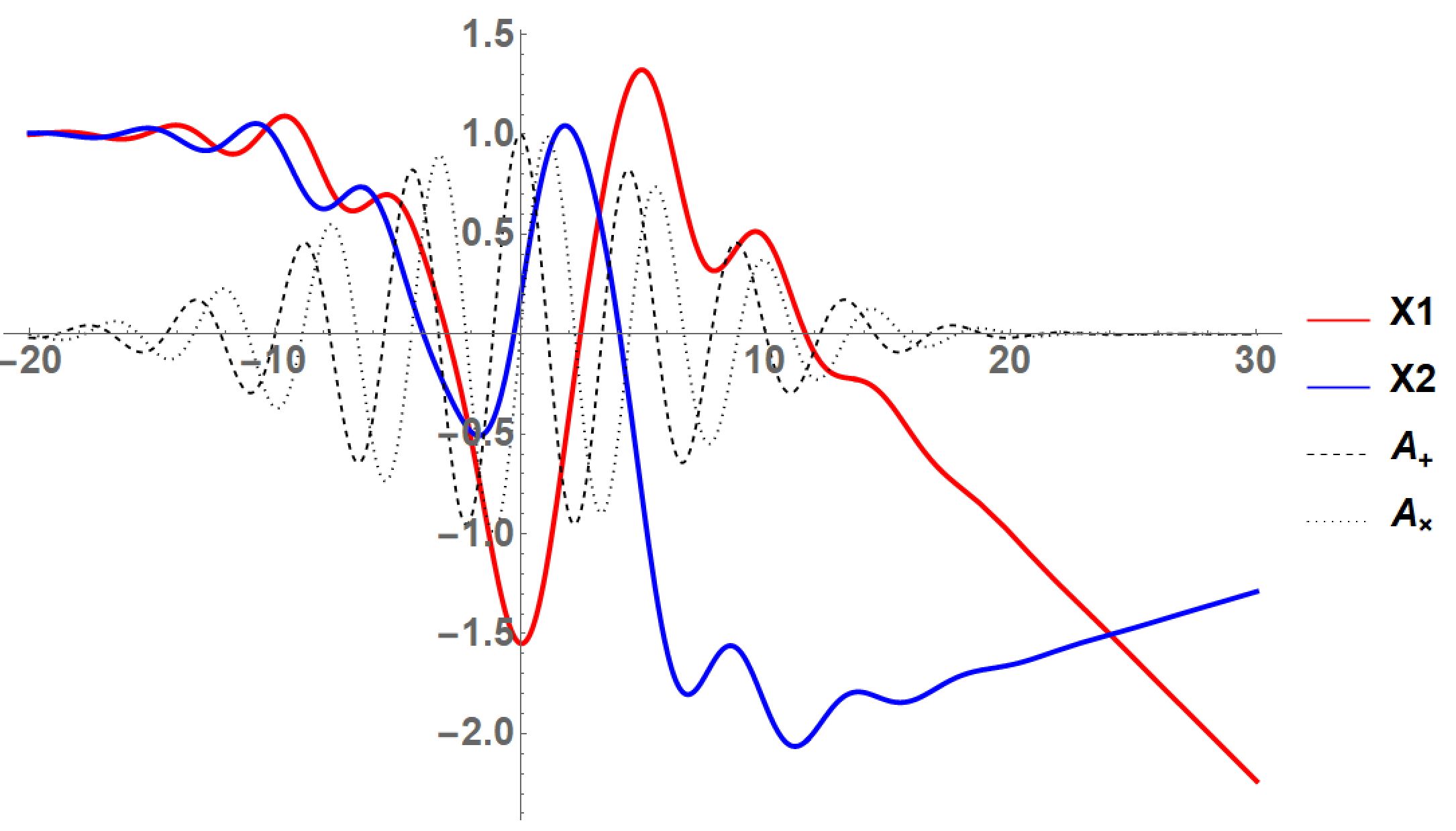}
\\
\null\vskip-14mm
\caption{\textit{\small In Brinkmann coordinates, the trajectory components \red{$X^1(u)$}, \blue{$X^2(u)$} of a
special geodesic [which is at rest in  the Beforezone], 
 $\bX_0=(\blue{\bf1},\red{\bf1})$, become straight lines in the Afterzone, as illustrated on the example of a circularly polarized sandwich wave.
}}
\label{specGeo11}
\end{figure}

\goodbreak
\begin{figure}[h]
\hskip-2mm
\includegraphics[scale=.33]{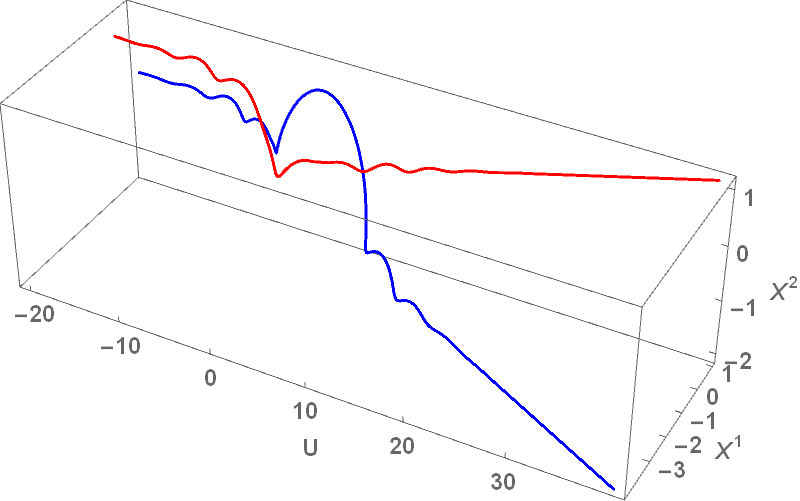}\;
\includegraphics[scale=.25]{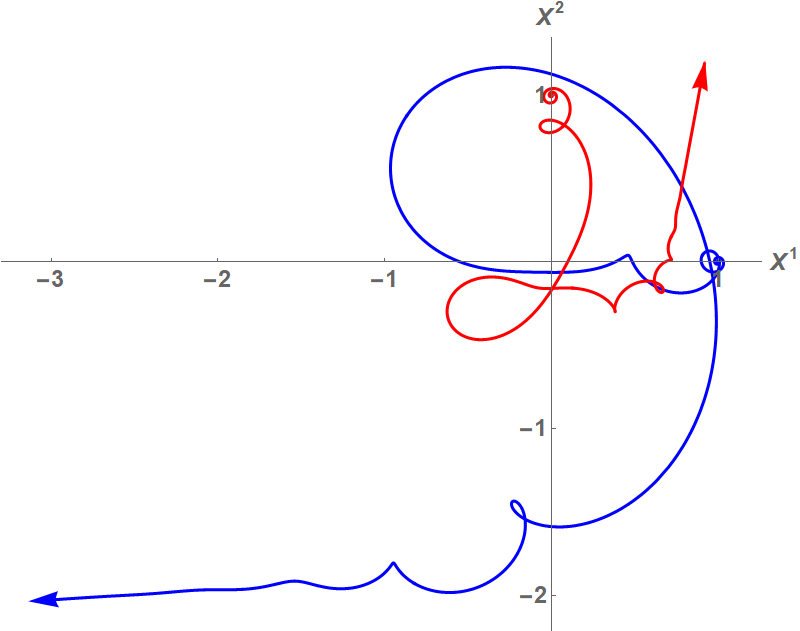}
\\
\null\vskip-8mm
\hskip6mm (a) \hskip 86mm (b)
\\
\caption{\textit{\small Brinkmann trajectories for initial conditions \red{$\bX^{(1)}_0=(0,1)$}
and \blue{$\bX^{(2)}_0=(1,0)$} in \eqref{X1X2init}.
(a) unfolded to $2+1$ dimensions.
(b) Projected to the transverse plane. The complicated motions in the Wavezone are smoothed out in the Afterzone where they follow straight trajectories with constant non-zero velocities.
 }}
\label{BspecGeo3d}
\end{figure}
\begin{figure}[h]
\includegraphics[scale=.5]{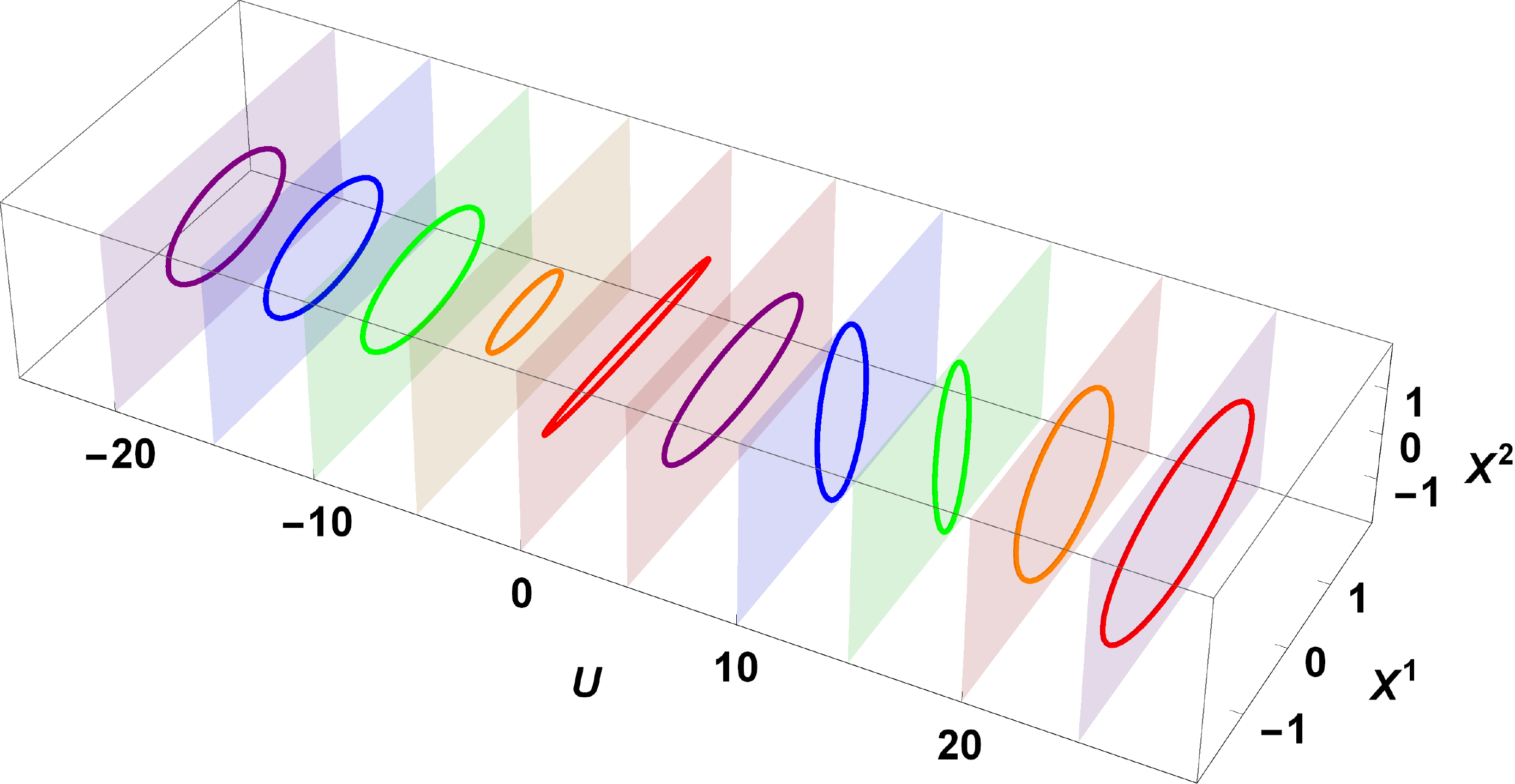}
\caption{\textit{\small Tissot diagram for 
 a circularly polarized sandwich wave \eqref{polgaussprof} shown in  FIG.\ref{CPPSProf}.
}}
\label{newTissot}
\end{figure}
\goodbreak

\kikezd{Symmetries}.
Distorted boosts  \eqref{genCarract}
are expressed by using the Souriau matrix ${\IS}$ in (\ref{Smatrix}).
Combining our previous results
allows us to plot them in BJR coordinates,
(FIG.\ref{impHfig}). The  implementation in the Afterzone \emph{differs substantially} from the Galilean one.
\begin{figure}[h]
\includegraphics[scale=.28]{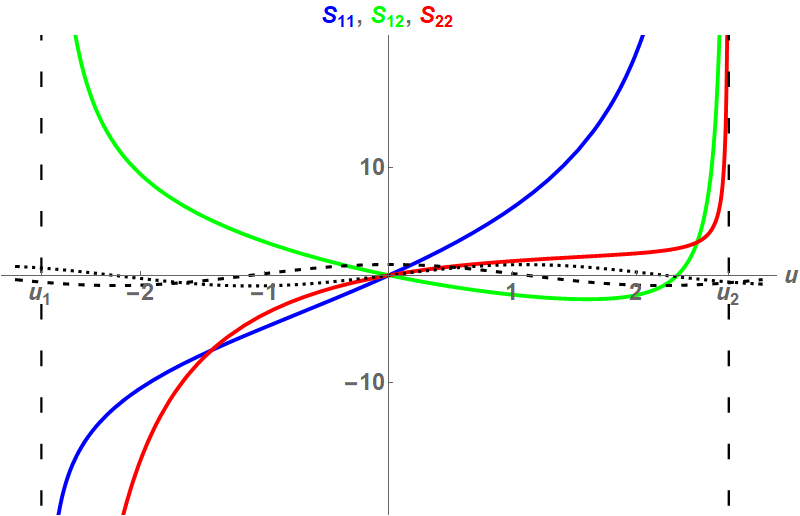}\;
\null\vskip-3mm
\caption{\textit{\small In BJR coordinates, boosts are implemented  through the Souriau matrix ${\IS}$ as in \eqref{genCarract}. We took $\bb=(1,1)$. The dotted and dashed lines in the background indicate the $\cA_+$ and $\cA_{\times}$ components of the Brinkmann profile \eqref{genBrink}.
${\IS}$ is defined  between two subsequent zeros $\mathbf{u_k}$ and $\mathbf{u_{k+1}}$ of the $\fa$ (or $P$)-matrix (see FIG.\ref{aPdet}).
Our plot shows what happens in $[-3,3]$  which contains
$I_2 = \big[\mathbf{u_2},\mathbf{u_{3}}\big]$
where $\mathbf{u_2} \approx -2.8$ and $\mathbf{u_3} \approx  2.7$ are the singular points which are the closest to the origin. $I_2$  contains the Wavezone.  All components of the Souriau matrix\, $\IS$  manifestly diverge when $u\to \mathbf{u_i}, i=2,3$.
}}
\label{impHfig}
\end{figure}

Covering the $u$-axis with BJR domains and gluing together the results in the subsequent coordinate patches yields the complicated figure FIG.\ref{impHfigLarge}, obtained by
  calculating the integrals $\displaystyle\int_{u_0}^u\! \!a^{-1}$  for $u_0\equiv {u_0}^{(k)},\,u \in I_k$. In detail, we calculated numerically
\benu
\item
on the left-hand part, for $u_0^{(1)}=-20,\, u < 3$ (which lies  in the Beforezone),

\item
in middle part, with $u_0^{(2)}=0,\, -3 < u < 3$ (which contains the Wavezone),

\item
in the right-hand part, for  $u_0^{(3)}=3 < u < 20$,
 (which lies  in the Afterzone).
\eenu
\goodbreak

Glueing the results obtained in the intervals $I_k$, (FIG.\ref{impHfigLarge}) indicates that at the contact points $u_k$ the Souriau matrix diverges from both sides.
\begin{figure}[h]
\null\hskip-6mm
\includegraphics[scale=.61]{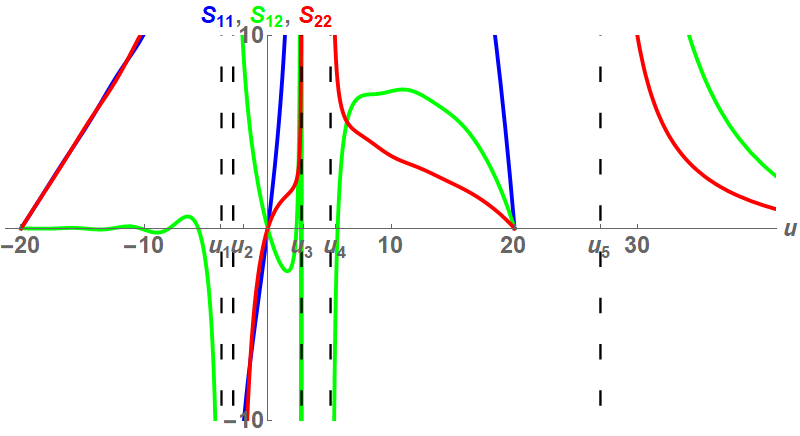}
\quad
\null\vskip-10mm
\caption{\textit{\small
The  implementation of a boost through the Souriau matrix is essentially
Galilean in the Beforezone $u < u_B$ but differs substantially from the latter both in the Wavezone and in the Afterzone.
}}
\label{impHfigLarge}
\end{figure}

\begin{figure}[h]
\includegraphics[scale=.39]{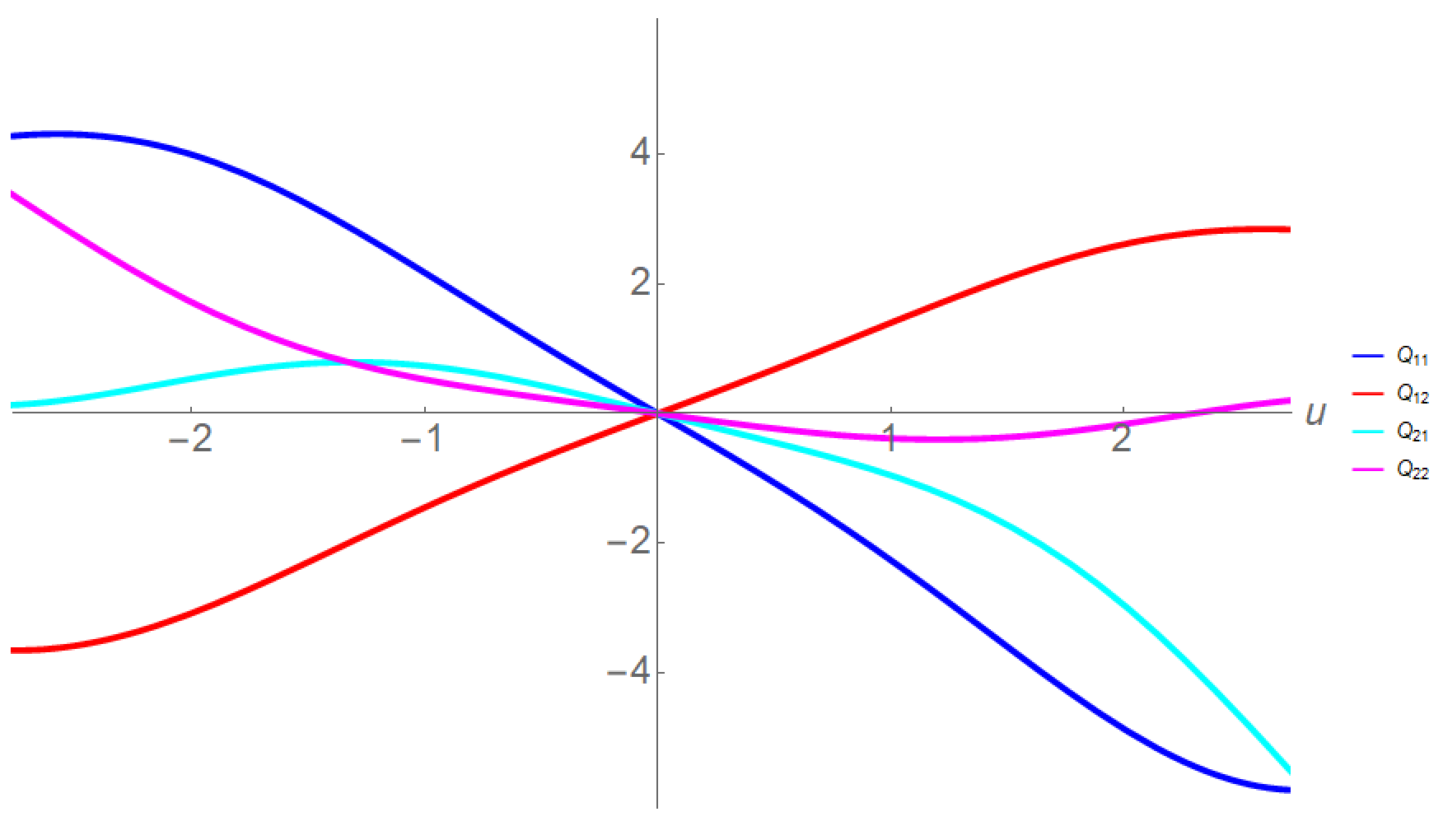}\;
\null\vskip-7mm
\caption{\textit{\small The ${\IQ}={\IP}{\IS}$-matrix \eqref{Qdef} which implements boosts in Brinkmann coordinates, \eqref{Bboost}, plotted in $[-3,3]$  as in FIG.\ref{impHfig}. To be compared with FIG.\ref{PHfig}.
}}
\label{CCPQmatrixfig}
\end{figure}

\section{Connecting the Before and Afterzones
}\label{LinkingSec} 

In both  flat regions the motion is free, 
determined by the constant velocity, $c$ in \eqref{flatchieq}. For our special geodesics $c = 0$ in the Beforezone, but it is some \emph{different, nonzero constant} in the Afterzone. The two separated flat zones are connected through the Wavezone, though, where the sandwich profile and therefore the motions are nontrivial and can indeed by quite complicated as seen in FIGs.~\ref{BspecP}-\ref{BspecGeo3d}.  

The  velocity in the Afterzone is then  determined by  
 the integrals \eqref{linpolveljump}.  
Since  the \SL equation can not be solved analytically in general, we have to resort to numerical
calculation: first one determines the trajectories (the initial position where chosen $X^1(-\infty)=1$ and $X^2(-\infty)=2$). Then insertion into \eqref{linpolveljump} yields the velocities, see FIGs. \ref{p1}-\ref{p2}-\ref{p3}-\ref{p4}.

\begin{figure}[h]
\includegraphics[scale=.23]{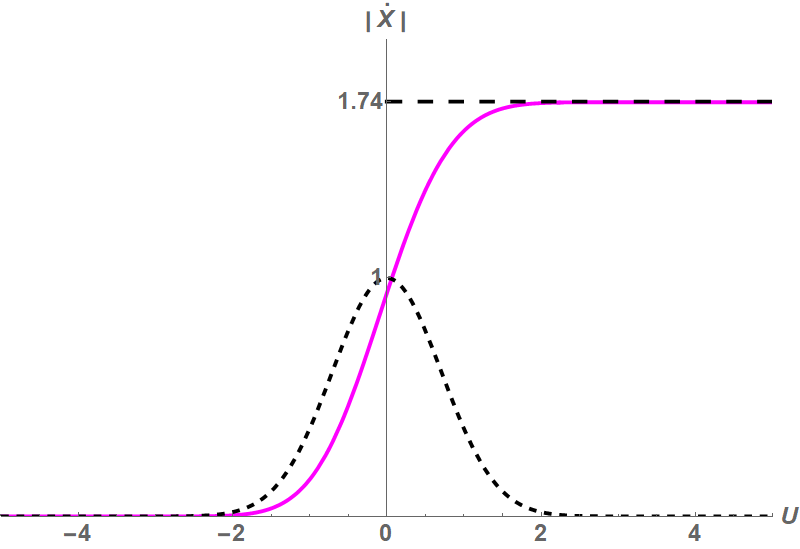}\quad
\includegraphics[scale=.3]{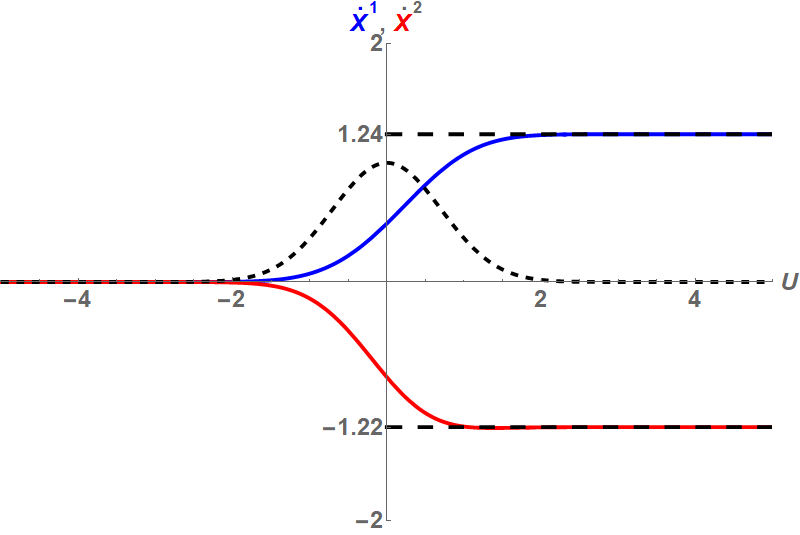}
\vskip-3mm\caption{\textit{\small Velocity jump between the flat Before and Afterzones for the \underline{simple Gaussian} profile \eqref{Gaussprof}.
}}
\label{p1}
\end{figure}
%
\begin{figure}[h]
\includegraphics[scale=.23]{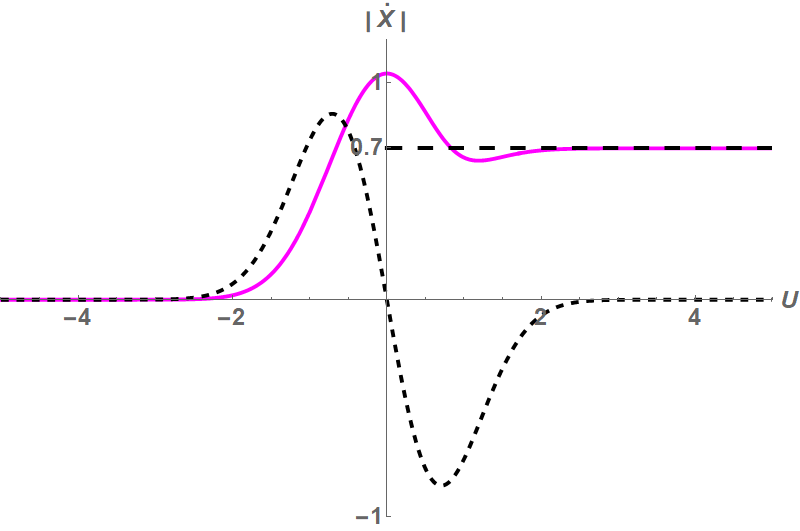}\quad
\includegraphics[scale=.28]{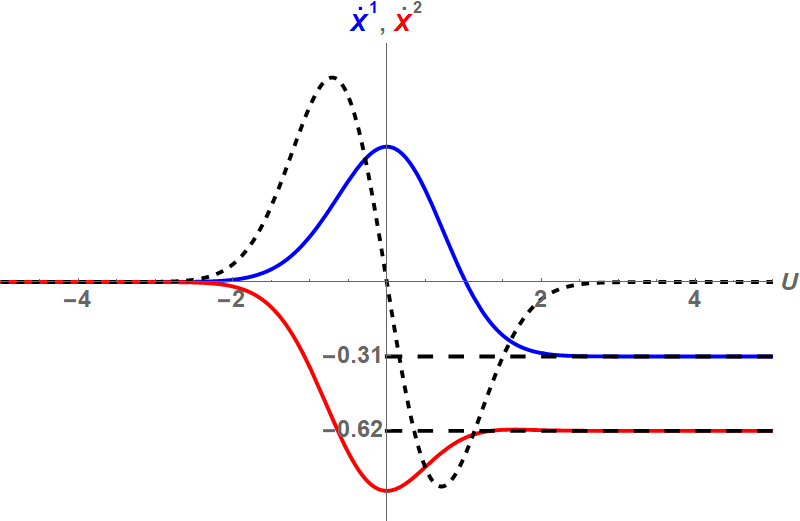}
\vskip-3mm\caption{\textit{\small Velocity jump for the first derivative of the Gaussian, \eqref{flybyprof}, representing \underline{flyby}.
}}
\label{p2}
\end{figure}
%
\begin{figure}[h]
\includegraphics[scale=.24]{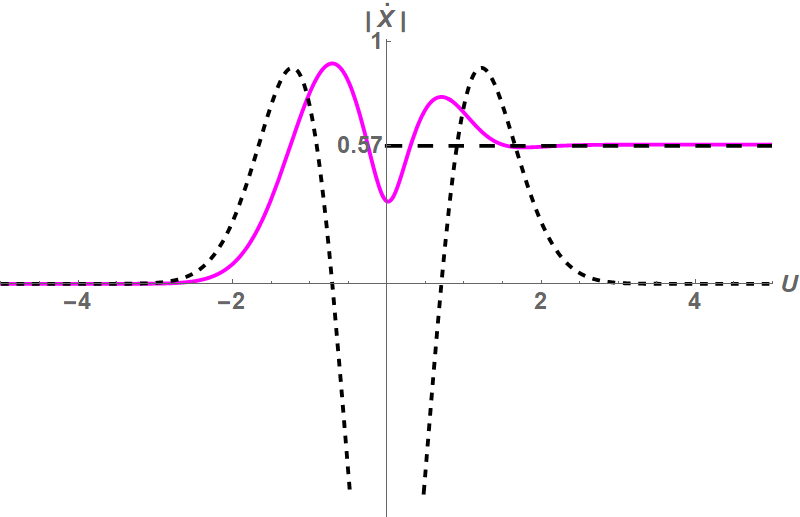}
\includegraphics[scale=.26]{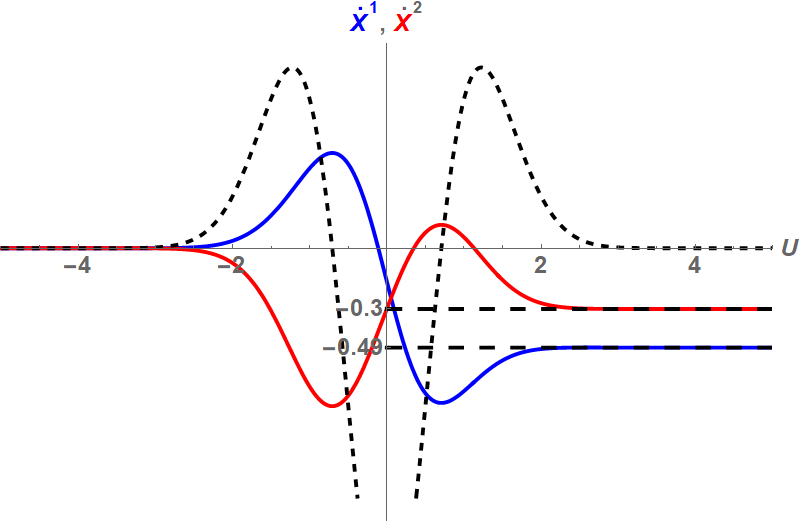}
\vskip-3mm\caption{\textit{\small Velocity jump for the second derivative of the Gaussian,  \eqref{Kd2}, FIG.\ref{Gauss2Geo}, proposed by Braginski and Thorn.
}}
\label{p3}
\end{figure}
\begin{figure}[h]\hskip-2mm
\includegraphics[scale=.23]{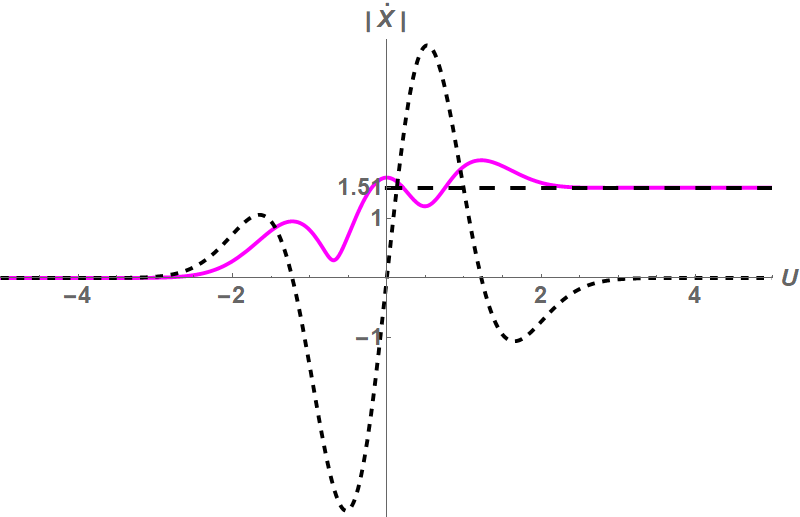}\;
\includegraphics[scale=.24]{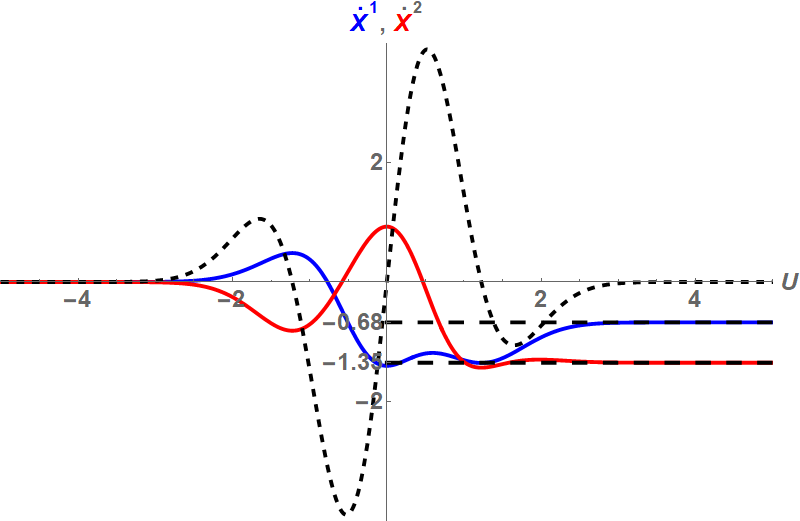}
\vskip-3mm\caption{\textit{\small Velocity jump for the third- derivative-of-the-Gaussian profile, \eqref{collapseprofile}
 in FIG.\ref{collapsegeofig}, proposed for \underline{gravitational collapse} \cite{GibbHaw71}.
}}
\label{p4}
\end{figure}

\section{Impulsive limit}\label{ImpSec}

The impulsive (or shock-wave) limit leads to subtle analytic questions \cite{Penr2,Steinbauer:1997dw,Kunzinger:1998nz,BarHog,PodSB,Podolsky:2022ase}, and to some controversy \cite{ImpMemory,SBComment}. A rigourous study goes beyond our scope here, therefore we limit ourselves with  a couple of intuitive statements. In fact, we discuss only one simple example, namely that of the \emph{linearly polarized Gaussian profile} \eqref{Gaussprof} with a scale factor 
$\lambda$,
\beq
 {\cal A}(U) = \frac{\lambda}{\sqrt{\pi}}\,e^{-\lambda^2 u^2}\,.
 \label{lambdaprof}
\eeq%
Squeezing the profile by letting $\lambda\to \infty$ yields
the impulsive profile shown in Fig.\ref{PL1-20} \footnote{
FIG.\ref{dotnfig} shows that for small $\lambda$, de behavior is substantially different from that in the impulsive ($\lambda\to\infty)$ case.},
\begin{equation}
\cA \equiv \cA_{+}(U) = 2k
\, \delta (U) \,.
\label{impPol}
\end{equation}
Conversely, $k$ is recovered as
\beq
k=\frac{1}{2}\int_{-\infty}^{+\infty}\!\!\cA(U)dU\,.
\label{NormA}
\eeq
In our FIGs.\ref{BspecP} - \ref{BspecGeo3d} we have choosen $k=\half$.

Taking the $\lambda\to\infty$ limit of the damped Gaussian profile \eqref{lambdaprof}  in  FIGs.\ref{PL1-20},
  the  $\IP$-matrix of an impulsive wave breaks
between the Before and the Afterzones by \cite{ImpMemory},
\beq
\Delta \dot{\IP}\equiv
\dot{\IP}(0+)=\fc_0\,.
\label{Pjump}
\eeq
But for a linearly polarized wave the  $\IP$-matrix
is also a trajectory, so this implies a velocity jump,
\beq
\medbox{
\dot{\bX}(0+)
=\fc_0\,\bX_0\,.
}
\label{BXveljumpBIS} 
\eeq
 
The memory effect in an impulsive \GW can also be studied in BJR coordinates. The reader is referred to \cite{ImpMemory,SBComment}
for details. 
\begin{figure}[h]
\includegraphics[scale=.27]{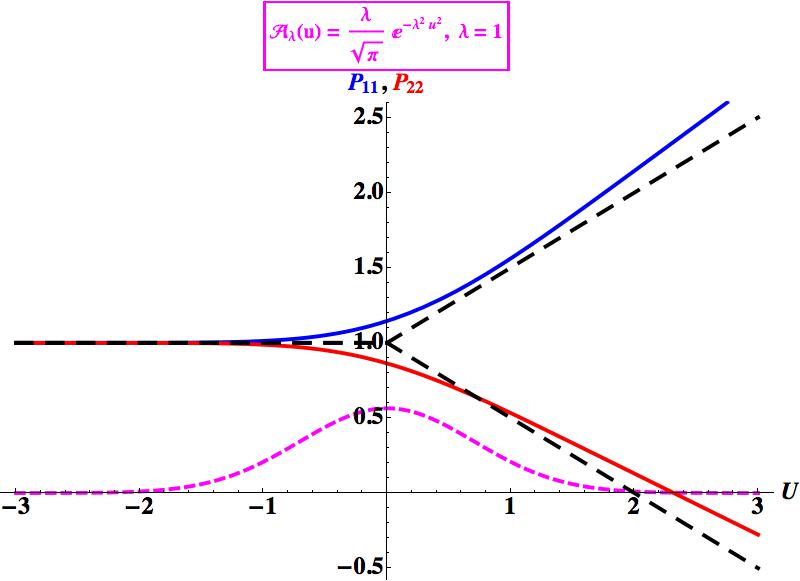}
\includegraphics[scale=.27]{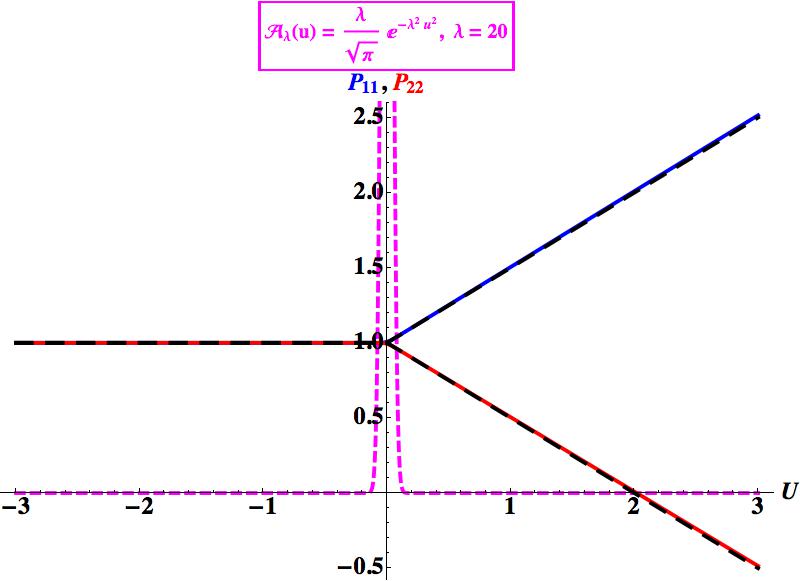}
\\ 
(a)\hskip 73 mm (b).
\\
\vskip-1mm
\caption{\textit{The matrix $\IP(u) = \diag\big(P_{11}(u),P_{22}(u)\big)$ [alias special geodesic]  for the linearly polarized Brinkmann  profile of the form \eqref{Gaussprof}, i.e., ${\cal A}(U) = (\lambda/\sqrt{\pi})e^{-\lambda^2 u^2}$ with $\lambda = 1\,, 20\,$.  When 
$\lambda\to\infty$ the profile tends to that of an impulsive wave, \eqref{impPol}.
 The numerical solutions  are plotted in solid \blue{\bf blue} and \red{\bf red} lines. For large $\lambda$, 
$\IP$ tends to the analytic impulsive profile depicted in {\bf dashed black lines}. The solution is regular for $u< k^{-1}=u_0$, which is also a zero of $\det(\chi)$.
}}
\label{PL1-20}
\end{figure}

\begin{figure}[h]
\includegraphics[scale=.25]{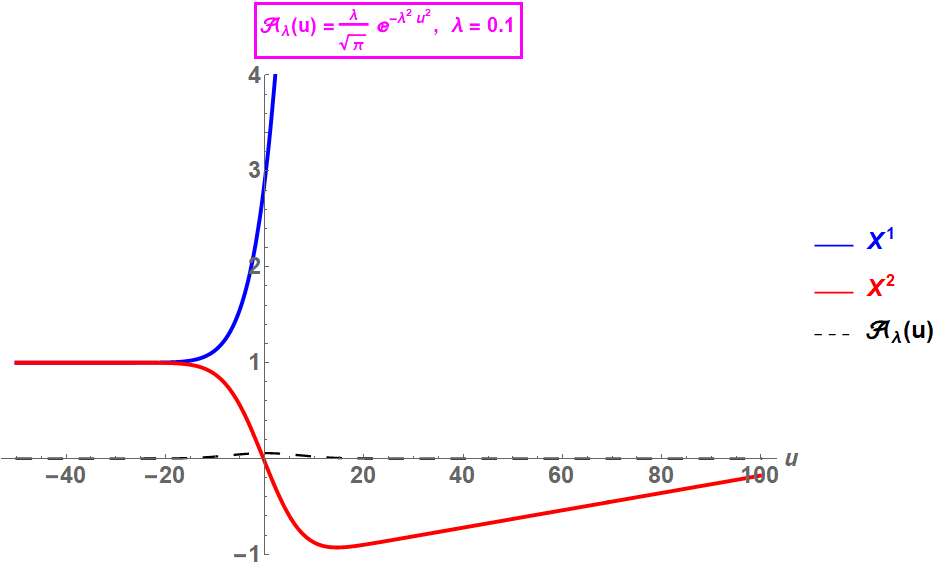}
\\ 
\vskip-1mm
\caption{\textit{
For small $\lambda$, de behavior is substantially different from that in the impulsive ($\lambda\to\infty)$ case.
}}
\label{dotnfig}
\end{figure}

\section{Conclusion}\label{Concl}

Gravitational waves could, in principle, be observed also through the \emph{Memory Effect}. In the `standard' version called the \emph{displacement effect} (DM) proposed by Zel'dovich-Polnarev and their collaborators \cite{ZelPol,BraTho,BraGri} the particles initially at rest will have an approximately \emph{vanishing relative velocity} \cite{Christo}.
 
 Another version (to which our paper is devoted), was proposed, earlier, by Ehlers and Kundt \cite{Ehlers}, and by Souriau \cite{Sou73}, finds instead that the particles move, after the passing of a burst of a plane gravitational wave, with \emph{constant but non-zero velocity} \cite{Sou73,BraGri,BoPi89})
 and is called therefore \emph{velocity effect} (VM). 
 
 Souriau \cite{Sou73} found, in particular, that using Baldwin-Jeffery-Rosen (BJR) coordinates \cite{BaJeRo} makes the description straightforward, eqn. \eqref{CarGeo}, -- but the price to pay for it is that the coordinates are only local: the become necessarily singular and the solutions obtained in adjacent intervals have to be glued together.  
 In the linearly polarized ``collapse'' case (sect.\ref{CollapseSec}), for example, we need 3 patches, and for the circularly polarized  one (sect.\ref{CPSSec}) we need 6 patches \footnote{The situation is reminescent of gauge theories as e.g. for a Dirac monopole \cite{Dirac31,WuYang75,SouPrequant}, with the ${\IP}$-matrix playing a r\^ole analogous to that of a gauge potential, the BJR profile $\fa$ that of the curvature tensor, and $\Ort(2)$  an internal symmetry group.}.
 
The question of symmetries is even more subtle. It has long been  known that the  \GW metric  admits generically a 5-parameter isometry group \cite{BoPiRo,exactsol}. Three of them are obvious (translations), however  two of them could be identified  only as solutions of a \SL equation \cite{Torre,SLC}. 

Souriau \cite{Sou73}, who followed his own ideas [as he always did], found an elegant \emph{explicit} expression in BJR coordinates, \eqref{genCarract}, in terms of the matrix  $\IS$ in \eqref{Smatrix} which allow us to identify
the symmetry as L\'evy-Leblond's Carroll symmetry with broken rotations \cite{Leblond,Carroll4GW} (what Souriau did not recognize). We just mention that the difference between the two, (displacement and velocity) types effects might well be related to the double Carroll structure, one associated with BMS \cite{BMS,Sachs61,Sachs62,ConfCarrBMS,PeHaStPRL,PeHaSt1611},
and the other with the internal \emph{dipole symmetry} of fractons considered in condensed matter physics \cite{Bidussi,LD,Figueroa} and references in them. Further applications of the Carroll symmetry include the cosmic Hall effect \cite{Marsot,HallonHole}. 
 
Souriau's expression can be pulled back to global Brinkmann coordinates. However:

$\bullet$ The solutions found in BJR coordinates are valid only in coordinate patches and requires us ``to glue them together''; 

$\bullet$ Their transcription to global Brinkmann coordinates requires to solve a \SL equation;

$\bullet$ Solving the equations of motion directly in Brinkmann coordinates leads to similar difficulties: it requires in particular to solve another Sturm-Liouville equation \cite{SLC}. 
Analytical solutions are therefore exceptional and one has to resort to numerical work \cite{Carroll4GW,SLC,Elbistan:2022plu}. 

The BJR solutions diverge at the contact points of the adjacent  intervals.
 FIG.\ref{impHfig} and FIG.\ref{CCPQmatrixfig} indicate, however, that multiplication by $\IP$, \eqref{RPgauge}, \emph{regularizes the trajectories}. 
Accordingly, the $\IQ=\IP\IS$ matrix yields smooth (although unusual) implementations of the boosts.
  
Impulsive waves whose wave-zone is so narrow that it can be approximated by a Dirac delta  yield explicit analytic solutions.
Unphysical as it might appear at first, the impulsive approximation is indeed quite realistic: think at those \GWs which, after having travelled at the speed of light for thousands or even millions of years, cross through the earth in a fraction of a second.

\smallskip
To conclude, our review is admittedly ``strip-cartoon-like''. We would however recall a poem written by the author of \cite{Alice} entitled ``The Ocean Chart"  \cite{Snark} :

\begin{quote}\textit{\narrower
[the Captain]  \dots had bought a large map representing the sea,\\
        Without the least vestige of land:\\
    	And the crew were much pleased when they found it to be\\
        A map they could all understand.
}\end{quote}

\begin{acknowledgments}\vskip-4mm
This paper is a substantially extended and completed version of the talk presented  at the ``Journ\'es Relativistes'', Tours, May 31, 2023 by PAH.
 Much of our results presented here come from joint work carried out with our late friend  C. Duval, and with G.~Gibbons to whom we are grateful for many insights and advices. 
Correspondence is acknowledged also to G. Barnich. We found inspiring the questions and suggestions of both of our referees. 
 ME was  supported by TUBITAK under 2236-Co-Funded Brain Circulation Scheme2 (CoCirculation2) with project number 121C356. PMZ was partially supported by the National Natural Science Foundation of China (Grant No. 11975320). 
\end{acknowledgments}
\goodbreak


\appendix
\section{{\bf The Bargmann framework}}\label{Appendix}

\renewcommand{\theequation}{\thesection.\arabic{equation}}
\renewcommand
\appendix{\appendix}{\setcounter{equation}{0}}

\newcommand{\brg}{\bar{\rg}}

Eisenhart \cite{Eisenhart} has shown that the dynamics of a conserva\-ti\-ve holonomic dynamical system with $n$ degrees of freedom can be trans\-cribed as \textit{geodesic motion} in a certain Lorentzian space-time of dimen\-sion~$n+1$. 
He starts with the space-time Lagrangian 
\begin{equation}
L=\half\,\rg_{\alpha\beta}\,\dx^\alpha\dx^\beta-V
\label{L}
\end{equation}
with coordinates  $(x^\alpha)=(x^1,\ldots,x^n,x^{n+1})$, where $t=x^{n+1}$ stands for the absolute time-coordinate; the quadratic form $(\rg_{\alpha\beta})$ as well as the potential function $V$ depend  arbitrarily upon $(x^\alpha)$. The quadratic form $(\rg_{\alpha\beta})$ is  {not} assumed to be non-degenerate as usual in non-relativistic mechanics, 
however the sub-matrix $(\rg_{ij})$ where $i,j=1,\ldots,n$ is required to represent local\-ly a  \textit{Rieman\-nian metric} on  each time-slice $t=\const$.

 Re\-writing the Lagran\-gian~(\ref{L}) as 
\beq
L=\half\,\rg_{ij}\,\dx^i\dx^j+A_i\dx^i+\half\rg_{tt}-V,
\eeq
 where $A_i=\rg_{it}$,  we end up with the Lagrange equations
\begin{equation}
\rg_{ij}\ddx^i+\Gamma_{jki}\,\dx^j\dx^k+\big(\partial_t\rg_{ij}+\partial_iA_j-\partial_jA_i\big)\dx^j+\partial_tA_i+\partial_i\big(-\half\rg_{tt}+V\big)=0
\label{LagrangeEquations}
\end{equation}
for all $i=1,\ldots,n$, where the $\Gamma_{jki}$ denote the  Christoffel symbols of the metric~$(\rg_{ij})$ of a $t=\const$ slice. The  equations (\ref{LagrangeEquations}) can be viewed as the geodesic equations of a special Lorentz metric on an $(n+2)$ dimensional extended space-time with coordinates $(x^\mu)=(x^1,\ldots,x^n,t,s)$ where $s=x^{n+2}$. 
The metric $\brg=\brg_{\mu\nu}\,dx^\mu{}dx^\nu$ introduced in \cite{Eisenhart,DBKP} reads in fact
\begin{equation}
\medbox{
\brg=\rg_{ij}\,dx^idx^j+2(A_i\,dx^i+ds)dt-2U\,dt^2
}
\label{brg}
\end{equation}
where the components $\brg_{ij}=\rg_{ij}$ and $\brg_{it}=A_i$ (for $i,j=1,\ldots,n+1)$ depend, along with $\brg_{tt}=-2U$, on the space-time coordinates $(x^\alpha)$ only. The metric (\ref{brg}) is  is indeed a \textit{Brinkmann metric} \cite{Brink} : it has a null, covariantly constant, nowhere vanishing vector field,
\begin{equation}
\xi=\frac{\partial}{\partial s}\,.
\label{xi}
\end{equation}
Such a pair $(\rg,\xi)$ has been called a \textit{Bargmann structure}  \cite{DBKP,DGH91,CDGH}.
Factoring out the foliation defined by $\xi$
yields a Newton-Cartan structure on the quotient \cite{DBKP} --- \ie, the structure of non-relativistic spacetime \cite{DGH91,DH-NC,Bekaert,Morand}.
The geodesic motion in the metric \eqref{brg} projects to that of a particle  with no spin in $(n+1)$ dimensional non-relativistic space-time \cite{DHP2}, 
as confirmed by spelling out the equations of motion:  the null geodesics of a 5-dimensional Bargmann metric \eqref{brg}  with $g_{ij}=\delta_{ij}$ project to 3 + 1 dimensional non-relativistic spacetime according to
\beq
\frac{d^2\bx}{dt^2} = -\bnabla U + \frac{d\bx}{dt} \times \bB
\where \bB = \rot \bA\,, 
\label{UAeqmot}
\eeq
which is the equation of motion for a spinless particle of unit mass in a potential combined  with a magnetic field
(or Coriolis force). 
This justifies our identification of the $\bar{g}_{it}\ (g_{it})$ component of the metric (\ref{LagrangeEquations}) with the gauge potential $A_i(\bx, t)$ cf. (\ref{gaugepot}).
 The quadratic coefficient of $dU^2$ in \eqref{genBrink} in particular, 
\beq
U(x^{i},t) = \sum_{i,j}\half K_{ij}(t)x^ix^j\,,
\label{oscipot}
\eeq
where $K_{ij}$ is symmetric represents a (generally anisotropic) harmonic oscillator with time-dependent frequencies\bblue cf. (\ref{Bprofile}) \eblue.

It is legitimate to consider the metric components $A_0 = - U$ and $A_i$ as the components of a vector potential: a gauge freedom, 
$A_{\mu} \to A_{\mu} + \p_{\mu}\sigma$ is associated with changing the ``vertical'' coordinate, $s \to  s + \sigma$.

The Bargmann framework is particularly convenient  to study the \emph{symmetries} of the underlying non-relativistic system \cite{DBKP,DGH91,CDGH} which is indeed  its {\sl ``raison d'\^etre"}~: they are given by those Killing vectors $X=(X^i)$ of the Bargmann metric \eqref{brg} which leave also $\xi=\p_s$ invariant,
\beq
L_Xg = 0
\aand
L_X \xi= 0\,.
\label{Bsymm}
\eeq
The Killing vector \eqref{xi}
generates, for all Bargmann manifolds, a conserved quantity identified as the \emph{mass} of the particle. 
In the flat case $g_{ij}=1, A_i = U = 0$, 
\eqref{Bsymm} yields \cite{DBKP,DGH91},
\beq
\barray{c}
\bomega\times{\bf r}
+\bbeta t+\bgamma
\\
\epsilon
\\
\bbeta\cdot\bx+\eta
\earray
\label{Bargvectfields}
\eeq
which span the one-parameter central extension by the mass of the Galilei group called the Bargmann group \cite{Barg54}, justifying the terminology.
In the quadratic case \eqref{oscipot} the isometries span the centrally extended Newton-Hooke group (possibly with broken rotations) \cite{ZHAGK}.

Remarkably, no similar Kaluza-Klein-type framework has been found so far for the 2-parameter central extension in the plane \cite{LLexo}.
 
\end{document}